\documentclass[twocolumn,trackchanges]{aastex701}
\usepackage{graphicx}
\usepackage{subfigure}
\usepackage{amsmath}
\usepackage{comment}
\usepackage{hyperref}

\newcommand{\Muv}{M_{\mathrm{UV}}}
\newcommand{\Msun}{\mathrm{M}_{\odot}}
\newcommand{\Mstar}{M_{\star}}
\newcommand{\Mhalo}{M_\mathrm{halo}}
\newcommand{\Myr}{\mathrm{Myr}}
\newcommand{\pc}{\mathrm{pc}}
\newcommand{\kpc}{\mathrm{kpc}}

\newcommand{\Zsun}{\mathrm{Z_{\odot}}}
\newcommand{\sfe}{\epsilon_\star}
\newcommand{\resbf}{m_{\rm baryon}\approx51,000~\Msun}

\newcommand{\JWST}{\textit{JWST}}

\received{\today}

\shorttitle{BonFIRE and CampFIRE}
\shortauthors{Samuel et al.}

\graphicspath{{./}{}}

\begin{document}

\title{Resolving galaxy formation in the early Universe with BonFIRE and CampFIRE}


\author[0000-0002-8429-4100]{Jenna Samuel}
\affiliation{Department of Astronomy, The University of Texas at Austin, Austin, TX 78712, USA}
\affiliation{Cosmic Frontier Center, The University of Texas at Austin, Austin, TX 78712, USA}
\email[show]{jenna.samuel@austin.utexas.edu}

\author[0000-0002-9604-343X]{Michael Boylan-Kolchin}
\affiliation{Department of Astronomy, The University of Texas at Austin, Austin, TX 78712, USA}
\affiliation{Cosmic Frontier Center, The University of Texas at Austin, Austin, TX 78712, USA}
\affiliation{Texas Center for Cosmology and Astroparticle Physics, Weinberg Institute, The University of Texas at Austin, Austin, TX 78712, USA}
\email{}

\author[0000-0002-1109-1919]{Robert Feldmann}
\affiliation{Department of Astrophysics, Universität Zürich, Zurich, CH-8057, Switzerland}
\email{}

\author[0000-0003-3729-1684]{Philip~F. Hopkins}
\affiliation{TAPIR, Mailcode 350-17, California Institute of Technology, Pasadena, CA 91125, USA}
\email{}

\author[0000-0003-4070-497X]{Guochao Sun}
\affiliation{CIERA and Department of Physics and Astronomy, Northwestern University, 1800 Sherman Ave., Evanston, IL 60201, USA}
\email{}

\author[0000-0003-0965-605X]{Pratik~J. Gandhi}
\affiliation{Department of Astronomy, Yale University, New Haven, CT 06520, USA}
\email{}

\author[0000-0003-2237-0777]{Alessandra Venditti$^{\dagger}$}
\affiliation{Department of Astronomy, The University of Texas at Austin, Austin, TX 78712, USA}
\affiliation{Cosmic Frontier Center, The University of Texas at Austin, Austin, TX 78712, USA}
\email{}
\begingroup
\renewcommand\thefootnote{}\footnotetext{$\dagger$ Cosmic Frontier Center Prize Fellow}
\endgroup

\author[0000-0002-6196-823X]{Xuejian Shen}
\affiliation{Department of Physics, Kavli Institute for Astrophysics and Space Research, Massachusetts Institute of Technology, Cambridge, MA 02139, USA}
\email{}

\author[0000-0003-0603-8942]{Andrew Wetzel}
\affiliation{Department of Physics \& Astronomy, University of California, Davis, CA 95616, USA}
\email{}

\author[0000-0002-3430-3232]{Jorge Moreno}
\affiliation{Department of Physics and Astronomy, Pomona College, Claremont, CA 91711, USA}
\affiliation{Carnegie Observatories, 813 Santa Barbara St., Pasadena, CA 91101, USA}
\email{}

\author[0000-0002-8984-0465]{Julian B.~Mu\~{n}oz}
\affiliation{Department of Astronomy, University of Texas at Austin, 2515 Speedway, Stop C1400, Austin, TX 78712, USA}
\affiliation{Cosmic Frontier Center, The University of Texas at Austin, Austin, TX 78712}
\email{}

\author[0000-0001-8855-6107]{Rachel K. Cochrane}
\affiliation{Jodrell Bank Centre for Astrophysics, University of Manchester, Oxford Road, Manchester M13 9PL, UK}
\email{}

\author[0000-0002-4900-6628]{Claude-Andr{\'e} Faucher-Gigu{\`e}re}
\affiliation{CIERA and Department of Physics and Astronomy, Northwestern University, 1800 Sherman Ave., Evanston, IL 60201, USA}
\email{}

\author[0000-0003-0212-2979]{Volker Bromm}
\affiliation{Department of Astronomy, The University of Texas at Austin, Austin, TX 78712, USA}
\affiliation{Cosmic Frontier Center, The University of Texas at Austin, Austin, TX 78712, USA}
\email{}

\author[0000-0001-8519-1130]{Steven L. Finkelstein}
\affiliation{Department of Astronomy, The University of Texas at Austin, Austin, TX 78712, USA}
\affiliation{Cosmic Frontier Center, The University of Texas at Austin, Austin, TX 78712, USA}
\email{}

\author[0009-0000-2266-1465]{Maria C. Straight}
\affiliation{Department of Astronomy, The University of Texas at Austin, Austin, TX 78712, USA}
\affiliation{Cosmic Frontier Center, The University of Texas at Austin, Austin, TX 78712, USA}
\email{}

\author[0000-0002-3531-4806]{Connor Painter}
\affiliation{Department of Astronomy, The University of Texas at Austin, Austin, TX 78712, USA}
\affiliation{Cosmic Frontier Center, The University of Texas at Austin, Austin, TX 78712, USA}
\email{}

\author[0000-0002-7541-9565]{Jonathan Stern}
\affiliation{School of Physics and Astronomy, Tel Aviv University, Tel Aviv 69978, Israel}
\email{}

\author[0000-0003-4298-5082]{James Bullock}
\affiliation{Department of Physics \& Astronomy, University of Southern California, Los Angeles, CA 90089, USA}
\email{}

\begin{abstract}
The abundance and rapid growth of galaxies at cosmic dawn revealed by the \textit{James Webb Space Telescope} challenges models of galaxy formation, motivating new simulations to uncover the processes driving early galaxy assembly. 
We present the first results from BonFIRE ($L\approx40$~cMpc, $m_{\rm baryon}\approx5\times10^4~\Msun$) and CampFIRE ($L\approx5$~cMpc, at both $m_{\rm baryon}\approx800~\Msun$ and $\approx6\times10^3~\Msun$), a suite of cosmological hydrodynamic simulations of early galaxy formation ($z\gtrsim6$) from the Feedback In Realistic Environments (FIRE) project, using the FIRE-3 model. 
We use a resampling procedure to combine the large statistics of BonFIRE with the higher resolution of CampFIRE and robustly predict galaxy properties over a wide dynamic range ($\Mstar\sim10^4$--$10^{10}~\Msun$). 
Galaxy formation in this suite emerges through clustered, bursty star formation, with halo-scale star formation efficiencies reaching 10--30\% in high-mass halos. 
A subset of low-mass halos also have surprisingly high efficiencies of $\gtrsim1\%$ and host ultra-compact galaxies with narrow age spreads. 
We predict galaxy UV luminosity functions at $9\lesssim~z\lesssim25$ in broad agreement with observations at $\Muv\gtrsim-19$, with a faint-end turnover at $\Muv\approx-14$, but we slightly overpredict the abundance of brighter galaxies. 
We find that UV luminosity variability in early galaxies is strongly mass-dependent, with halo-to-halo scatter dominating at low masses and contributing comparably to rapid temporal burstiness at $\Mhalo\gtrsim10^{10}~\Msun$. 
We also present first results from a simple Pop~III model with a top-heavy IMF, demonstrating broad agreement with independent Pop~III predictions and observational constraints.
\end{abstract}


\section{Introduction}\label{sec:intro}

Recent observations from the \textit{James Webb Space Telescope} (\JWST) have revealed a high-redshift galaxy population that challenges many pre-\JWST\ models of galaxy formation, particularly at $z\gtrsim10$. 
Across a range of observables, including ultraviolet (UV) luminosities, inferred stellar masses, and morphologies, early galaxies appear to assemble rapidly and exhibit complex, clumpy structures, often implying star formation efficiencies, burstiness, and UV emissivities that are difficult to reconcile with feedback-regulated models calibrated at lower redshift.
Proposed explanations span a wide range of physical processes that preferentially operate at high redshift, including enhanced star formation efficiencies, top-heavy or rapidly rotating stellar populations, stochastic variations in UV luminosity, and merger-driven growth, highlighting a growing tension between theoretical expectations and emerging observational constraints.

One of the first tensions to be recognized concerns the apparently too-rapid buildup of stellar mass at early times.
For instance, some galaxies have such high inferred stellar masses that they imply star formation efficiencies\footnote[1]{Throughout this work we use star formation efficiency in the halo-scale sense of stellar mass formed from the typically available gas mass: $\sfe\equiv \Mstar/(f_{\rm b}~\Mhalo)$.} significantly higher than typical feedback-regulated galaxies at low redshift
\citep[e.g.,][]{MadauDickinson2014,Finkelstein2023,Labbe2023,BK2023,Casey2024,Xiao2024,Shen2025}.
Although robustly measuring stellar masses is challenging, especially at high redshift where NIRCam probes only the rest-frame ultraviolet \citep{Sarrouh2024}, the large intrinsic UV luminosities of these galaxies pose a tension of their own. 
The observed UV luminosity function (UVLF) at $z\gtrsim10$ is up to an order of magnitude higher than most predictions from extrapolated empirical galaxy formation models \citep{Harikane2023,Helton2024,Finkelstein2024}, though some pre-\JWST\ predictions from simulations are indeed consistent with observations \citep[e.g.,][]{Ma2018b,Ma2019}.

The morphology of high-redshift galaxies has also turned out to be surprisingly complex. 
Many early galaxies exhibit clumpy and compact structures, suggesting that high-redshift star formation may have been driven by bursts occurring in massive star clusters \citep[e.g.,][]{Livermore2015,Atek2018,Neufeld2022,Adamo2024}. 
In fact, \citet{Harikane2024b} report that at least 70\% of galaxies at $z = 7$ are clumpy. 
They also note that compact galaxies exhibit stronger high-ionization lines compared to extended galaxies, potentially indicating different star formation processes or populations.
\citet{Mowla2024} also found evidence of early globular cluster formation in a lensed proto-Milky Way galaxy at $z = 8.3$, with indications of late-time bursts of star formation. 
Furthermore, \citet{Fujimoto2025a} identified a clumpy galaxy at $z = 6$ in a lensed \JWST\ field that additionally shows evidence of ordered rotation, with ALMA data revealing strong evidence for a rotating gas disk. 
A situation like this could potentially signal a weak feedback scenario wherein active star formation cannot fully disturb the disk.
Meanwhile, dynamically cold, rotationally supported galaxies have also been identified at $3\lesssim z\lesssim9$ via tracers like molecular gas, ionized gas, or optical morphology \citep{Kartaltepe2023,Rowland2024,Danhaive2025}. 
Collectively, these observations suggest that the earliest phases of galaxy formation may involve a broader diversity of physical processes and evolutionary pathways than predicted by many pre-\JWST\ models, while also highlighting the difficulty of uniquely identifying the physical mechanisms responsible for the rapid buildup and diversity of galaxies at high redshift.

A persistent challenge in interpreting any apparent agreement between theoretical predictions and observations is the presence of degeneracies in the physical processes that govern galaxy formation \citep[e.g.,][]{Finkelstein2023,Munoz2023,Jeong2025}. 
In particular, we focus on the fact that similar UV luminosity functions can arise from models with very different assumptions about star formation efficiency, stellar populations, feedback strength, and dust attenuation \citep{Katz2023,Katz2025}, making it difficult to determine which mechanisms drive the high luminosities observed at $z\gtrsim10$.

Many (but not all) degeneracies in high-redshift galaxy formation modeling can be broadly understood in terms of three interconnected physical pathways: local star formation efficiency, exotic stellar populations, and stochastic/halo-scale processes. 
At small scales, variations in the efficiency of star formation within dense gas can strongly impact galaxy luminosities and morphologies. 
Models invoking enhanced local efficiencies or feedback-free bursts allow gas to be rapidly converted into stars before feedback can regulate the process, producing short but intense star formation episodes \citep{Dekel2023,Li2024,Somerville2025}. 
A second pathway involves the properties of the stellar populations themselves: in low-metallicity environments, stars may exhibit enhanced rotation or follow a more top-heavy initial mass function, both of which increase the UV luminosity produced per unit stellar mass \citep{Liu_rotation2025}. 
Finally, processes operating on halo scales provide a third, complementary channel. 
The deeper potential wells and higher characteristic accelerations of early dark matter halos may increase gas densities and cooling rates, thereby enhancing the overall efficiency of baryon conversion into stars \citep{BK2024,Shen2024,Shen2025,Moreno2025}. 
Meanwhile, related scenarios emphasize stochastic processes such as mergers and variations in gas accretion rates, which can temporarily boost star formation without strongly altering global scaling relations \citep[e.g.,][]{Shen2023}.

Disentangling the relative contributions of these mechanisms (and more) is essential for identifying the physical origin of the rapid buildup of stellar mass in the early Universe, and it requires simulations that resolve the multiphase interstellar medium (ISM) and its coupling to large-scale halo dynamics.
In this paper, we introduce BonFIRE and CampFIRE, a novel suite consisting of a large-volume and two accompanying zoom-in simulations designed for this purpose. 
These simulations employ the latest generation of the FIRE model \citep[FIRE-3,][]{Hopkins2023} and include a simple treatment of Population~III star formation (Gandhi et al., in prep) with a top-heavy initial mass function (IMF), enabling us to simultaneously probe variations in star formation efficiency, stellar population physics, and their connection to dark matter halos.

We begin by reviewing the current state of high-redshift galaxy formation in FIRE-2 and related simulations.
Previous simulations from the FIRE project have successfully predicted multiple aspects of galaxy formation at high redshift. 
For example, studies using zoom-in simulations run with the FIRE-2 model have analyzed the morphologies, luminosities, and (post-processed) dust content of halos within the mass range $\Mhalo(z=5)\sim10^{8}$--$10^{12}~\Msun$ during $z=5$--$10$, finding broad agreement with contemporary observations \citep{Ma2018a,Ma2018b,Ma2019}. 
FIRE-2 simulations predict that star formation at early times is generally bursty due to cyclic feedback suppression, potentially explaining the unexpectedly high abundance of UV-bright galaxies \citep{Sun2023a,Sun2023b,Shen2023}. 
This burstiness in low-mass galaxies may be linked to the dispersion-dominated nature of the ISM and inner CGM \citep{Stern2021,Gurvich2023,Hopkins2023c,Sun2026}. 
The metallicities of high-redshift galaxies predicted by FIRE-2 are also in good agreement with recent \JWST\ observations, with the weak evolution of the mass–metallicity relation (MZR) at high redshift naturally arising from bursty star formation \citep{Marszewski2024,Marszewski2025}.

Furthermore, the FIREbox and FIREbox-HR simulations target galaxies in the high-redshift Universe at a resolution typical of zoom-in galaxies but in a (22.1 Mpc)$^3$ cosmological volume with FIRE-2 physics \citep{Feldmann2023,Feldmann2025}. 
FIREbox-HR reproduces the observed UV luminosity functions and UV luminosity densities, in broad agreement with \JWST\ data. 
In particular, the agreement was found to result from the weak dependence of star formation efficiency on halo mass in FIRE-2, which leads to an increasing contribution of lower-mass halos to the overall UV luminosity density with increasing redshift.

Despite these advances, the tension between the measured and predicted galaxy luminosities at very early times ($z>10$) remains a significant challenge for many theoretical models. 
While some large-volume simulations, such as those used in \citet{Keller2023} and \citet{McCaffrey2023}, report little discrepancy between theory and observations, the low mass resolution of these simulations may be insufficient to capture the hierarchical formation of galaxies, particularly in the early Universe. 
These models likely under-resolve the formation of the (low-mass) first galaxies, which could be crucial for understanding the rapid assembly of massive galaxies at high redshift.
Furthermore, the spectroscopic confirmation of some of the highest-redshift galaxies to date, such as GS-z14 at $z\sim14$, casts renewed doubt on the ability of current simulations to reproduce the highest-mass early galaxies \citep{McCaffrey2025}.

In light of these developments, it is clear that there is still much to be learned about galaxy formation in the early Universe \citep[e.g.,][]{BY2011,DF2018}. 
In particular, understanding the star formation efficiencies, ISM structure, and the rapid assembly of galaxies and elemental enrichment in the early Universe is critical for reconciling theoretical models with the latest observational results.

This work presents the first results from BonFIRE and CampFIRE, a new large-volume simulation suite of the early Universe that resolves the multiphase structure of the ISM and models the range of baryonic physics and stellar feedback processes necessary to produce a galaxy population in broad agreement with high-redshift observations from \JWST. 
BonFIRE models a (41.2 cMpc)$^3$ cosmological volume from $z=30$ to $z=9$. 
The two CampFIRE runs model a (5 cMpc)$^3$ subregion of BonFIRE at higher resolution, with one run pushed to $z=6$ to study the evolution of galaxies through the end of cosmic reionization.

This paper is organized as follows. 
Section~\ref{sec:sims} presents details of the BonFIRE and CampFIRE simulations, including the specific FIRE-3 physics and Population~III star modeling, and postprocessing. 
Section~\ref{sec:results} contains our main results, including the stellar mass-halo mass relation, star formation efficiency, galaxy morphology, UV luminosity, and Population~III star formation. 
We discuss the implications of our results in Section~\ref{sec:discussion}, and our conclusions can be found in Section~\ref{sec:conclusions}.

\begin{figure*}
    \centering
    \includegraphics[width=\textwidth]{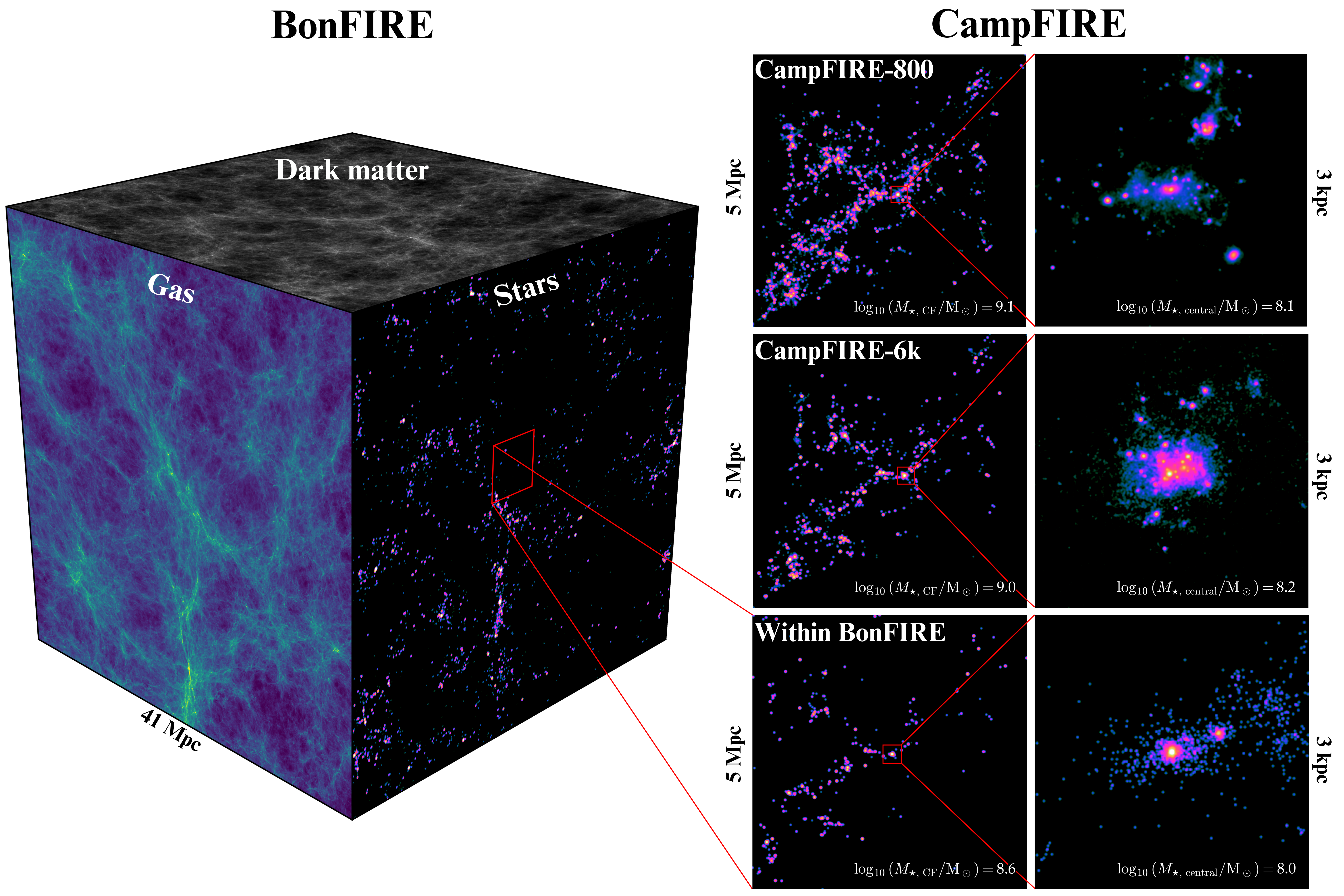}
    \caption{\textit{Left:} Stellar, gas, and dark matter density projected through a 5 cMpc slice in the BonFIRE volume. 
    \textit{Right:} Stellar density in the CampFIRE volume at the three resolutions simulated in the suite (left column, resolution increasing from bottom to top). 
    Stellar density around the central halo in the CampFIRE region at each resolution (right column).
    All labeled distances are comoving.}
    \label{fig:box_viz}
\end{figure*}

\section{Simulations}\label{sec:sims}

We present the BonFIRE, CampFIRE-6k, and CampFIRE-800 simulations, which comprise a new large-volume simulation suite of the early Universe using the Feedback in Realistic Environments \citep[FIRE,][]{Hopkins2014,Hopkins2018,Hopkins2023} galaxy formation framework\footnote[2]{For more information, publications, and visualizations from the FIRE collaboration, visit \url{https://fire.northwestern.edu/}.}.
We ran the simulations with the updated FIRE-3 models, which we describe further in Section~\ref{sec:sims_fire3}.
We generated initial conditions with \textsc{MUSIC} \citep{Hahn2011} using the cosmological parameters measured by \citet{PlanckCollaboration2018} for the six parameter standard $\Lambda$CDM model: 
$\Omega_{\rm m} = 1-\Omega_{\Lambda}= 0.3158$, 
$\Omega_{\rm b} = 0.0494$, $h = 0.6732$, $\sigma_8 = 0.8120$, and $n_{\rm s} = 0.96605$.

\subsection{BonFIRE}\label{sec:bonfire}

BonFIRE is a cosmological hydrodynamic simulation of high-redshift galaxy formation within a volume of $V=(41.2~\mathrm{cMpc})^3$ at approximately cosmic mean density (see Appendix~\ref{app:HOF}). 
BonFIRE contains $2\times2048^3$ particles at mass resolutions of $m_{\rm baryon}\approx51,\!000~\Msun$ and $m_{\rm DM}\approx273,\!000~\Msun$. 
Star and dark matter particles have Plummer-equivalent gravitational force softenings of $8~\rm{pc}~(\frac{10}{1+\mathit{z}})$ at $z\geq9$ (fixed to 8~pc at $z<9$) and $80~\rm{pc}~(\frac{10}{1+\mathit{z}})$ at $z\geq9$ (fixed to 80~pc at $z<9$), respectively. 
The gas cell kernel size and hydrodynamic smoothing length is adaptive, scaling to higher resolution at higher gas density down to a minimum of 0.03~pc. 
At $z=9$ the typical gas intercell spacing in BonFIRE varies from $3$ to $300~\rm{pc}$ (physical units) in the ISM, versus $1$ to $3~\kpc$ in the intergalactic medium (IGM).

Compared to FIREbox(-HR) \citep{Feldmann2023,Feldmann2025}, BonFIRE's volume is 6.5 times larger, and it can thus be used to examine the formation of more massive galaxies than are present in FIREbox. 
BonFIRE has roughly equal mass and spatial resolution compared to FIREbox, whereas, FIREbox-HR surpasses them both with about eight times better mass resolution and two times better spatial resolution. 
Because of the tradeoff between resolution and volume, BonFIRE has the same number of particles as FIREbox-HR ($2\times2048^3$). 
FIREbox was run to $z=0$ and FIREbox-HR to $z=6.3$, compared to BonFIRE's final redshift of $z=9$. 
We also remind the reader that FIREbox(-HR) was run with FIRE-2, whereas the BonFIRE and CampFIRE suite uses FIRE-3.

BonFIRE cost approximately 5.25 million CPU hours to run to $z=9$ ($\approx550$ Myr of cosmic time). 
The raw particle data output for BonFIRE across its 31 snapshots totals 27 TB. 
We discuss snapshot cadence in Section~\ref{sec:snapshots}.

\begin{deluxetable*}{lccc}
\tablecaption{Properties of the three simulations introduced in this work. We quote spatial resolutions in physical units at $z=9$. Gas smoothing is adaptive and we quote its minimum possible value. See Sections \ref{sec:bonfire} \& \ref{sec:campfire} for typical gas smoothing length ranges.}
\tablehead{
\colhead{Quantity} & \colhead{BonFIRE} & \colhead{CampFIRE-6k} & \colhead{CampFIRE-800}
}
\startdata
Box side length (cMpc)                     & 41  & 5   & 5   \\
Particle number                           & $2\times2048^3$ & $2\times500^3$ & $2\times1000^3$ \\
$m_{\rm baryon}$ ($M_\odot$)              & $5.1\times10^4$ & $6.3\times10^3$ & $8.0\times10^2$ \\
$m_{\rm DM}$ ($M_\odot$)                  & $2.7\times10^5$ & $3.4\times10^4$ & $4.7\times10^3$ \\
$\varepsilon_{\rm star}$ (pc)                      & 8   & 4   & 2   \\
$h_{\rm gas}^{\rm min}$ (pc)              & 0.03 & 0.2 & 0.1 \\
$\varepsilon_{\rm DM}$ (pc)               & 80  & 40  & 20  \\
Final redshift ($z$)                        & 8.9 & 6.0 & 8.9 \\
$N_{\rm galaxy}$ at final redshift\tablenotemark{a} & 38,146 (9,547) & 1,148 (615) & 1,654 (1,110) \\
Most massive galaxy $M_*$ ($M_\odot$)     & $5.9\times10^9$ & $1.2\times10^9$ & $1.2\times10^8$ \\
\enddata
\tablenotetext{a}{Values (in parentheses) indicate galaxies with $\geq1$ (10) star particles.}
\label{tbl:sim}
\end{deluxetable*}

\subsection{CampFIRE}\label{sec:campfire}

We reran an initially cubic (5 cMpc)$^3$ subregion of BonFIRE at two different higher resolutions using the zoom-in technique \citep{Onorbe2014} with the same physical models and cosmological parameters to test for convergence in galaxy properties at higher resolution and extend redshift coverage through reionization. 
The subregion simulations are named CampFIRE-6k and CampFIRE-800, with mass resolutions of $m_{\rm baryon}\approx6,300~\Msun$ and $m_{\rm baryon}\approx800~\Msun$, respectively. 
Star and dark matter particles in CampFIRE-6k have Plummer-equivalent gravitational force softenings of $4~\rm{pc}~(\frac{10}{1+\mathit{z}})$ at $z\geq9$ (fixed to 4~pc at $z<9$) and $40~\rm{pc}~(\frac{10}{1+\mathit{z}})$ at $z\geq9$ (fixed to 40~pc at $z<9$), respectively. 
At $z=9$ the typical gas intercell spacing (physical units) in CampFIRE-6k varies from 1 to $100~\rm{pc}$ in the ISM, versus $300~\rm{pc}$ to $1~\rm{kpc}$ in the IGM. 
Star and dark matter particles in CampFIRE-800 have gravitational force softenings of $2~\rm{pc}~(\frac{10}{1+\mathit{z}})$ at $z\geq9$ (fixed to 2~pc at $z<9$) and $20~\rm{pc}~(\frac{10}{1+\mathit{z}})$ at $z\geq9$ (fixed to 20~pc at $z<9$), respectively. 
At $z=9$ the typical gas intercell spacing (physical units) in CampFIRE-800 varies from $0.3$ to $100~\rm{pc}$ in the ISM, versus $150~\rm{pc}$ to $1~\rm{kpc}$ in the IGM.

We selected the CampFIRE region to maximize the amount of information obtained from it for minimal computational cost. 
We chose a region centered on a halo of $\Mhalo\sim10^{10}~\Msun$ (at $z\sim9$), without a halo of similar mass within $5~\rm{cMpc}$, in order to probe the formation of a single massive galaxy within an overdense peak of the dark matter distribution alongside a statistical sample of low-mass galaxies. 

The box side length of the CampFIRE region in the initial conditions is 5 cMpc, such that its initial volume is only $\approx0.2\%$ of the BonFIRE volume. 
Relative to BonFIRE, which is approximately cosmic mean density, the CampFIRE subregion is slightly overdense, with density contrast of
\begin{equation}
    \delta \equiv \frac{\rho_{\rm CF}}{\rho_{\rm BF}} - 1 \approx0.4~.
\end{equation}

CampFIRE-6k cost approximately 365,000 CPU hours to run to $z=6$ ($\approx1$ Gyr of cosmic time). 
The raw particle data output for CampFIRE-6k across its 56 snapshots totals $6.5~\rm{TB}$. 
CampFIRE-800 cost approximately 860,000 CPU hours to run to $z=9$ ($\approx550$ Myr of cosmic time). 
The raw particle data output for CampFIRE-800 across its 31 snapshots totals $11~\rm{TB}$.

Figure~\ref{fig:box_viz} shows the projected stellar, dark matter, and gas density at $z\sim9$ through a 5 cMpc slice of the BonFIRE volume (left) and in the projected stellar density in the CampFIRE subregion at the three resolutions we simulate (right). 
Stellar mass is distributed throughout the volume with some notable overdensities where massive galaxies and protoclusters exist. 
The stellar densities in the CampFIRE region display clear resolution dependence, whereby there are more galaxies at higher resolution.
The rightmost column of images are the projected stellar density of the central halo in the CampFIRE region, which further emphasize the resolution dependence of the buildup of stellar mass.

We summarize key properties for each of our three simulation runs in Table~\ref{tbl:sim}.

\subsection{Snapshot cadence}\label{sec:snapshots}

It is important to be able to examine galaxy dynamics and star formation and their effects on gas via stellar feedback with relatively high simulation snapshot cadence.
Dynamical timescales scale as $\rho^{-1/2}\sim(1+z)^{-3/2}$ and are shorter at high redshift, resulting in galactic dynamical timescales that are $\sim10$ Myr at $z\sim10$. 
Furthermore, high-redshift star formation in FIRE is bursty \citep[e.g.,][]{Muratov2015,AnglesAlcazar2017,Stern2021,Gurvich2023,Sun2023a,Sun2023b} and hence the UV luminosity of individual early galaxies may vary on timescales of $10-100$ Myr \citep{Sparre2017,Flores2021,Sun2023b}.
We therefore save full simulation outputs every $15~\rm{Myr}$ from $z=30$ to $z=9$ (spanning $\approx450~\Myr$, 31 snapshots in total), enabling dynamical studies of gas accretion, bursty star formation, stellar feedback, and mergers.
For CampFIRE-6k we extend the snapshot range with the same cadence to $z=6$ (56 snapshots spanning $\approx830~\Myr$ in total).
We note that most snapshots are not at integer redshifts, but we often round to the nearest integer in the text and figures for simplicity.

\subsection{FIRE-3 models}\label{sec:sims_fire3}

We ran BonFIRE and CampFIRE with the meshless finite mass (MFM) mode of the GIZMO \citep{Hopkins2015} hydrodynamics code\footnote[3]{For code documentation, visit \url{http://www.tapir.caltech.edu/~phopkins/Site/GIZMO.html}.} under the updated FIRE-3 model for galaxy formation physics \citep{Hopkins2023}.
Gravitational forces in GIZMO are solved using an improved version of the N-body \texttt{GADGET-3} Tree-PM solver \citep{Springel2005}, where the gravitational force softening of gas particles automatically adapts to their hydrodynamic smoothing length.

The FIRE simulations incorporate state-of-the-art numerical methods for fluid dynamics, star formation, and stellar feedback, which enable adaptive hydrodynamic force softening and ensure the conservation of mass, energy, and momentum \citep{Hopkins2015,Hopkins2018}. 
The model accounts for radiative heating and cooling of gas, thermochemical coupling, and non-equilibrium ionized/atomic/molecular/dust chemistry for 11 separately-evolved elements spanning a temperature range of approximately $1$--$10^{10}$~K \citep{Hopkins2023}. 
FIRE also explicitly includes a detailed set of stellar feedback processes: Type II (core-collapse) and Type Ia (white dwarf) supernovae, continuous stellar mass loss and momentum injection from O/B and AGB stellar winds, photoionization and photoelectric heating, and radiation pressure in multiple bands (FIR through ionizing).

We use the latest version of the code, FIRE-3, which features key updates to star formation and gas physics that are relevant for early galaxy formation \citep{Hopkins2023}. 
In FIRE-3, star formation occurs in gas that is self-gravitating, Jeans-unstable, and undergoing converging flow. 
Unlike in FIRE-2, star-forming gas is not required to meet an arbitrary density threshold (previously, $n\gtrsim10^3~\rm{cm}^{-3}$), nor must it be molecular. 
The star formation criteria are thus relaxed in FIRE-3 compared to FIRE-2. 
We emphasize that the identification of star-forming gas in FIRE-3 does not rely on possible chemical tracers of high density in the ISM like molecular gas, but rather on a set of direct \textit{a priori} conditions for gravitational collapse.

FIRE-3 introduces updated nucleosynthetic yield models for both core-collapse and Type Ia supernovae, an updated Type Ia rate (delay time distribution), and gas that is initialized \textit{without} a metallicity floor, at primordial elemental abundances.  
FIRE-3 also includes a more accurate treatment of the molecular fraction of gas and the cooling processes within molecular gas, which is crucial for modeling cooling in primordial gas \citep{Hopkins2023}. 
These updates lead to earlier star formation in low-mass halos and improved agreement between the $\Mstar$--$\Mhalo$ relation of FIRE-3 zoom-in simulations and that of empirical models at the low-mass end \citep[e.g.,][]{Nadler2020}, compared to FIRE-2.

Furthermore, FIRE-3 uses the updated metagalactic, spatially uniform UV background (UVB) from \citet{FaucherGiguere2020} to approximate the effects of global reionization.
The updated UVB reaches a midpoint in HI reionization in the intergalactic medium at $z\sim8$, later than the FIRE-2 version \citep[$z\sim10$,][]{FaucherGiguere2009} and in better agreement with constraints from \citep{PlanckCollaboration2018}.

All simulations in this work use the explicit IMF sampling scheme from \cite{Su2018} and \cite{Wheeler2019} wherein every star particle spawns a discrete number of massive O stars ($8$--$100~\Msun$) according to the chosen IMF. 
The spawning of massive O stars is distributed continuously in time, to prevent unphysical ``instantaneous'' initiation of massive star formation when a new star particle is formed, spread over a timescale set for each star particle to twice the free fall time of the gas cell (at formation) from which it formed, motivated by high-resolution simulations of individual star-forming clouds \citep{Grudic2022}.

FIRE-3 also includes updates to the gravity tree with more accurate opening and timestep criteria \citep{Grudic2020,Hopkins2023b}. 
We note that we do \textit{not} model magnetic fields in any of the simulations presented here, which could reduce star formation efficiency but also yield higher density cores in star-forming clouds \citep[e.g.,][]{Hennebelle2019}.

\subsubsection{Population III star model}\label{sec:sims_pop3}

Like most FIRE simulations, BonFIRE and CampFIRE employ a Kroupa IMF of the following form for metal-rich stars:
\begin{equation}
\phi(M) = A
 \begin{cases} 
      M^{-0.3}, & 0.01~\Msun < M < 0.08~\Msun, \\
      M^{-1.3}, & 0.08~\Msun < M < 0.5~\Msun, \\
      M^{-2.3}, & 0.5~\Msun < M < 100~\Msun  
   \end{cases}
\end{equation}
where $\phi(M)\equiv dN/dM$.

Note that in FIRE-3, stellar feedback (including stellar winds, radiation, and SNe yields) does depend on metallicity, as described in \citet{Hopkins2023}. 
However, the standard FIRE-3 models currently do not include an explicit model for Population~III (Pop~III) stars as distinct from ``very metal-poor'' Pop~II stars \citep[e.g.,][]{Bromm2013,KlessenGlover2023}. 
We therefore extend our star formation and evolution model following the simple Pop~III model in Gandhi et al. (in prep), which assumes that the IMF becomes top-heavy below a critical metallicity $Z_{\rm crit}=10^{-5}~Z_{\odot}$ \citep[e.g.,][]{Karlsson_RMP2013}.
Specifically, we follow \citet{Su2018,Wheeler2019}, and multiply all feedback rates and mass loss rates associated with massive stars (e.g.\ UV radiation, O/B winds, core-collapse supernova rates) at $Z < Z_{\rm crit}$ by an arbitrary ramp function $f_{Z} \sim 1+90~\log_{10}{(Z/10^{-5}~Z_{\odot})}$ (capped to a maximum at $Z_{\rm min}<10^{-7}~Z_{\odot}$) until the mass of the star particle is exhausted, chosen so the full mass of the star particle will go into core-collapse SNe (with 1~SNe per $\sim 10~{\rm M}_{\odot}$ formed) once the metallicity falls significantly below $Z_{\rm crit}$. 
The functional form and choice of $Z_{\rm crit}$ are arbitrary here, though motivated by different theoretical models (see \citealt{Karlsson_RMP2013}).

We do not further modify the feedback model or stellar yields from the fiducial FIRE-3 model, so we do not include any feedback channels that may be specific to Pop~III stars, such as pair-instability supernovae (PISNe) or hypernova \citep[e.g.,][]{Jaacks2018}, nor do we utilize different nucleosynthetic yields for the Pop III stellar winds and supernovae compared to default FIRE-3.

We do not differentiate between Pop~II and Pop~I stars in our modeling, so we refer to them collectively as Pop~II throughout this paper. 
Though we have not enforced an explicit form of the Pop~III IMF or yield model, our model approximates some of the most important baseline effects of Pop~III by assuming the Pop~III star particles have an implicitly top-heavy IMF. 
In comparison to our Pop~II star particles, this assumption leads to the following outcomes for Pop~III star particles: shorter lifetimes, a greater number of supernovae per star particle, and increased radiative/wind feedback per star particle. 
This means that Pop~III particles outshine their Pop~II counterparts in the UV at fixed age, and that Pop~III particles chemically enrich their surroundings more effectively than Pop~II.

The effects of varying the functional form of the ramp function and $Z_{\rm crit}$ that we use, as well as the effects of the Pop~III model on $z=0$ galaxies are studied in Gandhi et al.~(in prep), using a suite of zoom-in FIRE-3 simulations. 
We discuss Pop~III star formation and ``Pop~III galaxies" in our simulations in Section~\ref{sec:results_pop3}.

\subsection{Post-processing}

\subsubsection{Halo catalogs and merger trees}

We identify dark matter (sub)halos in our simulations at each snapshot with the \textsc{ROCKSTAR} halo finder \citep{Behroozi2013a}. 
We run \textsc{ROCKSTAR} on only the dark matter particles and assign star particles to (sub)halos in additional post-processing steps described below. 
We employ a default minimum threshold of 30 dark matter particles per halo to exclude spurious particle groupings and capture the halo mass function to the lowest resolved masses in each simulation. 
We construct merger trees using \texttt{CONSISTENT-TREES} \citep{Behroozi2013b}. 
We defer analysis of the gas and gas catalogs to future work.

\subsubsection{Star assignment}

We assign star particles to halos in the order of most massive halo to least massive halo. 
Our method is adapted from \cite{Necib2019} and \cite{Samuel2020}. 
We require star particles to be within a distance equal to 50\% of the halo radius ($R_{\rm 200m}$, where `200m' indicates a measurement relative to 200 times the mean matter density of the Universe) from the halo center, and to have halo-centric velocities less than five times the greater of either the halo's maximum circular velocity or the halo's velocity dispersion. 
We tested three different apertures for assigning stars to halos: $R_{\rm 200m}$, $0.5~R_{\rm 200m}$, and $0.15~R_{\rm 200m}$. 
We found that two important features of the galaxy population may depend on the choice of aperture: the number of low mass galaxies (particularly satellites) and the stellar masses of massive galaxies. 
The smallest aperture overlooks some low-mass compact galaxies entirely ($\sim25\%$ of all star particles go unassigned), while the largest aperture mis-assigns some of those stars to massive halos (though it does assign $\approx99\%$ of all star particles to some halo). 
The stellar masses of massive galaxies (and hence their star formation efficiencies and sizes) are significantly underestimated when using the smallest aperture. 
We chose the intermediate aperture of $0.5~R_{\rm 200m}$ to mitigate these effects.

At $z=9$, BonFIRE contains 9,547 galaxies that are resolved with at least 10 star particles ($\Mstar\gtrsim5\times10^5~\Msun$), and 38,146 galaxies with at least one star particle ($\Mstar\gtrsim5\times10^4~\Msun$). 
Nearly all of the galaxies in BonFIRE live in halos of $\Mhalo\gtrsim10^8~\Msun$ at $z\sim9$, which are resolved with $\gtrsim360$ dark matter particles (see Appendix~\ref{app:HOF}). 
This means that even if a halo at $\Mhalo\sim10^8~\Msun$ has only formed a single star particle, approximately $360$ gas cells have resolved its formation history, assuming it contained the cosmic baryon fraction worth of gas cells at some point \citep{Hopkins2018}. 
Thus, we consider the presence of a single star particle to be a first-order indicator of galaxy formation \citep{Moreno2025}, and the presence of $\gtrsim10$ star particles to indicate ``resolved'' galaxy formation. 
We do not mean that the internal dynamics, star formation history, or metal-mixing of a galaxy with only 10 star particles is resolved, simply that the existence of a galaxy is resolved \citep{Hopkins2018,Wheeler2019}. 
For analyses in which we examine internal dynamics we take a more conservative limit of $\gtrsim10^3$ star particles.

\subsubsection{UV luminosity and dust attenuation}\label{sec:sims_uv}

We calculate UV luminosity for a simulated galaxy by interpolating BPASS and Cloudy data tables \citep{StanwayEldridge2018,Xiao2018} according to each member star particle's age and metallicity and summing the particles' rest-frame 1500~\AA~luminosities, following \citet{Sun2023b}. 
The stellar spectral templates we use are from BPASS v2.2, assuming binary stars and BPASS's default broken power-law IMF with indices $\alpha_1 = -1.3$ and $\alpha_2 = -2.35$ over stellar masses between $0.1$--$0.5~M_{\odot}$ and $0.5$--$300~M_{\odot}$, respectively. 
We include nebular emission (including major rest-UV/optical lines and continuum) assuming ISM conditions ($n_\mathrm{H}=100~\mathrm{cm^3}$ and $U=0.01$) consistent with high-redshift observations \citep{Reddy2023}. 
For star particles that fall into our designated Pop III range ($\log(\rm{Z/Z_{\odot})<-5}$, see Section~\ref{sec:sims}) we use Yggdrasil spectral templates \citep{Zackrisson2011}, instead of BPASS, to account for the elevated UV luminosity of extremely metal-poor, young, massive stars. 
Specifically, we use the Yggdrasil Pop III.2 templates generated with a covering fraction of 0.5 to account for nebular emission. 
In all case s we linearly scale the SED with the star particle mass, to account for the difference between the template mass of $10^6~\Msun$ and the star particle mass.

We include an analytical dust correction to our UV magnitudes as a function of halo mass, with a functional form similar to the fit reported by \citet{Feldmann2025}. 
Based on \textsc{SKIRT} radiative transfer calculations for a sample of $\sim100$ BonFIRE galaxies at $z=9$ (see Appendix~\ref{app:dust}), we adopt the form
\begin{equation}
\Delta M_{\rm UV}(\Mhalo) = 1.03\exp\left[\frac{\log \Mhalo - 10.79}{0.38}\right].
\end{equation}
This correction reduces the UV brightness by an amount that increases steeply with halo mass and is applied uniformly at fixed $\Mhalo$. 
Compared to \citet{Feldmann2025}, our fit rises more steeply and at lower halo masses. 
The correction primarily affects the bright end of the UVLF, which we show throughout as solid (dust-corrected) versus dotted (uncorrected) curves.

We emphasize that this prescription is approximate, and that the underlying assumption of our correction is that dust mass scales with baryonic mass, and hence halo mass to first order. 
A fully self-consistent treatment of dust attenuation will require more detailed modeling of the gas content within the simulations in the future.

\subsubsection{Resolution correction}\label{sec:sims_rescorr}

In Appendix~\ref{app:restests}, we detail how galaxy properties such as stellar mass, size, UV luminosity, and stellar age dispersion are affected by resolution.
Here, we briefly describe our resolution correction resampling method where we combine the sample size of halos in BonFIRE with the high-resolution CampFIRE-800 galaxy properties to ameliorate resolution issues in low-mass galaxies and dark halos in BonFIRE.
We describe the method in detail in Appendix~\ref{app:resampling}.

Our resolution correction accounts for the effects of artificially short bursts of star formation on UV luminosity that occur in BonFIRE due to its limited mass resolution and the dependence of the pace of star formation on the ability of the simulation to resolve the Jeans mass of a self-gravitating cloud. 
We implement the correction by statistically resampling the stellar masses, UV luminosities, and sizes of galaxies within resolved halos in BonFIRE using properties from all galaxies with at least 1 star particle in CampFIRE-800.

\begin{figure*}
    \centering
    \begin{tabular}{cc}
    \subfigure{\includegraphics[width=0.51\textwidth]{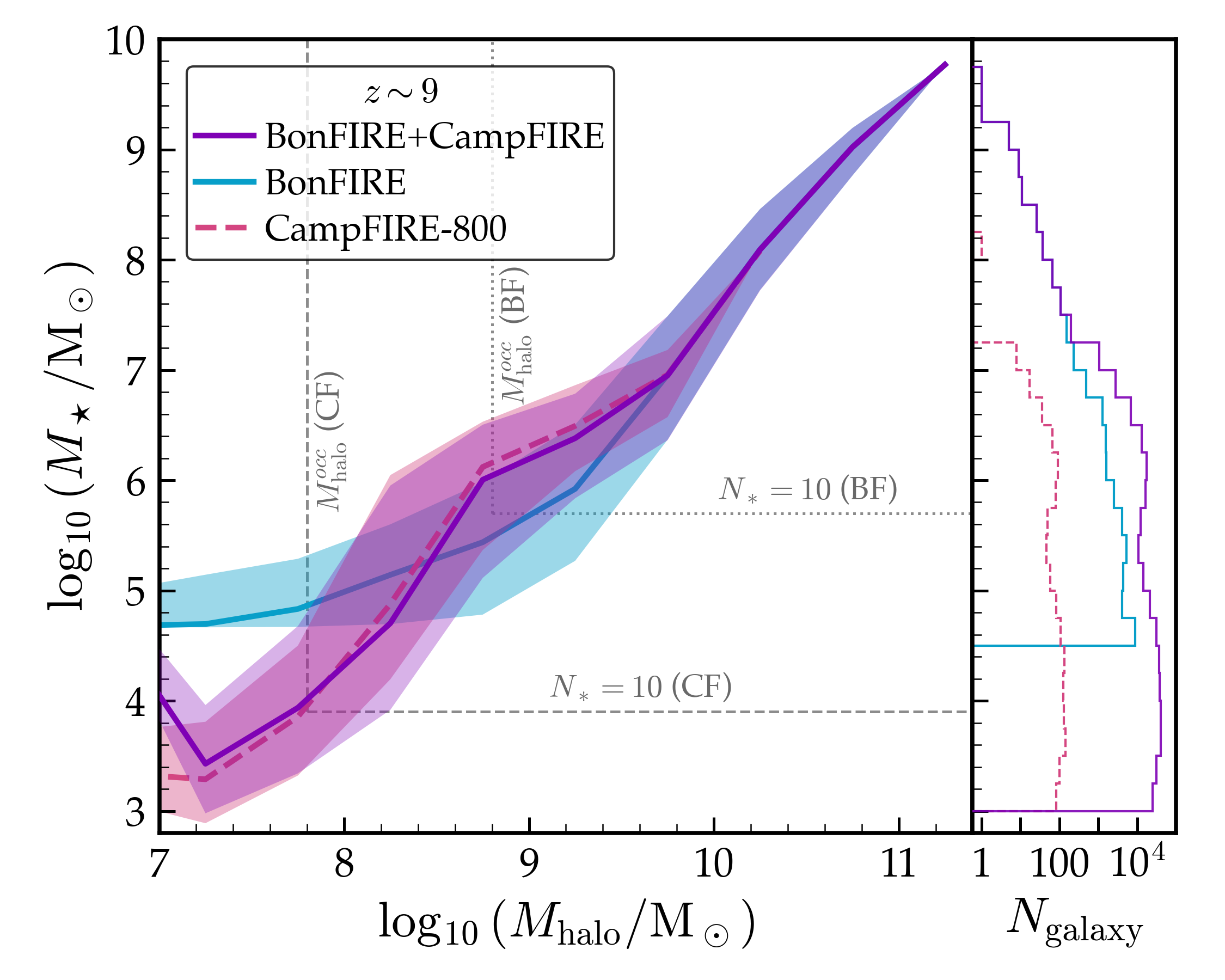}} &
    \subfigure{\includegraphics[width=0.48\textwidth]{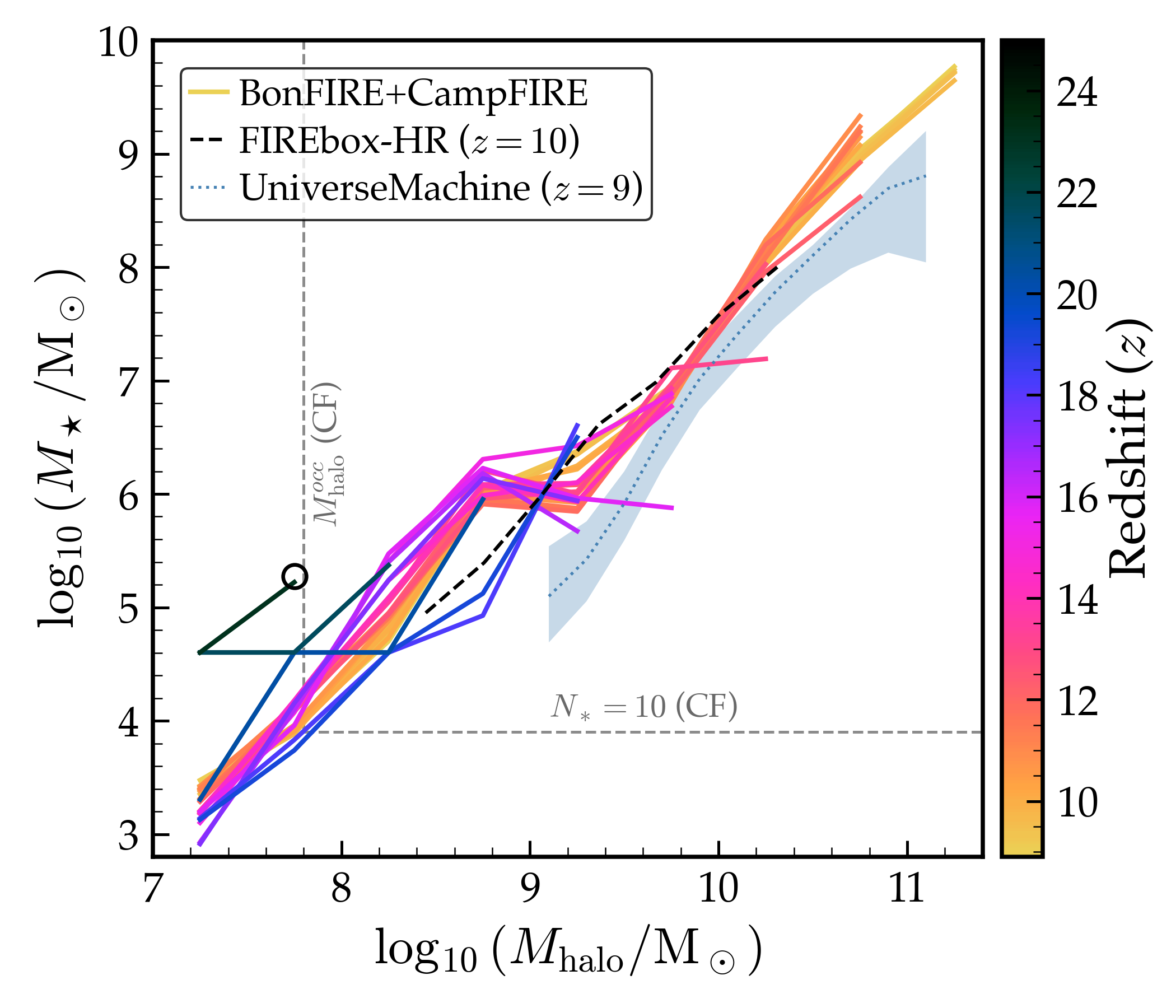}}
    \end{tabular}
    \caption{\textit{Left:} The stellar mass--halo mass (SMHM) relation in BonFIRE+CampFIRE, BonFIRE, and CampFIRE-800 at $z\sim9$. 
    Lines are median $\Mstar$ in each $\Mhalo$ bin and shaded regions show the 68\% scatter about the median. 
    BonFIRE+CampFIRE extends the coverage of the SMHM relation in the simulations down to $\Mstar\sim10^4~\Msun$ at $\Mhalo\sim10^8~\Msun$, whereas BonFIRE on its own suffers from a resolution-induced flattening in the relation at $\Mhalo\lesssim10^{9.5}~\Msun$. 
    The feature at $\Mhalo\sim10^{8.5}$--$10^{9.5}~\Msun$ and $\Mstar\sim10^6~\Msun$ represents a population of ultra-compact galaxies formed in efficient bursts and present throughout all simulations in the suite. 
    \textit{Right:} Evolution of the median BonFIRE+CampFIRE SMHM relation at $9\lesssim z\lesssim25$. 
    At $z\sim25$ there are only a few galaxies populating a single halo mass bin, which we show as an unfilled circle. 
    The BonFIRE+CampFIRE relation exhibits remarkably little evolution over the first $\sim500~\rm{Myr}$ of the Universe, as the variations at early times ($z\gtrsim20$) are due to small numbers of galaxies.
    We show results from FIREbox-HR \citep{Feldmann2025} at $z=10$ (black dashed line) and the empirical model UniverseMachine (UM; \citealt{Behroozi2019}) at $z=9$ (blue dotted line and shaded region).
    BonFIRE+CampFIRE probes $\sim1$ dex lower in $\Mstar$ than either FIREbox-HR or UM, and $\sim2$ dex higher than FIREbox-HR. 
    The median BonFIRE+CampFIRE relations are consistent with the other models (within typical scatter shown at left) at $\Mhalo\gtrsim10^9~\Msun$. 
    However, near the ultra-compact galaxy feature, we find that BonFIRE+CampFIRE produces galaxies that are 1--2~dex higher in stellar mass than galaxies in the other models.}
    \label{fig:smhm}
\end{figure*}

Our approach preserves the global star formation efficiency and UV luminosity distribution while mitigating numerical artifacts arising from resolution-dependence. 
The net effect is that the resampled ``BonFIRE+CampFIRE'' (B+C) galaxy population spans stellar masses of $\Mstar\sim10^4$--$10^{10}~\Msun$, allowing us to make more robust predictions for the faint end of the UVLF.
We verified that the correction preserves key quantities such as the total stellar mass within the BonFIRE volume, the $\Mstar$--$\Mhalo$ relation, and the $\Muv$--$\Mhalo$ relation across the overlapping halo mass range between BonFIRE and CampFIRE (see Figure~\ref{fig:smhm} and \ref{fig:MuvMhalo}).

\section{Results}\label{sec:results}

We present our results roughly in order of increasing post-processing complexity, beginning with relatively direct simulation outputs such as galaxy stellar masses and sizes, and progressing toward derived quantities that require additional modeling assumptions, such as UV luminosity functions. 
This structure is intended to first establish the core physical behavior of galaxy formation in BonFIRE and CampFIRE before introducing comparisons to observables that more significantly depend on resolution and model choice. 
Section~\ref{sec:results_mstar} describes the buildup of stellar mass and its dependence on halo properties. 
Section~\ref{sec:results_morph} describes galaxy morphologies. 
Section~\ref{sec:results_uv} presents the predicted UV luminosity functions. 
Section~\ref{sec:results_pop3} shows our predictions for Pop~III star formation. 
We begin below with a brief introduction to star formation in the simulations.

Star formation in BonFIRE proceeds in distinct stages that reflect the hierarchical growth of dark matter halos and the evolving ability of gas to cool and condense at early times. 
The first star formation in BonFIRE (at least one star particle) occurs in halos near the atomic cooling threshold ($M_{\rm halo} \sim 10^8~\rm{M_\odot}$) at $z \sim 25$, when the Universe is only about $100$--$150~\rm{Myr}$ old, producing short-lived bursts that rapidly enrich their surroundings. 
Star formation in CampFIRE begins slightly later, at $z \sim23$, as this smaller box does not capture the rarer, higher density regions of BonFIRE. 
This early onset of star formation aligns with theoretical expectations \citep[e.g.,][]{BrommLarson2004}, as the first stars are thought to form in minihalos ($\Mhalo\sim10^5$--$10^6~\Msun$) at $z\gtrsim20$--$30$ \citep{Haiman1996,Tegmark1997}. 
However, we do not resolve galaxy formation in minihalos in BonFIRE, and it is only somewhat resolved in CampFIRE-800.

As structure formation progresses, star formation becomes increasingly clustered within dense filaments feeding the most massive halos. 
By $z\sim9$, star formation in BonFIRE spans over three orders of magnitude in halo mass ($\Mhalo\sim10^8$--$10^{11}~\Msun$), from isolated proto-galaxies to highly clustered systems hosting tens to hundreds of stellar associations. 
This evolution highlights the transition from stochastic, burst-dominated activity at early epochs to the onset of sustained star formation within growing potential wells (see also \citealt{Pawlik2013}).
At $z\sim9$, star formation in CampFIRE-800 spans $\Mhalo\sim10^7$--$10^{10}~\Msun$, and we use this increased low-mass coverage to resolution-correct galaxy properties in BonFIRE (see Section~\ref{sec:sims_rescorr} and Appendix~\ref{app:resampling}).
We refer to the resolution-corrected dataset as ``BonFIRE+CampFIRE" and we use it to present our main results throughout this work.

\subsection{Stellar mass}\label{sec:results_mstar}

The left panel of Figure~\ref{fig:smhm} shows the BonFIRE+CampFIRE stellar mass-halo mass (SMHM) relation at $z\sim9$. 
We delineate resolution limits in the figure as follows. 
The vertical grey dashed (dotted) line marks the approximate halo mass resolution limit in CampFIRE-800 (BonFIRE), below which halo occupation statistics become unreliable (see Appendix~\ref{app:HOF}). 
Similarly, the horizontal dahsed (dotted) grey line marks the $N_*=10$ star particle limit in CampFIRE-800 (BonFIRE). 
BonFIRE's SMHM relation without any resolution correction exhibits a resolution-induced flattening below $\Mhalo\lesssim10^{9.5}~\Msun$, near where the halo occupation fraction and $N_*=10$ star particle limits for BonFIRE intersect. 
In contrast, BonFIRE+CampFIRE extends the stellar mass--halo mass relation down to $\Mstar\sim10^4~\Msun$ at $\Mhalo\sim10^8~\Msun$, about 1.5 dex lower in stellar mass than BonFIRE alone.

The bend visible in BonFIRE+CampFIRE's SMHM relation at $\Mhalo\sim10^{8.5}~\Msun$ is due to the presence of two morphologically distinct populations of galaxies: the familiar track showing the correlation of an increasing trend in stellar mass as a function of halo mass for $\Mhalo \lesssim 10^{11}~\Msun$, and a horizontal feature at $\Mstar\sim10^{6.5}~\Msun$, corresponding to ultra-compact galaxies with star formation histories dominated by a single starburst; we discuss the latter population further in Section~\ref{sec:results_ucg}.
We note here that both of these populations are well-resolved in CampFIRE-800 with $\gtrsim100$ star particles, and similar populations exist across the full simulation suite.

The right panel of Figure~\ref{fig:smhm} shows the evolution of the SMHM relation in BonFIRE+CampFIRE over $9\lesssim z\lesssim25$.
The SMHM relation in BonFIRE+CampFIRE is remarkably constant with redshift once enough galaxies have formed to produce sufficient statistics, consistent with results from FIRE-2 \citep{Ma2018b,Feldmann2025}.
The simulations exhibit a tight, approximately monotonic relation in which stellar mass increases with halo mass, with modest ($\lesssim0.5$ dex) evolution in normalization at fixed halo mass over time. 
Notably, the bend in the relation (visible in the left panel as well) persists across all redshifts probed by the simulations down to $z\sim9$, indicating the persistent formation and/or survival of ultra-compact galaxies.

\begin{figure*}
    \centering
    \includegraphics[width=\textwidth]{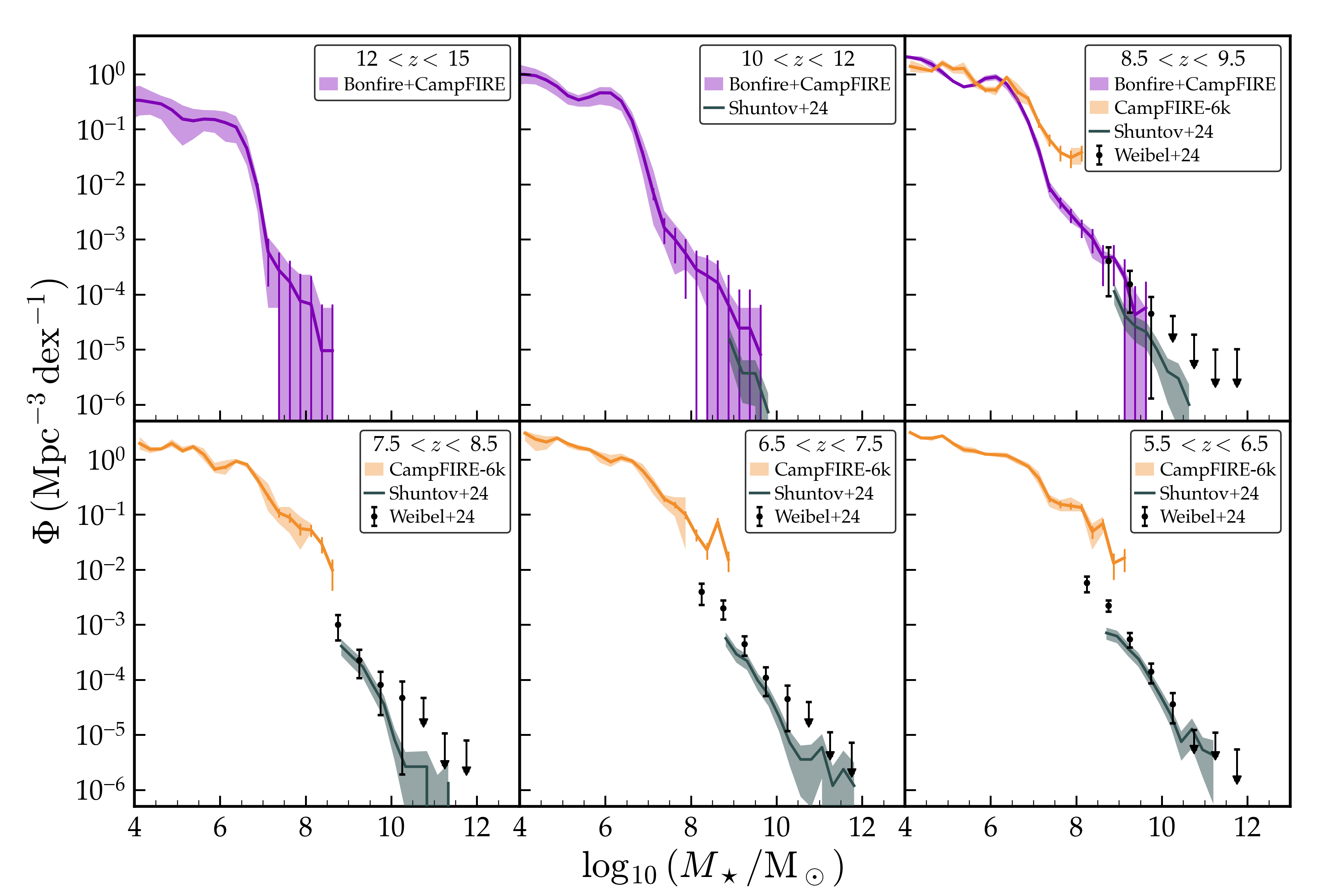}
    \caption{
    The stellar mass function (SMF) in BonFIRE+CampFIRE and CampFIRE-6k over $6\lesssim z\lesssim15$. 
    Shaded regions for simulations are the minimum and maximum values of the SMF over the corresponding redshift interval in each panel, whereas, lines and errorbars are the coadded SMF and $1~\sigma$ Poisson errors. 
    BonFIRE+CampFIRE predicts a sharply rising SMF at $\Mstar\lesssim10^{7}~\Msun$, indicating a large population of low-mass galaxies at $z\gtrsim9$. 
    The BonFIRE+CampFIRE SMF then flattens at $\Mstar\lesssim10^6~\Msun$. 
    CampFIRE-6k allows us to extend our SMF predictions down to $z=6$, showing an overall similar shape to the BonFIRE+CampFIRE SMF. 
    We compare to available observations at $6\lesssim z\lesssim10$ (grey and black) from \citet{Weibel2024,Shuntov2025}.
    The BonFIRE+CampFIRE SMF agrees with observations in the massive range ($\Mstar\sim10^9-10^{10}~\Msun$) at $z\sim9$. 
    However, CampFIRE-6k overpredicts the observations at $\Mstar\sim10^8~\Msun$ during $z\lesssim8$. 
    }
    \label{fig:smf_comparison}
\end{figure*}

We compare our results to two other SMHM relations from the literature: the average from the FIREbox-HR simulation \citep{Feldmann2025} at $z=10$, and the median from the empirical UniverseMachine (UM) model at $z=9$ \citep{Behroozi2019}. 
At $\Mhalo\lesssim10^9~\Msun$, BonFIRE+CampFIRE produces $\sim0.2$--1~dex larger stellar masses compared to FIREbox-HR. 
Both BonFIRE+CampFIRE and FIREbox-HR lie about $0.1$--$1$ dex higher than UM likely because FIRE directly resolves the multiphase ISM and dense star-forming gas, which can lead to higher instantaneous star formation efficiencies than empirical models like UM, where star formation is constrained by low-redshift observations.

The increased amplitude of the SMHM in BonFIRE+CampFIRE compared to FIREbox-HR could arise from the updated FIRE-3 prescriptions and simple Pop~III model (Section~\ref{sec:sims_pop3}) in the BonFIRE and CampFIRE suite. 
FIRE-3 produces stronger early star formation in low-mass, low-metallicity galaxies due to later reionization, more accurate metal-free cooling, weaker early stellar mass-loss feedback at low metallicities, more energy-conserving supernovae, and less restrictive star formation criteria \citep{Hopkins2023}. 
Combined with rapid early enrichment from our simple Pop~III model, the physics prescriptions in our suite could reasonably lead to larger stellar masses in low-mass halos at fixed (early) times.

We now examine the stellar mass function (SMF) of galaxies in the suite and compare it with observations.
Figure~\ref{fig:smf_comparison} shows the evolution of the SMF across $6\lesssim z\lesssim15$, comparing results from the BonFIRE+CampFIRE and CampFIRE-6k simulations with recent \JWST\ observations.
For each redshift interval, the shaded regions indicate the minimum and maximum SMF across snapshots, while the lines and errorbars show the coadded SMF with $1~\sigma$ Poisson uncertainties. 
For each redshift bin, we coadd SMFs by summing galaxy counts in bins of $\Delta \log_{10}(\Mstar/\Msun) = 0.25$~dex across snapshots and dividing by the total comoving volume of those snapshots: 
\begin{equation}
\Phi_{\rm stack}(\Mstar) =
\frac{\sum_i N_i(\Mstar)}{\left(\sum_i V_i\right)\Delta \Mstar}.
\end{equation}
This assumes that the snapshots are independent realizations, which is not strictly true, but it enhances the volume coverage of our results without affecting our main conclusions. 
We show Poisson errors on our stacks, which we propagate from the galaxy number counts in each stack \citep{Gehrels1986}.

To account for the overdensity of the CampFIRE high-resolution region ($\delta \approx 0.4$), we apply a bias correction to the SMF. 
We compute the halo bias $b(M,z)$ using the \citet{Tinker2010} fitting function and assign each galaxy a weight $w = [1 + b(M,z)~\delta]^{-1}$ to correct for the enhanced abundance of halos in the region. 
We then construct the SMF by summing these weights within stellar mass bins, thereby recovering a distribution representative of the cosmic mean.

The BonFIRE+CampFIRE SMF exhibits a steep rise at $\Mstar\lesssim10^{7}~\Msun$, indicating a large population of low-mass galaxies at $z\gtrsim9$, followed by a flattening below $\Mstar\lesssim10^{6}~\Msun$, which likely reflects both physical and numerical limits in galaxy formation efficiency. Among the former may be the transition from efficient atomic to inefficient molecular cooling halo regimes \citep[e.g.,][]{Jaacks2019}.

At $z\sim9$, the combined BonFIRE+CampFIRE SMF captures the abundance of the most massive galaxies, consistent with the observed SMFs from \citet{Shuntov2025} and in good agreement with \citet{Weibel2024}, while extending predictions down to $\Mstar\sim10^{4}~\Msun$. 
The CampFIRE-6k simulation extends these predictions to lower redshift, down to $z=6$. 
However, at $z\lesssim8$, CampFIRE-6k overpredicts the abundance of galaxies at $\Mstar\sim10^{8}~\Msun$, which may reflect either residual overdensity of the simulated subregion or potential incompleteness in the observational constraints at these masses.

\begin{figure*}
    \centering
    \begin{tabular}{cc}
    \subfigure{\includegraphics[width=0.51\textwidth]{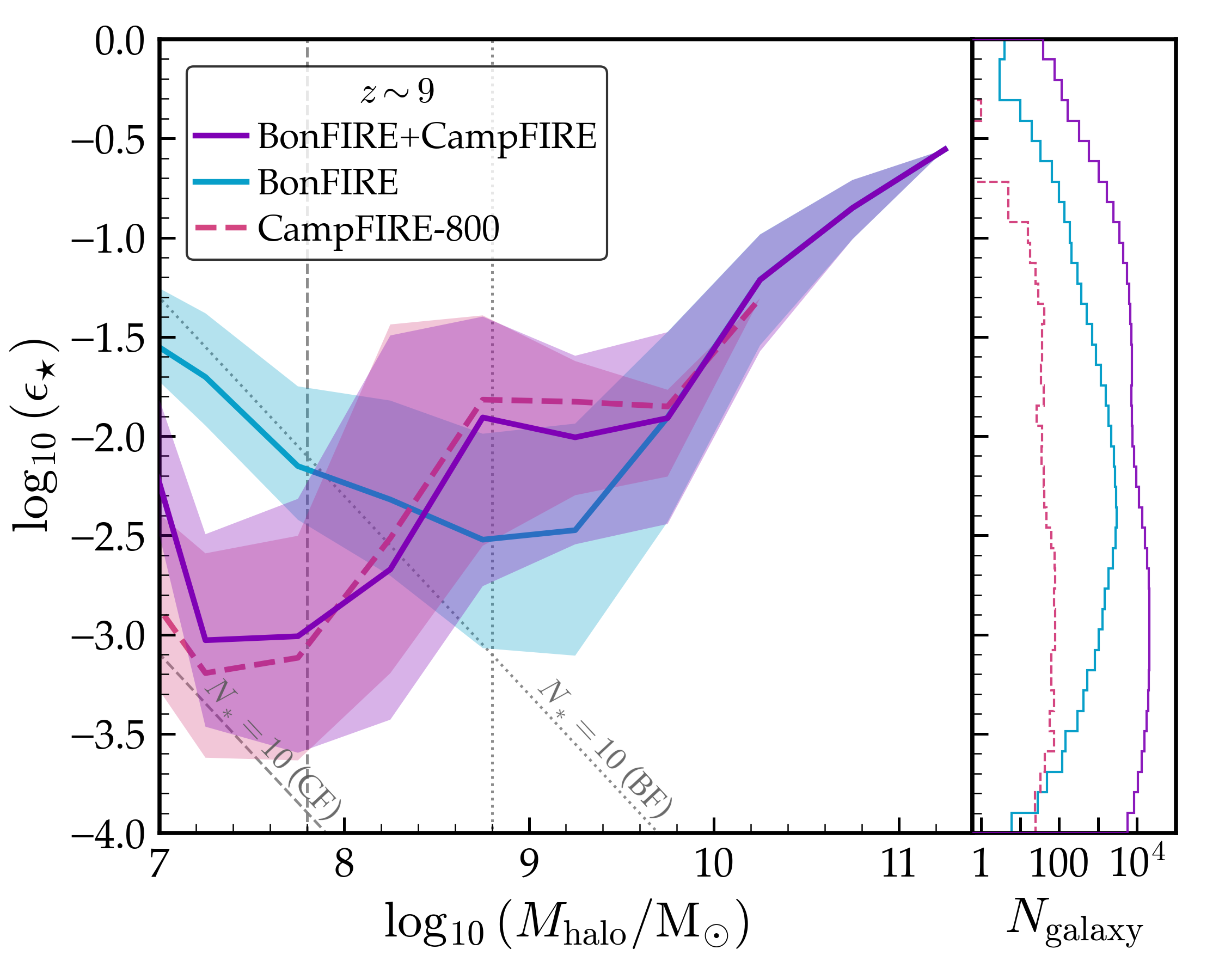}} &
    \subfigure{\includegraphics[width=0.48\textwidth]{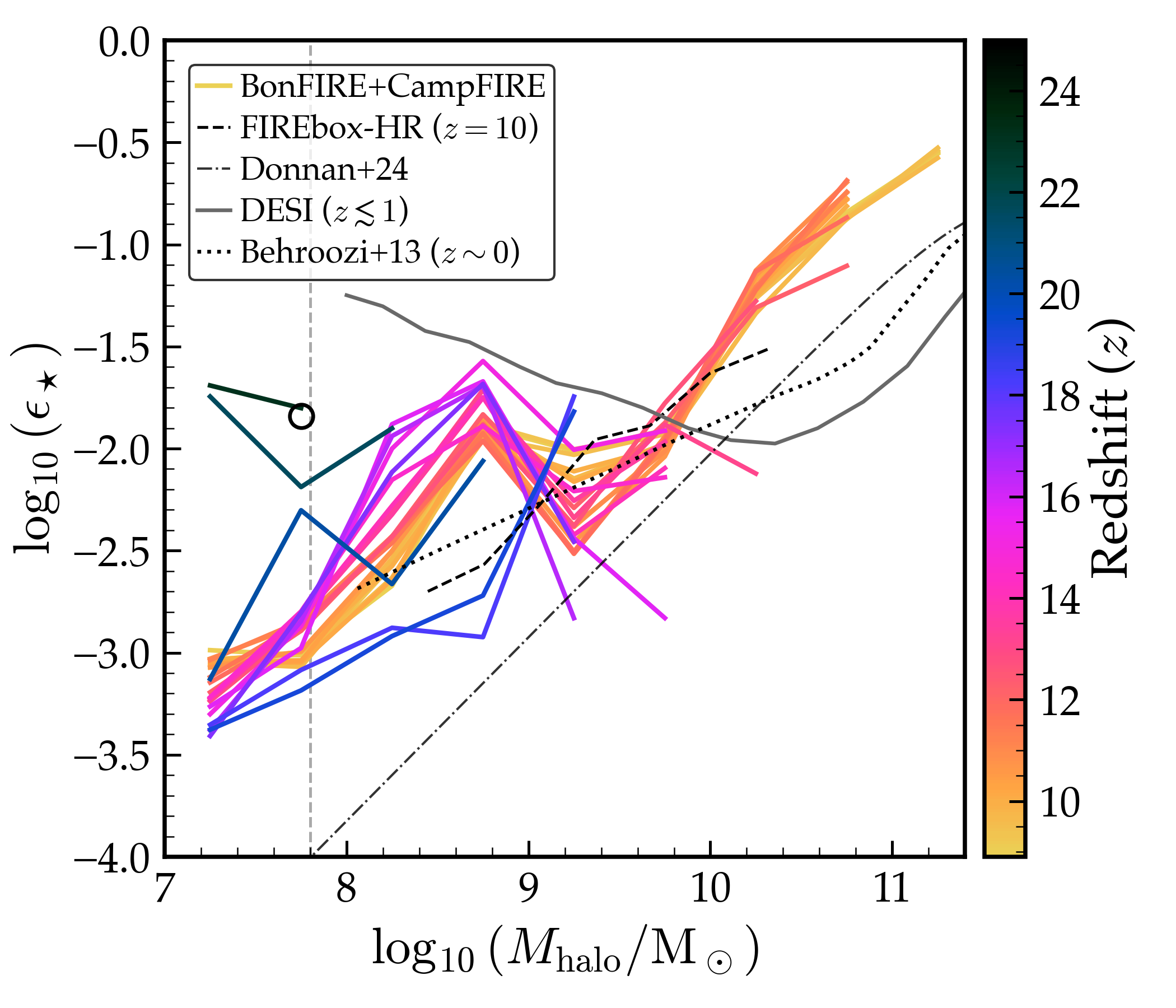}}
    \end{tabular}
    \caption{\textit{Left:} The halo-scale star formation efficiency (SFE, $\sfe$) in BonFIRE+CampFIRE, BonFIRE, and CampFIRE-800 as a function of halo mass at $z\sim9$. 
    As in Fig.~\ref{fig:smhm}, the shaded regions and lines show the 68\% ranges and medians. 
    We only show $\sfe$ for halos that have an associated galaxy, down to a single star particle.
    We mark the 100 star particle limits as diagonal lines (dot-dashed for BonFIRE, and dashed for CampFIRE-800), and the halo occupation limit for CampFIRE-800 as a grey shaded region.
    SFE uniformly rises with halo mass in the well-resolved regime ($\Mhalo\gtrsim10^9~\Msun$), but below that there is a mixture of resolved and unresolved galaxies that spreads to both high and low SFE.
    The location of resolution effects on this relation varies, but the uptick in SFE at $\Mhalo\sim10^{8.5}~\Msun$ and the downturn through $\Mhalo\sim10^8~\Msun$ are robustly resolved in CampFIRE-800.
    \textit{Right:} The evolution of the median BonFIRE+CampFIRE $\sfe$-$\Mhalo$ relation over $9\lesssim z\lesssim25$.
    We predict a $\sfe$-$\Mhalo$ relation that does not significantly evolve down to $z\sim9$.
    The dashed black line shows results from FIREbox-HR, which are consistent with our results at $\Mhalo\sim10^{9-10}~\Msun$, but do not show the same uptick in SFE at lower halo masses.
    The dot-dashed black line shows the model of \citet{Tacchella2018} fit to observations of high-redshift galaxies from \citet{Donnan2024}.
    }
    \label{fig:sfe_vs_Mhalo}
\end{figure*}

\subsubsection{Star formation efficiency}\label{sec:results_sfe}

Given the increased stellar mass coverage and the presence of a distinct population of ultra-compact galaxies in the BonFIRE and CampFIRE suite, we now examine the halo-scale star formation efficiency (SFE) of galaxies to assess how efficiently early halos convert gas into stars across different mass and morphological regimes.
We calculate SFE as $\sfe\equiv \Mstar/(f_{\rm b}~\Mhalo)$, such that it represents an integrated SFE over the lifetime of a halo.
This is essentially the same information as in Figure~\ref{fig:smhm}, but we derive some useful insight on early galaxy formation from this viewpoint.

The left panel of Figure~\ref{fig:sfe_vs_Mhalo} shows the SFE for galaxies in BonFIRE+CampFIRE, BonFIRE, and CampFIRE-800 at $z\sim9$. 
The vertical grey dashed (dotted) line marks the limit of halo occupation in CampFIRE-800 (BonFIRE), and the two diagonal lines mark the $N_*=10$ star particle limits in CampFIRE-800 (lower line) and BonFIRE (upper line). 
We find that galaxies with prolonged star formation follow the expected trend of increasing SFE with halo mass \citep[e.g.,][]{Tacchella2018}. 
In particular, the most massive halos ($\Mhalo \gtrsim 10^{10}~\Msun$) at $z\sim9$ reach $\sfe \sim 0.1$--0.3, consistent with values inferred for massive galaxies in recent \JWST\ observations \citep{Chworowsky2024}. 
Toward lower masses, this population exhibits a decline in SFE, reaching $\sfe \sim 0.001$--$0.01$ around $\Mhalo\sim10^{9}~\Msun$.

In contrast, galaxies dominated by a single, short burst of star formation occupy halos with $\Mhalo\sim10^{7}$--$10^{9}~\Msun$ and follow an inverse trend, with SFE increasing toward lower halo masses. 
These systems can achieve comparably high efficiencies ($\sfe\gtrsim0.1$) at the low-mass end, potentially indicating a distinct mode of star formation. 
Despite their differing trends, both populations converge to similar, more typical efficiencies of $\sfe\sim0.001$--0.01 near $\Mhalo\sim10^{9}~\Msun$.

We see clear resolution dependence in the $\sfe$--$\Mhalo$ relation along the finite-particle number threshold ($\lesssim10$ star particles, diagonal lines in Figure~\ref{fig:sfe_vs_Mhalo}). 
For example, in this regime in BonFIRE, the stellar mass of compact galaxies can build up too rapidly once the local Jeans mass threshold is reached, leading to brief ($\sim1$ Myr), intense starbursts that consume available gas on dynamical timescales, because the minimum ``unit" of resolved self-gravitating gas is limited to a few gas cells. 
This causes an upturn in SFE at lower halo masses, visible where the $\sfe$--$\Mhalo$ relation begins to follow the dot-dashed diagonal resolution line in Figure~\ref{fig:sfe_vs_Mhalo}. 
Higher-resolution runs in the CampFIRE suite confirm that this effect primarily reflects finite resolution rather than a fundamental change in star formation physics: when the Jeans length is better resolved at fixed mass, star formation proceeds more gradually, with longer depletion times ($\sim10~\Myr$), and the upturn in SFE is correspondingly shifted to lower halo masses.

Importantly, we note that the population of ultra-compact galaxies at $\Mstar\gtrsim10^6~\Msun$ (responsible for the visible bends in the SMHM and $\sfe$--$\Mhalo$ relations just below $\Mhalo\sim10^9~\Msun$) is present and well-resolved in the CampFIRE runs with $\gtrsim10^3$ star particles ($\gtrsim20$ in BonFIRE), \textit{distinct} from the resolution-limited low-mass galaxies. 
Thus, though this population suffers from some SFR numerical resolution issues, it is robustly predicted across all of our simulations. 
We further discuss resolution dependence within this population in Section~\ref{sec:results_ucg} and and more generally in Appendix~\ref{app:restests}.

The right panel of Figure~\ref{fig:sfe_vs_Mhalo} shows the evolution of the median $\sfe$--$\Mhalo$ relation in BonFIRE+CampFIRE at $9\lesssim z\lesssim25$. 
As with the SMHM relation, the $\sfe$--$\Mhalo$ relation evolves only weakly at $\Mhalo\gtrsim10^9~\Msun$. 
At lower halo masses, there is a persistent bend just below $\Mhalo\sim10^9~\Msun$, due to the large population of ultra-compact galaxies at all snapshots.

\begin{figure*}
\centering
\includegraphics[width=0.85\textwidth]{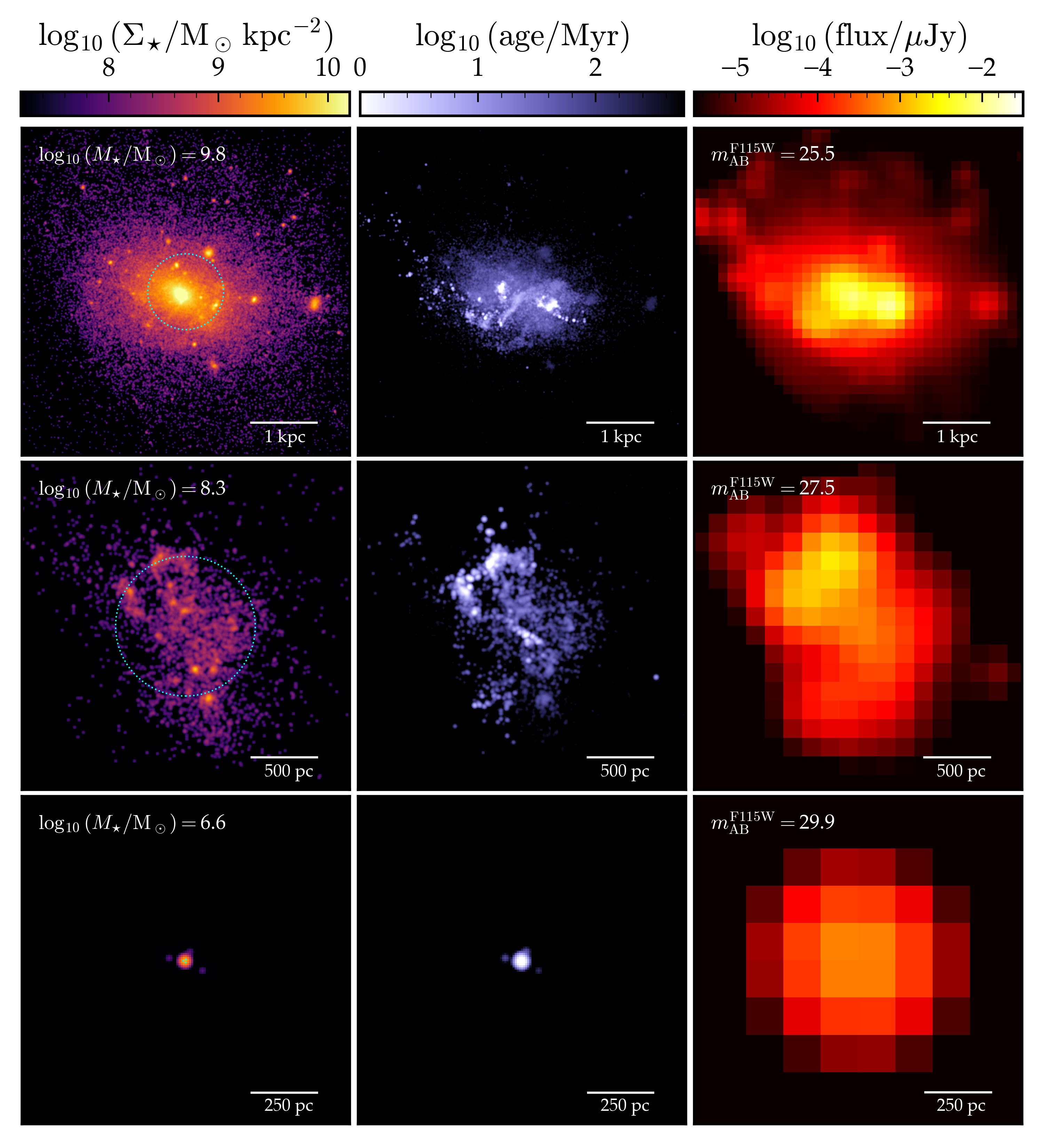}
\caption{Images of three galaxies representing a diversity of morphologies in BonFIRE at $z=9$. 
\textit{Left:} True stellar surface density showing substructure down to scales of $\sim10~\pc$. 
Dotted cyan circles are the radius enclosing 50\% of a galaxy's stellar mass ($R_{\star,~ 1/2}$). 
\textit{Middle:} Map of stellar age highlighting the youngest stars in the lightest colors. 
\textit{Right:} Mock \JWST\ NIRCam F115W image showing rest-frame UV emission. 
Large clumps with $\sim0.1$--$1~\kpc$ sizes are visible.}
\label{fig:multipanel_morph}
\end{figure*}

Compared to FIREbox-HR, BonFIRE+CampFIRE shows broadly similar trends at intermediate halo masses but exhibits higher star formation efficiencies at $\Mhalo\gtrsim10^{10}~\Msun$, particularly at lower redshifts. 
Relative to empirical and semi-empirical models such as UniverseMachine and \citet{Behroozi2013c}, our results predict significantly elevated efficiencies at high redshift across nearly all halo masses, with $\sim1$~dex offset but similar slopes at $\Mhalo\gtrsim10^{10}~\Msun$. 
Interestingly, the model of \citet{Tacchella2018} fit to observations of high-redshift galaxies from \citet{Donnan2024} is similar in normalization and shape to the low-redshift empirical model of \citet{Behroozi2013c}, possibly suggesting higher SFE in our simulations compared to observed high-redshift galaxies.
In contrast, the high-mass, low-redshift DESI constraints \citep{Xu2026} lie well below our relations, reflecting the strong evolution in star formation efficiency from early cosmic times to the present day. 
The upturn seen in the DESI relation at low masses, if real, may represent a low-redshift analog of the knee in BonFIRE+CampFIRE, though they are likely shaped by different feedback and gas accretion regimes.

Compared to other recent high-redshift galaxy formation models, our results occupy a distinct regime of SFE, predicting elevated SFE in high-mass halos (as expected) \textit{and uniquely} in low-mass halos. 
Simulations such as THESAN and THESAN-ZOOM generally show a monotonic rise in $\sfe$ with halo mass, reaching $\sfe\sim0.05$--0.1 at $\Mhalo\sim10^{10}~\Msun$, consistent with feedback suppressing star formation more efficiently in low-mass halos \citep{Kannan2022,Shen2025}. 
In contrast, BonFIRE+CampFIRE extends high efficiencies to lower masses, likely due to the FIRE-3 treatment of star formation in dense, metal-poor gas \citep{Hopkins2023}, and our simple Pop~III model. 
In these systems, feedback may act to primarily terminate star formation after the initial burst rather than regulating it continuously, leading to high halo-scale SFE in low-mass halos.

In comparison, large-volume simulations such as IllustrisTNG and SIMBA rely on effective ISM models and fixed high-density star formation thresholds \citep{Pillepich2018,Dave2019}, which tend to delay or suppress star formation in low-mass halos and produce lower SFE in this regime, consistent with semi-empirical constraints at low redshift \citep{Behroozi2013c}. 
As a result, these simulations do not (and should not be able to) produce the same population of small/faint yet efficient galaxies that we find in BonFIRE+CampFIRE at $\Mhalo\lesssim10^9~\Msun$, demonstrating how differences in small-scale star formation and feedback physics may shape the efficiency of early galaxy formation.

\subsection{Stellar morphology}\label{sec:results_morph}

We examine the morphology of high-redshift galaxies in the BonFIRE and CampFIRE suite of simulations first by exploring the physical extent of their stellar mass and light. 
Figure~\ref{fig:multipanel_morph} shows the stellar density, age, and UV emission maps for a sample of representative galaxies in BonFIRE at $z\sim9$. 
The UV maps are mock \JWST\ NIRCam F115W images, smoothed to reflect the NIRCam PSF following \cite{Sun2023b}, where we do not account for dust attenuation. 
There are numerous star clusters visible in the stellar density maps of the massive galaxies, as well as a diffuse stellar component. 
The stellar age and UV emission maps highlight that NIRCam is mostly sensitive to the youngest stars ($\lesssim10~\Myr$) and does not spatially resolve individual star clusters (without the aid of magnification from gravitational lensing), suggesting that recent star formation may be obscuring the morphology and amount of older stellar populations at high redshift \citep{Narayanan2024}. 
The true morphology of high-redshift galaxies, particularly with regard to their detailed substructure, therefore remains unresolved in JWST NIRCam observations without the aid of gravitational lensing.

The last row of Figure~\ref{fig:multipanel_morph} shows an example of a typical ultra-compact galaxy in BonFIRE, which are present throughout the BonFIRE and CampFIRE suite. 
These are dense, compact galaxies where the stellar mass is dominated by a single star cluster. 
A striking feature of the BonFIRE and CampFIRE simulation suite is the emergence of such ultra-compact, single-burst galaxies—systems in which nearly all stellar mass forms within a single, short ($\lesssim10$~Myr) episode of star formation. 
We devote the following subsection (\ref{sec:results_ucg}) to a deeper exploration of the ultra-compact galaxy population.

\begin{figure}
    \centering
    \includegraphics[width=\columnwidth]{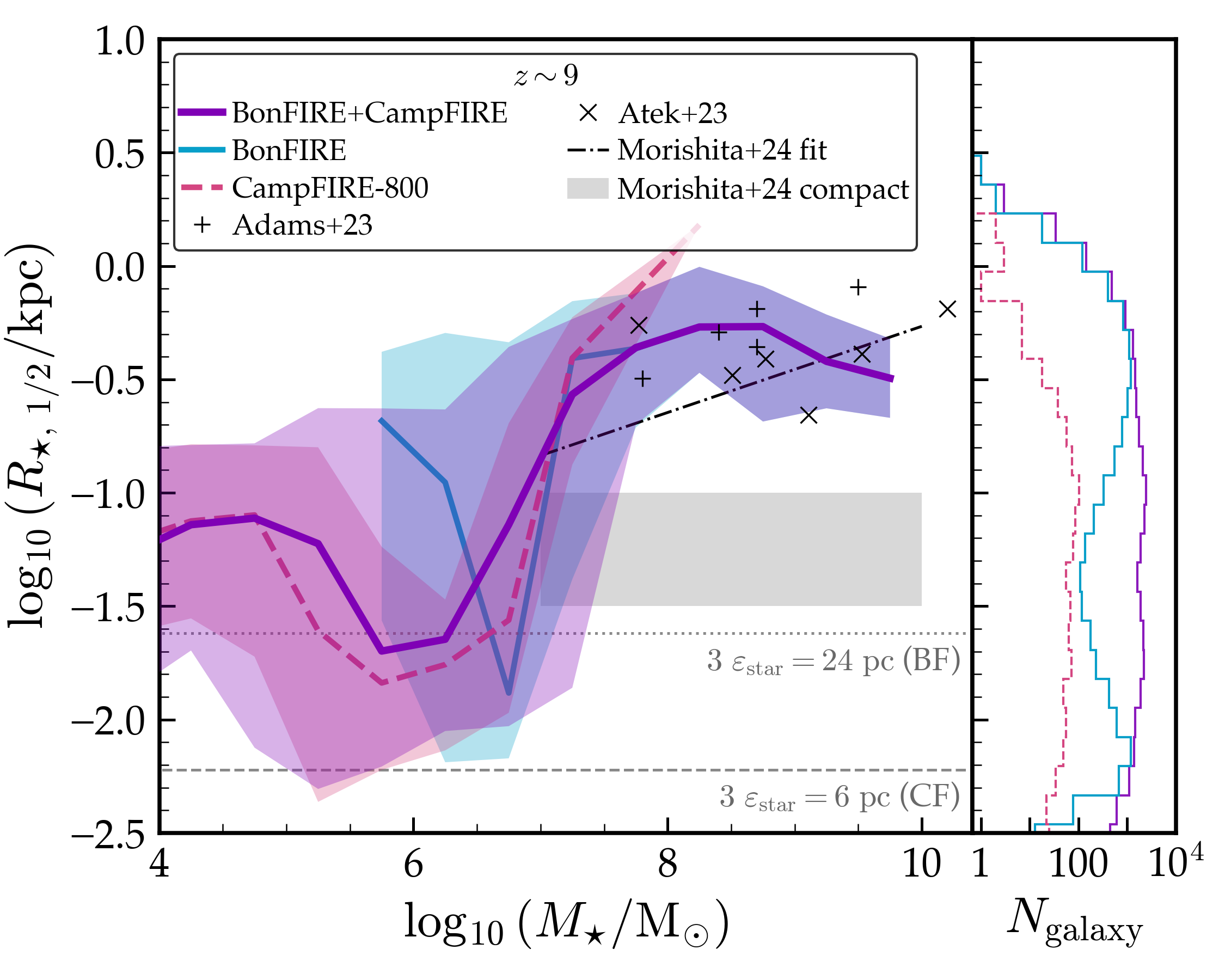}
    \caption{The size--mass relation for galaxies in BonFIRE+CampFIRE, BonFIRE, and CampFIRE-800 at $z\sim9$. 
    Colored lines and shaded regions show medians and 68\% scatter. 
    We show simulation results down to the 10 star particle limit. 
    The horizontal lines show approximate spatial resolution limitations on galaxy sizes for the two base simulations. 
    There is a prominent drop in median galaxy sizes at $\Mstar\lesssim10^7~\Msun$, corresponding to a large population of galaxies with small sizes ($\lesssim100$ pc). 
    Points are measured sizes of high-redshift candidates from lensed \citep{Atek2023} and unlensed \citep{Adams2023} fields. 
    The dot-dashed black line and grey shaded region are the best-fit relation of high-redshift galaxies and a population of compact galaxies from the same sample, respectively \citep{Morishita2024}.}
    \label{fig:size_vs_mass}
\end{figure}

Figure~\ref{fig:size_vs_mass} shows the size--mass ($R_{\star,~1/2}$--$\Mstar$) relation for galaxies in BonFIRE+CampFIRE, BonFIRE, and CampFIRE-800 at $z\sim9$. 
We calculate the size of a galaxy as the spherical three-dimensional radius enclosing 50\% of a galaxy's stellar mass ($R_{\star,~1/2}$). 
The horizontal grey lines mark galaxy size resolution limits, which we approximate as three times the gravitational force softening of stars ($3~\varepsilon_{\rm star}$) in each simulation. 
This leads to a lower limit on robust galaxy sizes of $\sim8$~pc in CampFIRE-800 and $\sim24$~pc in BonFIRE.

Once again, the population of ultra-compact galaxies is visible as a distinct feature in the size--mass relation. 
Ultra-compact galaxies ($\Mstar\lesssim10^7~\Msun$) drag the relation down with their typical sizes of $\lesssim100~\pc$, similar to massive star clusters \citep{Grudic2019,Kruijssen2026}. 
Whereas, the more typical population of larger galaxies with prolonged star formation extends over the full stellar mass range. 
At $\Mstar\gtrsim10^7~\Msun$, the simulations all predict that galaxy sizes increase and then level off with stellar mass, with typical half-light radii of $\sim0.1$--$1~\kpc$.

We compare to observational measurements of massive high-redshift galaxy sizes from \cite{Adams2023} and \cite{Atek2023}. 
\citet{Adams2023} use \JWST\ NIRCam imaging of the SMACS 0723 ERO field to identify $9\lesssim z\lesssim12$ galaxies and measure their sizes via profile fitting, finding compact half-light radii typically $\sim0.5-1$ kpc for massive systems. 
Whereas, \citet{Atek2023} analyze strongly lensed \JWST\ NIRCam galaxy candidates behind SMACS 0723 out to $z\sim16$ (though we only compare to their $9\lesssim z\lesssim12$ sample), deriving sizes of $\sim0.5$ kpc from lensing-corrected half-light radii. 
The high-mass end of our simulated size-mass relation broadly agrees with the observations. 
However, we note that this is not strictly a one-to-one comparison, as observational measurements report half-light radii, whereas we report intrinsic half-mass radii. 
Previous work with FIRE simulations has shown that these two size metrics can be off by up to a factor of a few and that the half-light radius is sensitive to feedback and dust prescriptions in radiative transfer modeling \citep{Parsotan2021,Cochrane2023,Klein2024,Wheeler2025}. 
However, we leave more detailed modeling of light-weighted galaxy sizes for future work.

Furthermore, \citet{Morishita2024} analyzed a large sample of 341 galaxies at $5<z<14$ and measured their sizes using profile fitting to derive half-light radii from public \JWST\ data. 
They find a clear size–luminosity relation in which galaxy sizes decrease with increasing redshift and decreasing luminosity, with typical half-light radii of $\sim0.1-1$ kpc across their sample. 
This trend is broadly consistent with the sizes of galaxies in BonFIRE+CampFIRE at fixed halo mass and redshift. 
Notably, \citet{Morishita2024} also identify a population of extremely compact systems with sizes $\lesssim0.1$ kpc, which may represent observational counterparts to the ultra-compact galaxies in our simulations, although their compact sample extends to higher stellar masses than we predict. 
However, it is possible that by measuring galaxy sizes from UV imaging we could recover small sizes for galaxies in their mass range.

\begin{figure}
    \centering
    \includegraphics[width=\columnwidth]{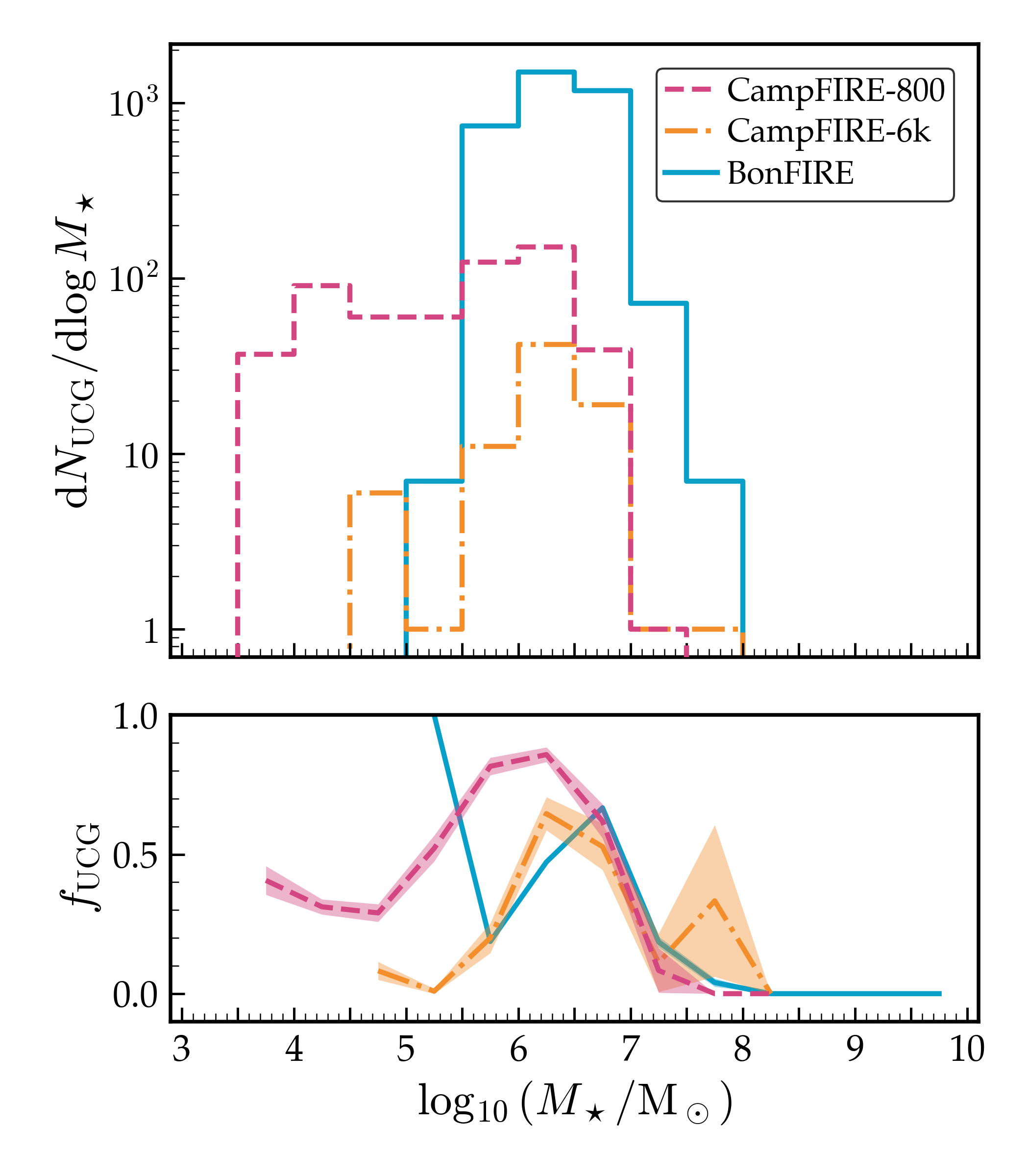}
    \caption{
    \textit{Top:} The stellar mass functions of ultra-compact galaxies (UCGs) at $z\sim9$ in BonFIRE (blue solid), CampFIRE-6k (orange dot-dashed), and CampFIRE-800 (pink dashed). 
    \textit{Bottom:} The fraction of galaxies classified as UCGs, $f_{\rm UCG}$, in each mass bin. 
    Shaded regions indicate Poisson uncertainties. 
    Though the absolute abundance of UCGs varies significantly with resolution, the stellar mass range over which UCGs dominate and the qualitative mass dependence of $f_{\rm UCG}$ are broadly consistent across simulations.}
    \label{fig:ucg1}
\end{figure}

\begin{figure*}
    \centering
    \includegraphics[width=\textwidth]{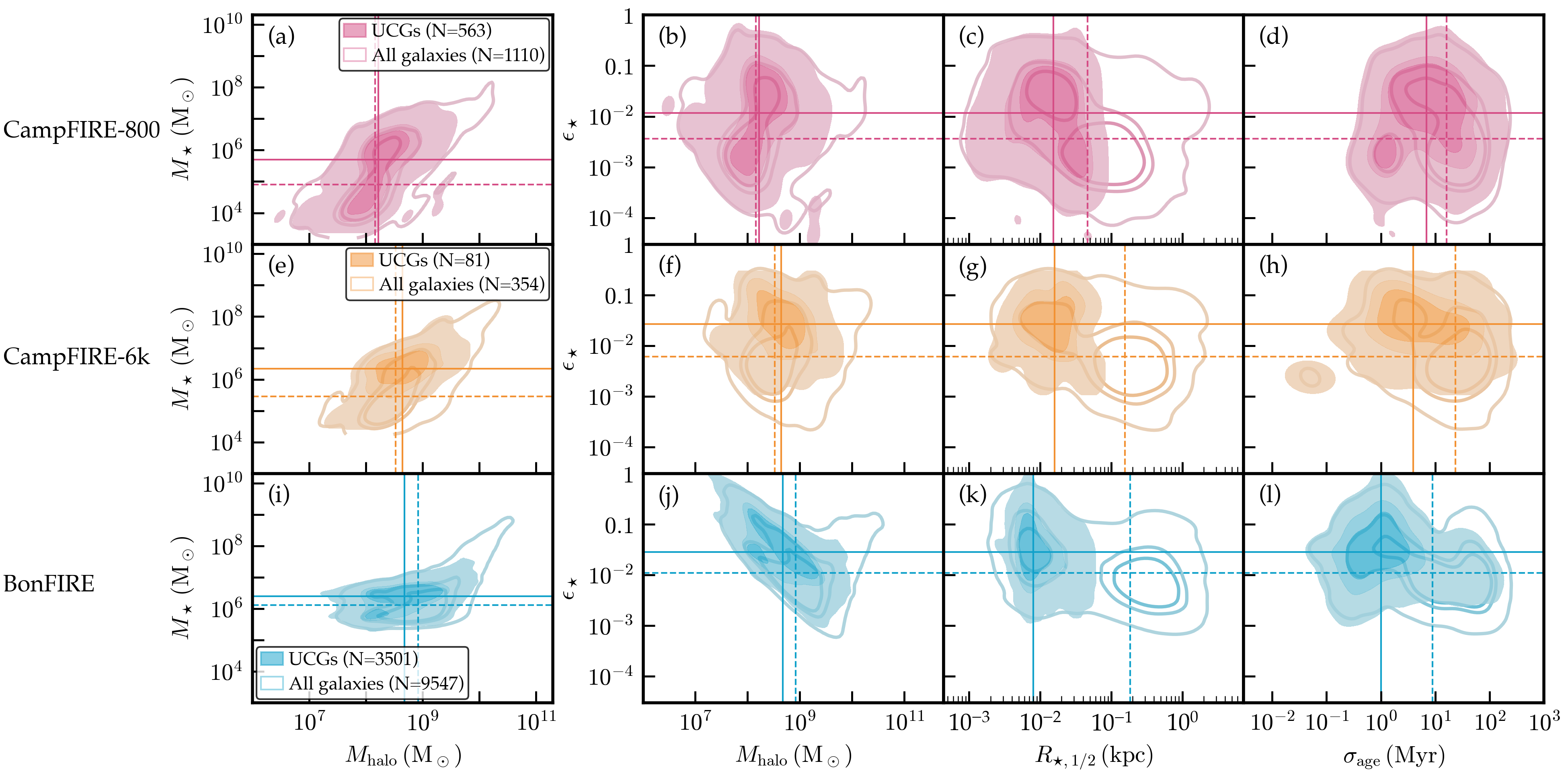}
    \caption{
    Joint distributions of stellar mass, halo mass, star formation efficiency, stellar half-mass radius, and stellar age dispersion for galaxies in CampFIRE-800 (top row, pink), CampFIRE-6k (middle row, orange), and BonFIRE (bottom row, blue) at $z\sim9$. 
    Filled contours show ultra-compact galaxies (UCGs), and unfilled contours show the total galaxy population. 
    Contours represent enclosed probability levels (75/90/99\%) of the two-dimensional Gaussian kernel density estimate. 
    Solid crosshairs mark median values for UCGs, and dashed crosshairs mark medians for the total galaxy population. 
    Across all resolutions, UCGs occupy regions of high star formation efficiency, low age dispersion, and compact stellar size at fixed halo mass compared to all galaxies, with modest resolution dependence in their median structural and star formation properties.}
    \label{fig:ucg2}
\end{figure*}

\subsubsection{Ultra-compact galaxies}
\label{sec:results_ucg}

In this section, we focus on the population of star-cluster-sized galaxies in the BonFIRE and CampFIRE suite, which we refer to as ultra-compact galaxies (UCGs). 
UCGs are common across all simulations in the BonFIRE and CampFIRE suite, and they typically reside in halos at $\Mhalo \sim 10^8$--$10^{9.5}~\Msun$, roughly in the range of atomic cooling halos. 
Using all of the simulations in our suite, we demonstrate that UCGs are not simply a consequence of low resolution, and are a distinct morphological population of galaxies at high redshift within our models.

The top panel of Figure~\ref{fig:ucg1} shows the stellar mass function of UCGs in BonFIRE, CampFIRE-6k, and CampFIRE-800 at $z\sim9$. 
Here, we have plotted UCGs meeting the following criteria: $R_{\star,~1/2}\leq50~\pc$ and $N_{*}\geq10$. 
These selection criteria prescribe a different lower stellar mass limit on UCGs for each simulation, given their different mass resolutions, but they also ensure that we are selecting somewhat resolved and highly compact systems. 
We find that all three simulations show a common peak at $\Mstar\sim10^6$--$10^7~\Msun$, with the higher resolution CampFIRE runs showing tails to lower masses. 
In particular, CampFIRE-800, our highest-resolution simulation, shows a prolific population of UCGs at low masses down to $\Mstar\sim10^4~\Msun$.

The bottom panel of Figure~\ref{fig:ucg1} shows the fraction of galaxies in each mass bin that belong to the UCG category. 
We limit the population of all galaxies that we compare against to those with $N_{*}\geq10$. 
Notably, across all simulations, the UCG fraction peaks at $\approx0.7-0.8$ near $\Mstar\sim10^6$--$10^7~\Msun$, regardless of resolution or volume. 
In CampFIRE-800, the extended tail of UCGs at $\Mstar\lesssim10^5~\Msun$ accounts for less of the total galaxy population, but is still significant at a UCG fraction of $\approx0.25$. 
Together, these trends suggest that UCGs constitute a major channel of early galaxy formation over a broad mass range, particularly near $\Mstar\sim10^{6.5}~\Msun$, where they dominate the galaxy population.

Figure~\ref{fig:ucg2} shows that UCGs occupy a narrow and well-defined region of morphological and star formation parameter space that is clearly separated from the bulk of the galaxy population. 
We plot the Gaussian kernel density estimates (KDEs) for the UCG population and the total galaxy population for each base simulation in the BonFIRE and CampFIRE suite (rows) in four different planes (columns). 
We highlight the medians of the distributions in each panel with crosshairs for easy comparison of resolution trends; solid crosshairs are medians for UCGs and dashed crosshairs are medians for the total galaxy population.

Beginning with the left column (panels a, e, and i), we find that in the $\Mstar$--$\Mhalo$ plane the median stellar masses of the UCG populations remain mostly stable at $\Mstar\sim2\times10^6~\Msun$. 
The median stellar mass of UCGs drops slightly in CampFIRE-800 due to its extended tail of lower-mass UCGs (which are not resolved in BonFIRE), but the same locus at $\Mstar\sim10^6~\Msun$ is visible. 
The median halo mass of UCGs changes from $\Mhalo\sim10^9~\Msun$ in BonFIRE to $\Mhalo\sim10^8~\Msun$ in CampFIRE-800, likely reflecting the numerical limits of halo occupation (Appendix~\ref{app:HOF}) and the larger variety of halos present in the BonFIRE volume rather than a physical shift in the UCG population. 
The relative stability of the stellar mass medians and their position well above the stellar mass resolution limits in the corresponding simulations (typical UCGs have $\sim20$ star particles even in the lowest-resolution simulation, BonFIRE), indicate that the characteristic masses of UCGs are not driven solely by resolution.

In the second column (panels b, f, and j), the $\sfe$--$\Mhalo$ relation shows that the full galaxy population (especially at the massive end) follows an expected trend of increasing $\sfe$ with increasing $\Mhalo$. 
Whereas, UCGs display high $\sfe$ over a more narrow range in $\Mhalo$ across all three base simulations, with the highest density in the UCG KDEs clearly offset to higher $\sfe$ than halos of similar mass along the expected trend. 
The median star formation efficiency of UCGs exceeds that of the full population by $\sim0.5$~dex, with typical $\sfe\gtrsim0.01$ and some of the lowest-mass UCGs reaching $\sfe\gtrsim0.1$.

In the $\sfe$--$R_{\star,~1/2}$ plane (third column; panels c, g, and k), the median sizes of UCGs remain clustered at $R_{\star,~1/2}\sim10~\pc$ (and the clear separation between UCGs and the full population of galaxies is by definition) with modest systematic shift to larger sizes at higher resolution. 
In contrast, we see an opposite resolution dependence in $R_{\star,~1/2}$ in the full population of galaxies, whereby higher resolution leads to modestly smaller typical sizes. 
We note that the sizes of UCGs in BonFIRE appear to cluster around the stellar force softening ($\varepsilon_{\rm star,~\rm{BF}}\sim8~\pc$), likely reflecting the much smaller gas softening in their dense formation environments ($\lesssim1~\pc$) and denoting a critical limitation in BonFIRE's ability to resolve the morphology of UCGs as they evolve.

Finally, in the $\sfe$--$\sigma_{\star,~\rm{age}}$ plane (right column; panels d, h, and l), UCGs consistently show low median stellar age dispersions (defined as the standard deviation of stellar ages, $\sigma_{\rm age}$) of $\lesssim10~\Myr$. 
However, there is clear resolution dependence in stellar age dispersion for UCGs at $\Mstar\sim10^6~\Msun$, increasing from $\sim1~\Myr$ at the lowest resolution to $\sim5~\Myr$ at the highest resolution. 
This dependence indicates that the star formation histories (and hence UV luminosities) of UCGs are sensitive to resolution, motivating our resolution correction (Section~\ref{sec:sims_rescorr}). 
The overall small stellar age dispersions of UCGs imply that the majority of their stellar mass is assembled over short timescales, consistent with star formation dominated by a single burst.

These trends indicate that, although both the total galaxy population and the UCG population exhibit moderate systematic resolution dependence, the defining properties of UCGs (high $\Mstar$ and $\sfe$ for a given halo mass, plus compact sizes and low age dispersions compared to the general populations) are robust to resolution. 
These offsets are visible in all projections shown in Figure~\ref{fig:ucg2} and persist across nearly two orders of magnitude in halo mass, indicating that UCGs are not simply an extension of the low-size tail of the galaxy distribution but instead represent a distinct galaxy population with potentially unique formation histories. 
This is further supported by the fact that UCGs typically contain $\gtrsim10$ and often $\gtrsim100$ star particles, placing them above the nominal resolution limits in all simulations.

Our results on UCGs are consistent with simulations from the EDGE project, which show that dense, compact stellar systems form in low-mass halos through centrally concentrated gas inflows and short-timescale starbursts \citep{Taylor2025}. 
In their simulations, rapid gas collapse in shallow potential wells produces dense, compact stellar systems with sizes of order $\sim$10–50 pc and elevated star formation efficiencies, bridging the gap between classical globular clusters and dwarf galaxies. 
This formation pathway naturally produces objects with high effective star formation efficiencies and small physical sizes, consistent with the structural and SFE offsets we identify for UCGs.

A similar picture emerges from recent simulations of star cluster formation at cosmic dawn by \cite{Williams2025}, which find that compact stellar systems form in $\sim10^7$–$10^9,\Msun$ halos via rapid ($\lesssim10$ Myr), burst-dominated star formation. 
These systems exhibit pc-scale to $\sim$10–50 pc sizes, narrow stellar age distributions, and high star formation efficiencies, and account for a significant fraction of early stellar mass assembly. 
Together, these results support the interpretation that UCGs are not a numerical artifact but instead reflect a physically distinct, burst-driven mode of galaxy formation that is common in low-mass halos at high redshift.

While our present analysis focuses on $z\gtrsim9$, preliminary inspection of formation histories and stellar metallicity distributions down to $z\sim6$ in CampFIRE-6k (not shown) suggests that UCGs are preferentially metal-poor and that the rate at which new UCGs form declines toward lower redshift. 
This behavior is qualitatively consistent with the decreasing availability of dense, low-metallicity gas and the increasing importance of stellar feedback and environmental effects at later times.

\begin{figure*}
    \centering
    \begin{tabular}{cc}
    \subfigure{\includegraphics[width=0.51\textwidth]{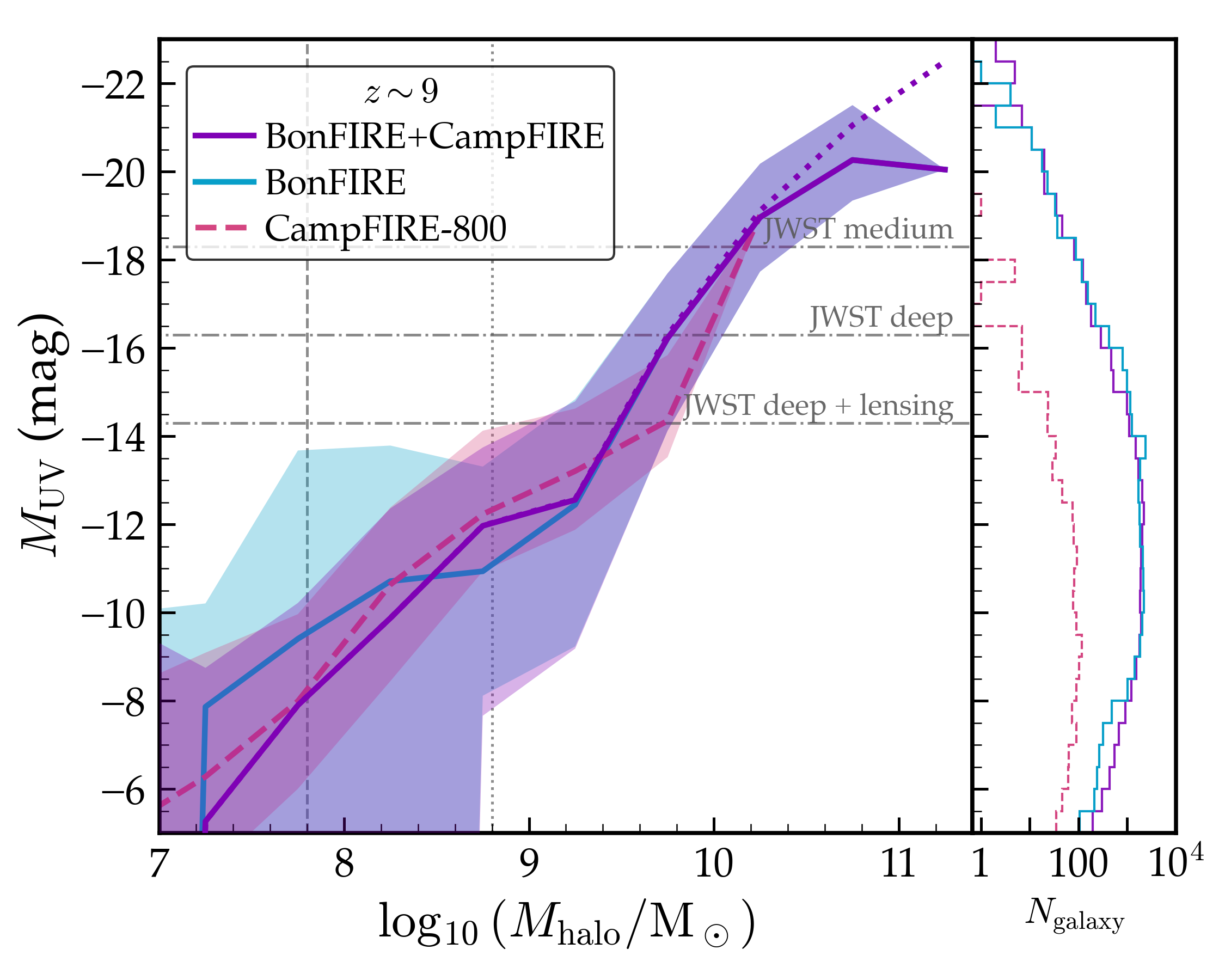}} &
    \subfigure{\includegraphics[width=0.48\textwidth]{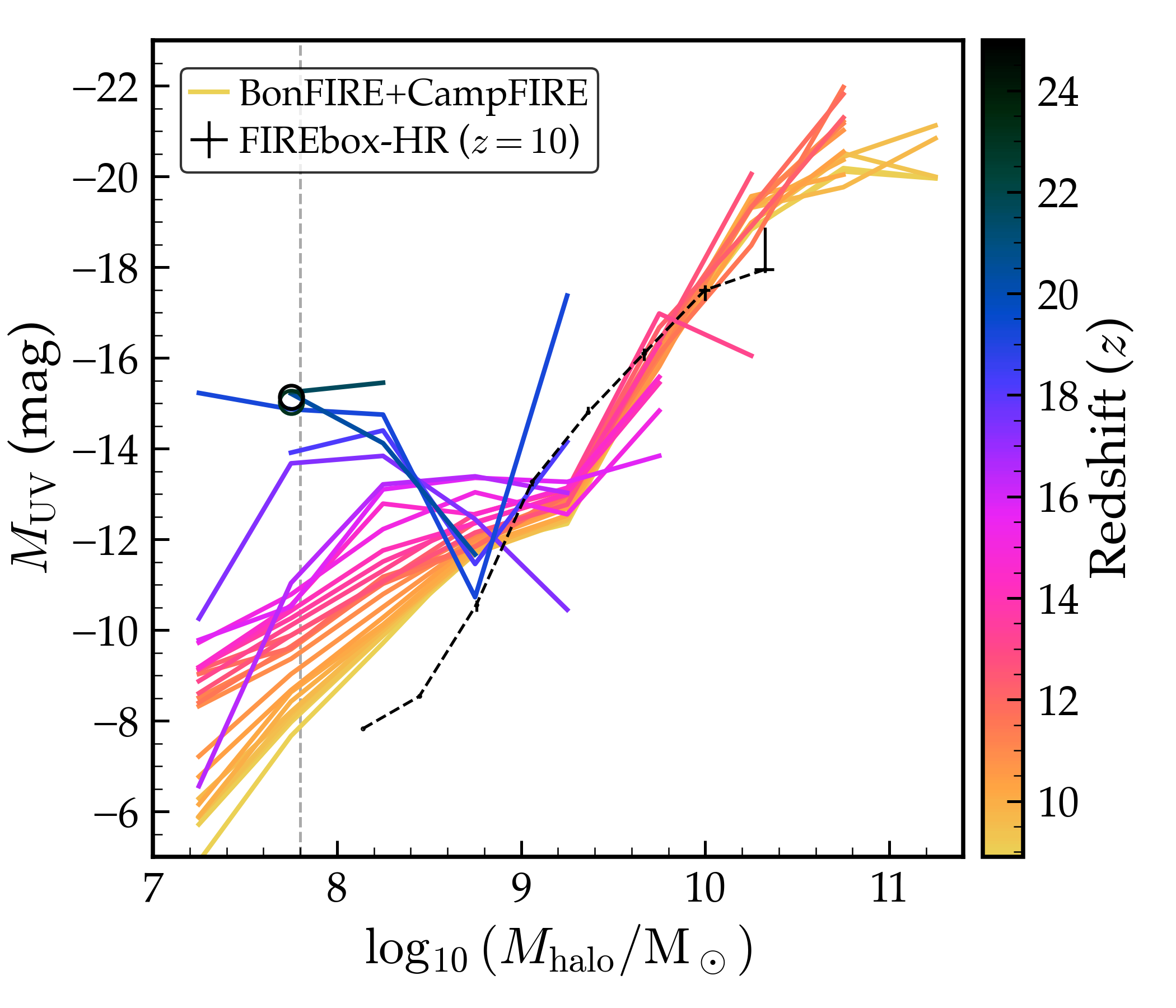}}
    \end{tabular}
    \caption{\textit{Left:} The $\Muv$--$\Mhalo$ relation in BonFIRE+CampFIRE, BonFIRE, and CampFIRE-800 at $z\sim9$. 
    Lines are median $\Muv$ in each $\Mhalo$ bin and shaded regions show the 68\% scatter about the median. 
    Solid and dashed colored lines include dust, while the dotted purple line shows dust-free results. 
    The spread in the galaxy population is large at all halo masses, but especially at $\Mhalo\lesssim10^{9.5}~\Msun$. 
    \textit{Right:} The evolution of the median trend in $\Muv$--$\Mhalo$ in BonFIRE+CampFIRE over $9\lesssim z\lesssim25$. 
    The median trend does not significantly evolve at $\Mhalo\gtrsim10^9~\Msun$, but galaxies in lower-mass halos tend to be brighter at earlier times. 
    We show the median trend from FIREbox-HR using FIRE-2 physics for reference \citep{Feldmann2025}.} 
    \label{fig:MuvMhalo}
\end{figure*}

\subsection{UV luminosity}\label{sec:results_uv}

Rest–frame UV luminosities of high-redshift galaxies serve as a key benchmark for galaxy formation and understanding the role of galaxies in reionization. 
In this section we explore $\Muv$ as a function of halo mass, compare the UV luminosity function (UVLF) to other models and observations, and then examine the variability of $\Muv$. 
Recall from Section~\ref{sec:sims_uv} that we report UV magnitudes at the rest-frame wavelength of 1500~\AA~and we apply an analytical correction for dust attenuation to our UV magnitudes, which is based solely on halo mass. 
Throughout the figures pertaining to UV luminosity, we show the dust-corrected results for BonFIRE+CampFIRE as a solid purple line, and the results without dust attenuation as a dotted purple line.

The left panel of Figure~\ref{fig:MuvMhalo} shows the median relationship between absolute UV magnitude and halo mass ($\Muv$--$\Mhalo$) in BonFIRE+CampFIRE, BonFIRE, and CampFIRE-800 at $z\sim9$. 
As in other figures, the colored lines and shaded regions are medians and 68\% scatter, and the vertical grey lines are the halo occupation limits for CampFIRE-800 and BonFIRE. 
The horizontal grey lines mark approximate observational limits for different depths of \JWST\ surveys.

We find consistent results across the simulation suite, where the median $\Muv$ essentially monotonically increases with $\Mhalo$ until dust attenuation becomes significant ($\sim0.5$--$2$~mag) at $\Mhalo\gtrsim10^{10}~\Msun$. 
Scatter in CampFIRE-800 is consistently $\sim1$--$2$~mag at all halo masses, whereas, scatter BonFIRE+CampFIRE and BonFIRE increases to $\gtrsim3$~mag at $\Mhalo\lesssim10^9~\Msun$. 
The larger scatter in the latter data sets is likely due to a combination of lower resolution and a wider variety of low-mass halos/environments within the larger BonFIRE volume.

In the right panel of Figure~\ref{fig:MuvMhalo}, we show the evolution of the median $\Muv$--$\Mhalo$ relation in BonFIRE+CampFIRE at $9\lesssim z\lesssim25$. 
We show that the median $\Muv$--$\Mhalo$ relation does not significantly evolve at $\Mhalo\gtrsim10^9~\Msun$, except to flatten at the bright end at later times ($z\lesssim10$) once halos have grown enough to have significant dust.

However, the median $\Muv$--$\Mhalo$ relation at $\Mhalo\lesssim10^9~\Msun$ has two features worth noting: it is not a pure power law relation and it experiences significant evolution in the normalization and slope at $\Mhalo\lesssim10^9~\Msun$. 
The dark colored lines for $z\gtrsim17$ indicate the dominating presence of the previously discussed population of ultra-compact galaxies (UCGs) at $\Mstar\sim10^5$--$10^7~\Msun$ that form rapidly in roughly single starbursts within atomic cooling halos (Section~\ref{sec:results_ucg}). 
They appear brighter than galaxies in halos of similar halo mass that form at later times because of the highly clustered and short-lived bursts in which their stars formed and the boosted UV luminosities of Pop~III stars. 
These objects quickly fade and contribute to the large scatter to low luminosities visible in the left panel at $\Mhalo\sim10^8$--$10^9~\Msun$. 
The flattening of these lines is largely a combination of low number statistics, resolution (as the resampling correction is ineffective at early times), and our choice to show the relations only for halos that are occupied by galaxies; including dark halos would induce a steeper slope.

We compare our predictions for the median $\Muv$--$\Mhalo$ relation to FIREbox-HR (dust-corrected and binned similarly) at $z=10$ \citep{Feldmann2025} in the right panel of Figure~\ref{fig:MuvMhalo}. 
BonFIRE+CampFIRE spans a wider dynamic range in both halo mass and luminosity than FIREbox-HR, owing to BonFIRE’s 6.5 times larger volume at the massive end and CampFIRE's 10 times higher mass resolution at the faint end. 
For a fair comparison, we focus on the BonFIRE+CampFIRE results at $9\lesssim z\lesssim10$ (lightest colors).

Both simulations show the expected trend that more massive halos host brighter galaxies, but the detailed slopes differ across halo mass. 
At the high-mass end ($\Mhalo\gtrsim10^{10}~\Msun$), FIREbox-HR exhibits a pronounced flattening that sets in at $\sim0.5$~dex lower halo mass than in BonFIRE+CampFIRE. 
This difference persists even in dust-free comparisons, suggesting it is not solely driven by attenuation effects. 
Instead, it may reflect intrinsically burstier star formation in massive halos under FIRE-3 with a simple Pop~III model relative to FIRE-2, boosting peak UV luminosities at fixed halo mass despite broadly similar $\Mstar$--$\Mhalo$ relations.

At intermediate masses ($\Mhalo\sim10^9$--$10^{10}~\Msun$), FIREbox-HR galaxies are typically $\sim1$--2~mag brighter than those in BonFIRE+CampFIRE, reversing the trend seen at both higher and lower masses. 
This transition likely reflects differences in the timing and regulation of star formation, with the delayed onset of star formation in FIRE-2 \citep{Hopkins2023} potentially producing brighter galaxies in this mass range at fixed redshift.

At the low-mass end ($\Mhalo\lesssim10^9~\Msun$), BonFIRE+CampFIRE galaxies are $\sim0.5$--2~mag brighter at fixed halo mass. 
This offset is likely driven by the population of ultra-compact galaxies (UCGs), which undergo rapid, short-lived ($\lesssim10~\Myr$) star formation episodes that elevate their instantaneous UV luminosities. 
As a result, the low-mass $\Muv$--$\Mhalo$ relation in BonFIRE+CampFIRE reflects not only average star formation efficiency but also the enhanced burstiness and compactness of early star-forming systems in the FIRE-3 model.

\begin{figure*}
\centering
\includegraphics[width=\textwidth]{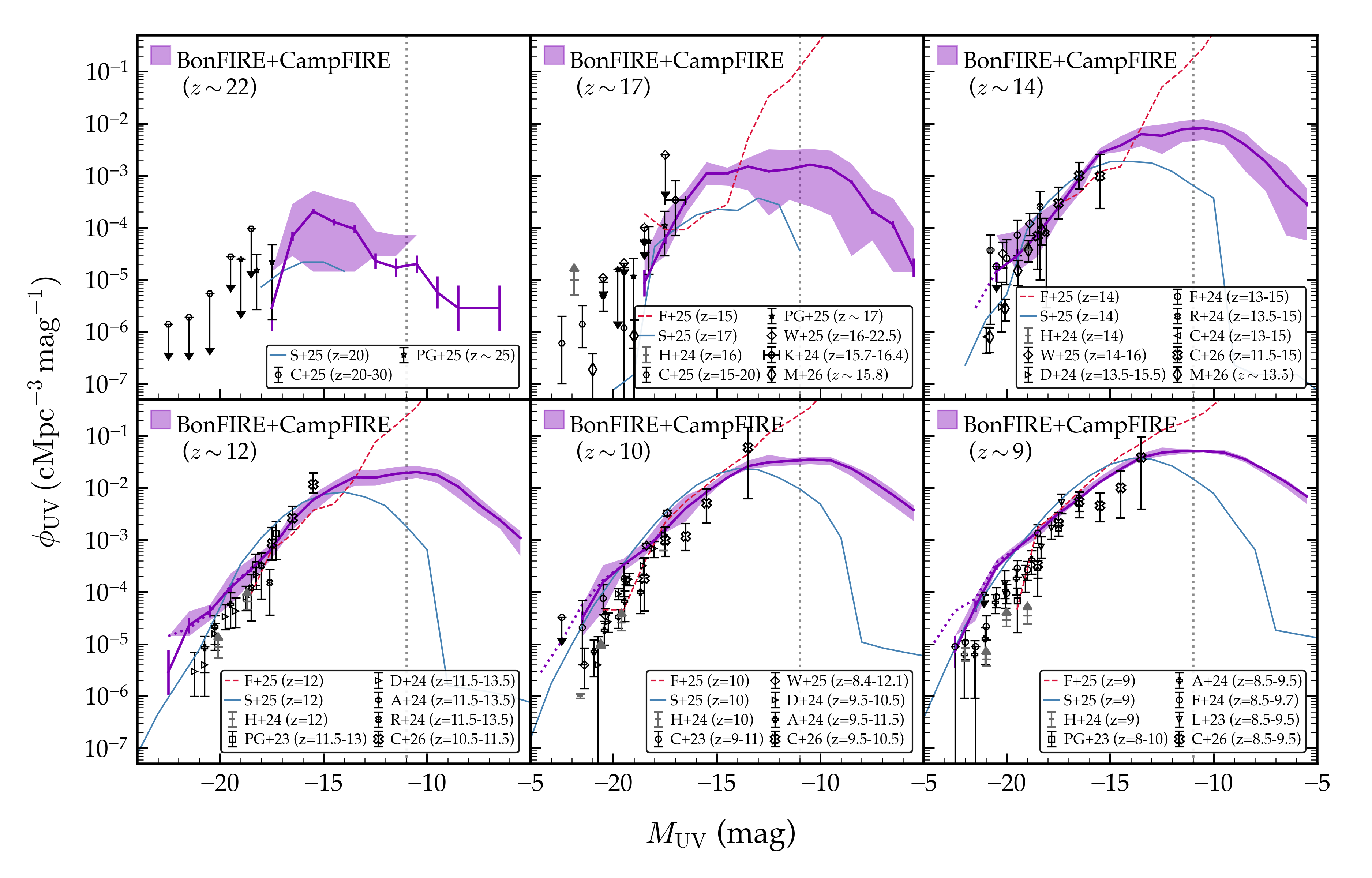}
\caption{
The BonFIRE+CampFIRE UV luminosity functions (UVLFs) at $9\lesssim z\lesssim25$. 
In each panel, we show a purple line representing the UVLF stacked over $\sim60~\Myr$ with the average redshift indicated in the top left legend. 
We include a dust correction in our fiducial results, but we show results without dust as dotted purple lines (with minimal effects at faint magnitudes). 
The purple band is the maximal and minimal (dust-corrected) UVLF over the same time period. 
We mark a resolution convergence limit with the grey dotted line at $\Muv=-11$. 
We compare to various spectroscopically-confirmed high-redshift galaxies \citep[grey points,][]{Harikane2024a} and photometric candidates \citep[black points,][]{Leung2023,PerezGonzalez2023,Adams2024,Casey2024,Donnan2024,Finkelstein2024,Robertson2024,Castellano2025,Kokorev2025,PerezGonzalez2025,Whitler2025,Chemerynska2026,McLeod2026}. 
Other colored lines are theoretical models from \cite{Somerville2025} and FIREbox-HR \citep{Feldmann2025}. 
}
\label{fig:uvlf_multipanel}
\end{figure*}

\subsubsection{UV luminosity functions}

Using BonFIRE+CampFIRE, we self-consistently predict the UVLF at $9\lesssim z\lesssim25$, leveraging the statistical sample of halos in BonFIRE and resampling their UV magnitudes from the high-resolution CampFIRE-800 run as described in Section~\ref{sec:sims_rescorr} to correct for resolution–induced discreteness at the faint end (details in Appendix~\ref{app:resampling}). 
The BonFIRE+CampFIRE UVLFs span $-23\lesssim \Muv \lesssim-5$ and capture the emergence and evolution of galaxies driving the ionizing photon budget in the first 500 Myr of cosmic history.

Based on a comparison of the UVLFs of BonFIRE, CampFIRE-6k, and CampFIRE-800 within the CampFIRE subregion, we assess resolution convergence across the suite (see Appendix~\ref{app:restests}). 
BonFIRE appears converged down to $\Muv \sim -15$, while CampFIRE-6k extends this limit to $\Muv \sim -13$. 
By ansatz, we adopt $\Muv=-11$ (vertical dotted grey lines in Figure~\ref{fig:uvlf_multipanel}) as a conservative convergence limit for CampFIRE-800, and therefore for the combined BonFIRE+CampFIRE results shown here.

Our analytical dust attenuation correction (a function halo mass, Equation~\ref{eq:schechter}), primarily affects the bright end. 
In particular, dust can shift the normalization of the UVLF down by $\sim0.1-0.5$ dex at $\Muv \lesssim -20$, but it has a negligible impact at fainter magnitudes. 
We show dust-free results as dotted purple lines for BonFIRE+CampFIRE.

Figure~\ref{fig:uvlf_multipanel} shows the BonFIRE+CampFIRE UVLFs as purple bands/lines in six redshift bins. 
We indicate the median redshift of each group of snapshots in a panel in the top left legends. 
We construct the bands by taking the minimum and maximum number density in each magnitude bin across five snapshots spanning $\sim60~\Myr$. 
The lines show our stacked results for each redshift bin. 
For each redshift bin, we stack UVLFs by summing galaxy counts in bins of $\Delta M_{\rm UV} = 1$ mag across snapshots and dividing by the total comoving volume of those snapshots, similar to our procedure for the SMF in Section~\ref{sec:results_mstar}. 
We show Poisson errors on our stacks as vertical lines, which we propagate from the galaxy number counts in each stack.

At $z\gtrsim11$ (top three panels and bottom left panel of Figure~\ref{fig:uvlf_multipanel}), the bright end ($\Muv\lesssim-17$) grows steadily in abundance as increasingly massive halos assemble and sustain high star formation rates ($\gtrsim1-10~\Msun~\rm{yr}^{-1}$). 
In contrast, at $z\lesssim10$ (bottom middle and right panels) the bright end largely stabilizes in number density, as also seen in Figure~\ref{fig:MuvMhalo}, indicating that the most massive systems may have reached a quasi-equilibrium between gas accretion and feedback regulation. 
During this later phase, a systematic excess of bright galaxies in BonFIRE+CampFIRE relative to previous FIRE-2 models and some observational measurements becomes more apparent. 
We comment further on this excess below.

The UVLF exhibits a smooth shape that steepens toward higher redshift and shows a faint-end turnover at $\Muv \gtrsim -14$. 
The UVLF remains well populated down to and beyond our estimated resolution convergence limit at $\Muv \approx -11$, implying that faint galaxies remain abundant even at $z>10$ and likely contribute significantly to reionization. 
The location of the turnover does not significantly evolve with redshift, suggesting that it is set by a combination of feedback-regulated star formation and resolution limits. 
We quantify this behavior further in Section~\ref{sec:results_uv_fits} by fitting our UVLFs.

The scatter in the UVLF (from variation over $\sim60~\Myr$ in each panel) decreases markedly over time. 
At $z\gtrsim15$, the number density at fixed $\Muv$ varies by up to $\sim1$–2 dex, possibly reflecting highly stochastic star formation and diverse assembly histories at early times. 
By $z\sim9$, this scatter decreases to $\lesssim0.1$ dex, indicating that galaxy growth becomes more regulated and uniform, but also reflecting a much larger number of galaxies contributing to the UVLF. 
This trend is most evident at the faint end and mirrors the decreasing scatter in $\Muv$ at fixed halo mass (Figure~\ref{fig:MuvMhalo}). 
We explore the origin of this evolution further in Section~\ref{sec:results_uv_var}.

We compare the BonFIRE+CampFIRE UVLFs with recent \JWST\ measurements in Figure~\ref{fig:uvlf_multipanel}.
We show galaxies with spectroscopically-confirmed redshifts from \citep{Harikane2024a} as grey points, and high-redshift photometric candidates as black points \citep{PerezGonzalez2023,Adams2024,Casey2024,Donnan2024,Finkelstein2024,Robertson2024,Castellano2025,Kokorev2025,PerezGonzalez2025,Whitler2025,Chemerynska2026,McLeod2026}.

At $9\lesssim z\lesssim12$, the simulations agree well with the observed UVLF over $-20 \lesssim \Muv \lesssim -14$, reproducing both the slope and normalization within $\sim0.2$ dex. 
At brighter magnitudes ($\Muv \lesssim -20$), BonFIRE+CampFIRE modestly overpredicts the abundance of galaxies by $\sim0.3$–0.5 dex. 
This discrepancy decreases toward higher redshift, where the simulations and observations converge at $z\gtrsim13$. 
We speculate that the high star formation efficiencies, lack of dust, and enrichment from Pop~III star formation at earlier times in our simulations may cause this bright-end excess, which we discuss further in Section\ref{sec:discussion}.

At the faint end ($\Muv \gtrsim -14$), BonFIRE+CampFIRE predicts a somewhat shallower slope than many current observational inferences, though this difference remains within the large systematic uncertainties associated with completeness corrections and lensing magnification. 
In particular, recent GLIMPSE constraints from \citet{Chemerynska2026}, which probe higher redshifts closer to those examined here, remain broadly consistent with our predictions within observational uncertainties. 
However, \citet{Atek2026} find no evidence for a turnover in the GLIMPSE UVLF down to $\Muv\sim-12$ at $z\sim7$, suggesting a steeper faint-end slope than predicted by our simulations. 
We note that the comparison is not direct though, as our predictions focus on $z\gtrsim9$, where the faint-end UVLF remains less well constrained observationally. 
If future observations confirm a similarly steep faint-end slope at $z\gtrsim9$, this could suggest that our simulations underproduce low-luminosity galaxies.

Overall, the similarities between our simulated UVLFs and observations in terms of normalization and shape at $9 \lesssim z \lesssim 25$ indicates that BonFIRE+CampFIRE captures the dominant physical mechanisms governing early galaxy formation, with some tension at the extreme bright end.
We emphasize that many of the observational constraints, especially at $z\gtrsim14$, rely on photometric samples that are subject to significant uncertainty. 
Some fraction of these sources may be low-redshift interlopers, which would shift them to lower-redshift bins and/or reduce their inferred luminosities, improving agreement with our`  ` simulations.

Comparing to FIRE-2 UVLFs from FIREbox-HR \citep{Feldmann2025}, we find broad agreement in overall normalization and shape at the bright end ($-19 \lesssim \Muv \lesssim -15$) at $9\lesssim z\lesssim12$, but with systematic offsets that grow toward higher luminosity. 
In particular, BonFIRE+CampFIRE exceeds FIRE-2 predictions by up to $\sim0.5$–1 dex at $\Muv \lesssim -20$ and exhibits a flatter bright-end slope, indicating a larger population of luminous galaxies. 
At the faint end, however, BonFIRE+CampFIRE shows a significant deficit relative to FIREbox-HR and FIRE-2 zoom-in simulations \citep{Ma2019}, with differences of $\gtrsim1$ dex at $\Muv\gtrsim-15$. 
We note that \citet{Sun2023a} also analyzed the high-redshift UVLF and the effects of burstiness on it using the same FIRE-2 zoom-in simulations as \citet{Ma2018b,Ma2019}, but we show only the UVLF from \citet{Ma2019} here for simplicity. 
This reflects a much earlier turnover in our UVLFs, which occurs near $\Muv\sim-14$, compared to $\Muv\sim-8$ in FIRE-2.

The bright-end excess and brighter turnover that we find in our simulations relative to FIRE-2 likely reflect the trends we identify in the $\Muv$--$\Mhalo$ relation (Figure~\ref{fig:MuvMhalo}). 
At high halo masses, the lack of flattening in the BonFIRE+CampFIRE $\Muv$--$\Mhalo$ relation relative to FIREbox-HR leads to systematically brighter galaxies and enhances the abundance of luminous systems. 
At low halo masses, ultra-compact, burst-dominated galaxies shift the population to brighter $\Muv$ at fixed mass, while the brighter turnover reflects the combined effects of stellar feedback and limited resolution in suppressing sustained star formation in small halos. 
In contrast, the intermediate-mass regime where FIREbox-HR galaxies are brighter may contribute to localized agreement in the UVLF at $\Muv\sim-15$.

We note that the top panels (spanning $13\lesssim z\lesssim25$) include many galaxies where we are unable to resample their properties from the CampFIRE-800 simulation because of low numbers of matching galaxies in our resampling scheme.
Therefore, the top panels are more susceptible to resolution issues, particularly burstiness and numerically-driven UV brightness in compact and quickly-forming galaxies.
This effect is apparent as an enhancement in the number density of galaxies at $-16\lesssim\Muv\lesssim-15$, a regime where the BonFIRE+CampFIRE results also predict a higher number density of galaxies compared to FIREbox-HR \citep{Feldmann2025}.
We caution the reader not to over-interpret the difference between the FIRE-2 and FIRE-3 large volume runs here, and in general to treat the top row as more speculative predictions, given resolution limitations.

Comparison to the similarly-shaped UVLF of the density-modulated star formation efficiency (DMSFE) model of \citet{Somerville2025} provides additional insight on our results. 
DMSFE predicts a comparably rapid buildup of the bright end at high redshift, but in some cases exceeds our UVLF by up to $\sim0.5$–1~dex for galaxies with $\Muv\lesssim-18$ at $z\lesssim12$. 
This behavior reflects the model’s assumption that star formation efficiency increases strongly with gas density, leading to a significant population of luminous galaxies in rare, overdense regions, even at early times. 
At the same time, DMSFE generally predicts a lower normalization at the faint end compared to BonFIRE+CampFIRE, as lower-density environments are less efficient at forming stars in their model.

Overall, BonFIRE+CampFIRE occupies an intermediate regime between FIREbox-HR, which predicts a high faint-end normalization, and DMSFE, which enhances the bright end while suppressing low-mass galaxy formation. 
This behavior likely reflects the FIRE-3 treatment of low-metallicity gas, where clustered star formation proceeds early and efficiently, and our simple Pop~III model, which boosts luminosities of the first galaxies in our simulations. 
Bursty and clustered star formation then continues to occur in low-mass halos, keeping the faint end populated.
Meanwhile, the early enrichment from Pop~III stars fuels efficient cooling and star formation in massive halos. 
As a result, BonFIRE+CampFIRE produces both a substantial faint galaxy population and a modest bright-end excess, implying that star formation is somewhat suppressed in low-mass halos and concentrated in rare, massive systems. 
These differences highlight the role of resolution, small-scale star formation, and metal enrichment physics in shaping the $\Muv$--$\Mhalo$ relation and, in turn, the UVLF.

\begin{figure}
    \centering
    \includegraphics[width=\columnwidth]{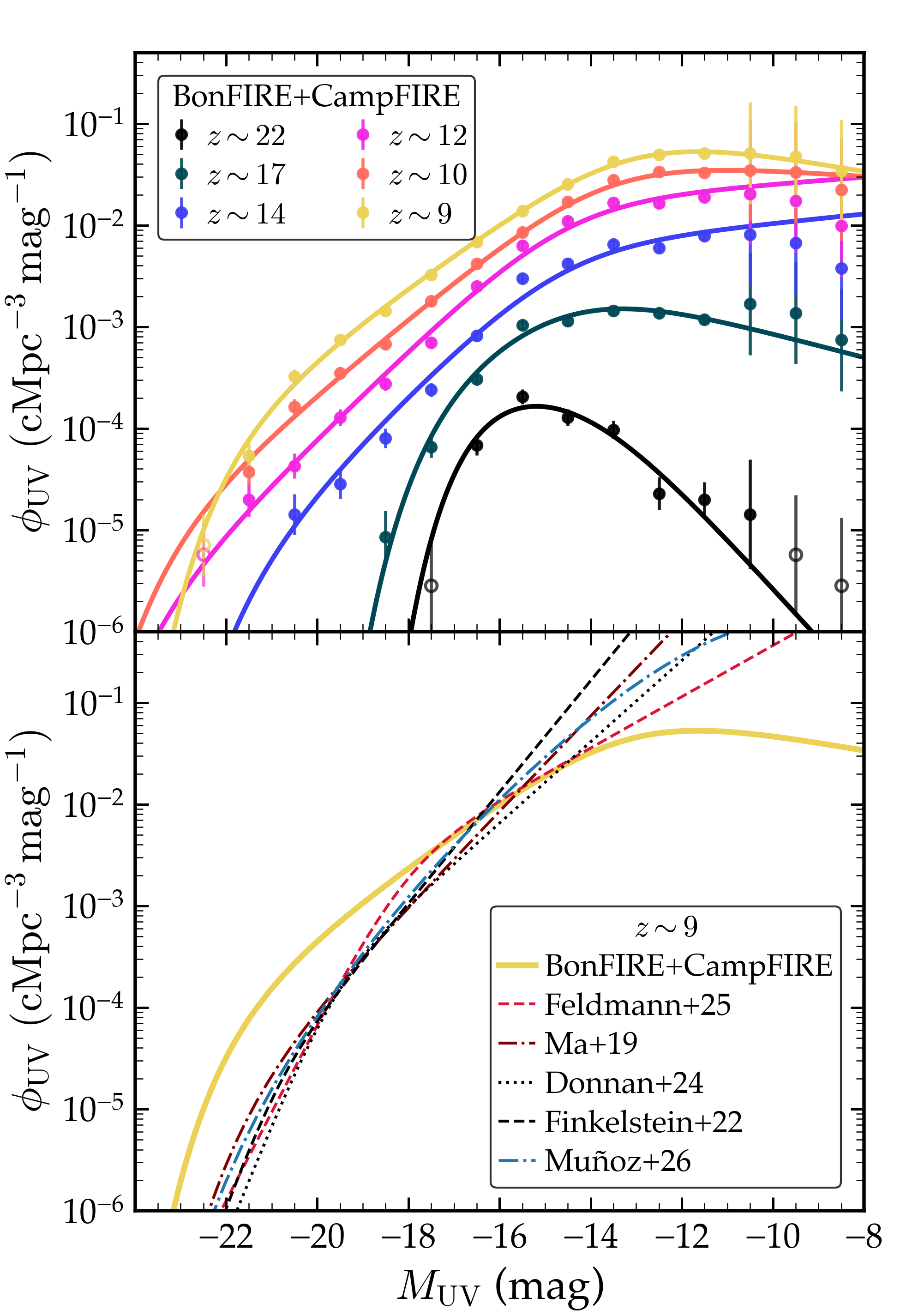}
    \caption{\textit{Top:} Stacked BonFIRE+CampFIRE UVLFs in six redshift bins (points) with best-fit Schechter with turnover models (solid curves, Eq.~\ref{eq:schechter}). 
    Redshift bins are the same as in Figure~\ref{fig:uvlf_multipanel}. 
    Filled symbols indicate bins included in the fit, and open symbols are bins with low number counts that we exclude from fitting. 
    \textit{Bottom:} Comparison of our best-fit UVLFs at $z\sim9$ with predictions from FIREbox-HR \citep[DPL with mass-dependent scatter][]{Feldmann2025}, FIRE-2 zoom-in simulations \citep{Ma2019}, an empirical model \citep{Munoz2026}, and observational inferences using \JWST\ data \citep{Finkelstein2022,Donnan2024}.}
    \label{fig:uvlf_fits}
\end{figure}

\subsubsection{Fits to BonFIRE+CampFIRE UVLFs}\label{sec:results_uv_fits}

To quantify the shape of the BonFIRE+CampFIRE UVLFs, we fit them using a Schechter function with a low-luminosity turnover, providing a smooth transition at the faint end. 
As a reminder, we apply an analytical dust correction to UV magnitudes of galaxies (Section~\ref{sec:sims_uv}).

We stack snapshots into the same six redshift bins as in Figure~\ref{fig:uvlf_multipanel}, with each bin covering five snapshots and spanning a total of $60~\rm{Myr}$. 
In addition to the Poisson errors in Figure~\ref{fig:uvlf_multipanel}, we also include a small global systematic error floor of 0.05 dex on all our data points to provide flexibility in fitting, as well as an additional 0.5 dex uncertainty for bins fainter than $M_{\rm UV}=-11$ to account for resolution limitations. 
We omit from the fitting procedure bins with fewer than three galaxies in the stack (omitted points are unfilled in Figure~\ref{fig:uvlf_fits}).

We fit a modified Schechter function with a smooth faint-end turnover, relying on standard modified Schechter parameterizations used to model faint-end suppression in high-redshift UV luminosity functions \citep[e.g.,][]{Atek2018}. 
Our suppression factor takes the form $(1 + 10^{0.4(\Muv-M_{\rm turn})})^{-1}$, which preserves the standard Schechter behavior at the bright end while introducing a characteristic turnover scale. 
In this parameterization, the luminosity function is suppressed by a factor of two at $\Muv = M_{\rm turn}$ and transitions smoothly to a shallower slope at fainter magnitudes. 
Compared to more general modified Schechter forms that include an additional curvature parameter (such as $\beta$ in \citet{Atek2018}), our form reduces the number of free parameters by fixing the shape of the turnover ($\beta=1$). 
We adopt the following parameterization for the Schechter function with faint--end turnover: 
\begin{align}
    \phi(M_{\mathrm{UV}}) = & \; 0.4 \ln(10) ~ \phi_*
    \left[ 10^{0.4(M_* - M_{\mathrm{UV}})} \right]^{\alpha + 1} \nonumber \\
    & \times \exp\!\left[-10^{0.4(M_* - M_{\mathrm{UV}})}\right] \nonumber \\
    & \times \left[ 1 + 10^{0.4(M_{\mathrm{UV}} - M_{\mathrm{turn}})} \right]^{-1}.\label{eq:schechter}
\end{align}
This functional form allows us to quantify the evolution of the characteristic magnitude ($M_*$), normalization ($\phi_*$), and faint-end slope ($\alpha$), as well as the turnover magnitude ($M_{\rm turn}$) that signals the transition to unresolved and/or feedback-suppressed star formation. 
We comment on less successful attempts to fit other functional forms in Appendix~\ref{app:fits}.
The log-likelihood is
\begin{equation}
\ln \mathcal{L}
= -\frac{1}{2}
\sum_i
\left[
\frac{\log_{10}\phi_i - \log_{10}\phi_{\rm model}(M_i)}
{0.434~\sigma_{\phi,i}/\phi_i}
\right]^2,
\end{equation}
where the sum runs over magnitude bins.
We adopt the following uniform priors and enforce that the turnover magnitude is fainter than the characteristic magnitude:
\begin{align}
-3.5 &< \alpha < -0.5, \\
-12 &< \log_{10}\phi_* < -1, \\
-26 &< M_* < -16, \\
-16 &< M_{\rm turn} < -7, \\
M_{\rm turn} &> M_*.
\end{align}

For each stacked UVLF, we first perform a coarse grid search over parameter space to identify a high--probability starting point. 
We refine the choice of model parameters via maximum--likelihood optimization, sampling the posterior distribution using Markov Chain Monte Carlo (MCMC) with the \texttt{emcee} sampler \citep{ForemanMackey2024}.

We present the best--fitting UVLF model parameters in Table~\ref{tab:schechter_params} as medians and 68\% credible intervals over the marginalized posterior distributions.

Figure~\ref{fig:uvlf_fits} illustrates our UVLF fits and compares them to other models and fits to observations.
The top panel shows the best-fit UVLFs to the BonFIRE+CampFIRE results at $9 \lesssim z \lesssim 22$, illustrating the redshift evolution in both normalization and shape. 
The fits exhibit a steadily increasing normalization toward lower redshift, along with a persistent faint-end turnover at $\Muv \sim -14$. 
In the bottom panel, we compare our $z\sim9$ results to fits from observations and other theoretical models. 
Although all models broadly agree at intermediate magnitudes, BonFIRE+CampFIRE stands out by predicting both an excess of bright galaxies and a more pronounced turnover at the faint end.

\begin{table}
\centering
\caption{Best-fit parameters of our Schechter function with faint-end turnover (Eq.~\ref{eq:schechter}) in each redshift bin. 
The normalization is given as $\log_{10}(\phi_* / \mathrm{Mpc}^{-3}~\mathrm{mag}^{-1})$. Uncertainties are 68\% credible intervals from MCMC.}
\begin{tabular}{c c c c c}
\hline
$z$ & $\alpha$ & $\log_{10}\phi_*$ & $M_*$ & $M_{\rm turn}$ \\
\hline
22 & $-0.76^{+0.17}_{-0.18}$ & $-3.32^{+0.13}_{-0.12}$ & $-15.92^{+0.29}_{-0.41}$ & $-14.14^{+1.08}_{-0.61}$ \\
17 & $-1.70^{+0.14}_{-0.08}$ & $-3.26^{+0.11}_{-0.12}$ & $-17.08^{+0.18}_{-0.20}$ & $-14.39^{+0.80}_{-0.43}$ \\
14 & $-2.11^{+0.05}_{-0.04}$ & $-5.13^{+0.49}_{-1.00}$ & $-21.38^{+0.94}_{-2.12}$ & $-14.70^{+0.30}_{-0.21}$ \\
12 & $-2.08^{+0.04}_{-0.04}$ & $-5.62^{+0.51}_{-0.43}$ & $-23.61^{+1.09}_{-0.95}$ & $-14.57^{+0.29}_{-0.25}$ \\
10 & $-1.92^{+0.05}_{-0.05}$ & $-4.81^{+0.46}_{-0.48}$ & $-23.22^{+1.04}_{-1.17}$ & $-13.72^{+0.36}_{-0.37}$ \\
9 & $-1.80^{+0.05}_{-0.06}$ & $-3.75^{+0.23}_{-0.38}$ & $-21.65^{+0.44}_{-0.82}$ & $-13.18^{+0.42}_{-0.43}$ \\
\hline
\end{tabular}
\label{tab:schechter_params}
\end{table}

Our best-fit UVLF parameters are, however, broadly consistent with recent \JWST\ measurements at $9\lesssim z \lesssim 13$ from CEERS and JADES \citep{Finkelstein2023,Bouwens2023,Robertson2023}, as well as with predictions from high-resolution FIRE-2 galaxy formation simulations \citep{Ma2018b,Ma2019,Sun2023a,Feldmann2025} and an empirical model for bursty star formation \citep{Munoz2026}. 
In particular, we recover steep faint-end slopes ($\alpha \approx -1.8$ to $-2.0$) and characteristic magnitudes ($M_* \approx -21$ to $-24$) consistent with observational constraints over this redshift range, although our values of $M_*$ tend toward the brighter end of the allowed range, indicative of our bright-end excess. 

We further find evidence for a consistent low-luminosity turnover at $\Muv \approx -13$ to $-15$, comparable to the scales at which simulations predict suppressed star formation in low-mass halos due to feedback and inefficient gas cooling \citep[e.g.,][]{Wise2014,Ocvirk2016,Katz2025}. 
This turnover range is well above our nominal resolution limit of $\Muv=-11$ in CampFIRE-800 (see Appendix~\ref{app:restests}), so we consider it resolved. 
Our turnover lies well below most current \JWST\ detection limits, and existing lensing-based measurements at $z\gtrsim9$ such as from GLIMPSE \citep{Chemerynska2026} only tentatively probe flattening at $\Muv\gtrsim-15$, with significant systematic uncertainties.
Our inferred turnover at $\Muv\sim-13$ to $-15$ therefore represents a concrete prediction for future ultra-deep and lensing-assisted surveys, which will be required to distinguish between continued steep faint-end growth and a true flattening of the UVLF at $z\gtrsim9$. 
Finally, the turnover scale is broadly consistent with expectations from near-field cosmology, where suppressed star formation in low-mass halos is required to reproduce the satellite populations of Local Group galaxies \citep[e.g.,][]{Weisz2014,BK2015}, suggesting a connection between the faint end of the high-redshift UVLF and present-day dwarf galaxy formation.

\begin{figure}
\centering
\includegraphics[width=\columnwidth]{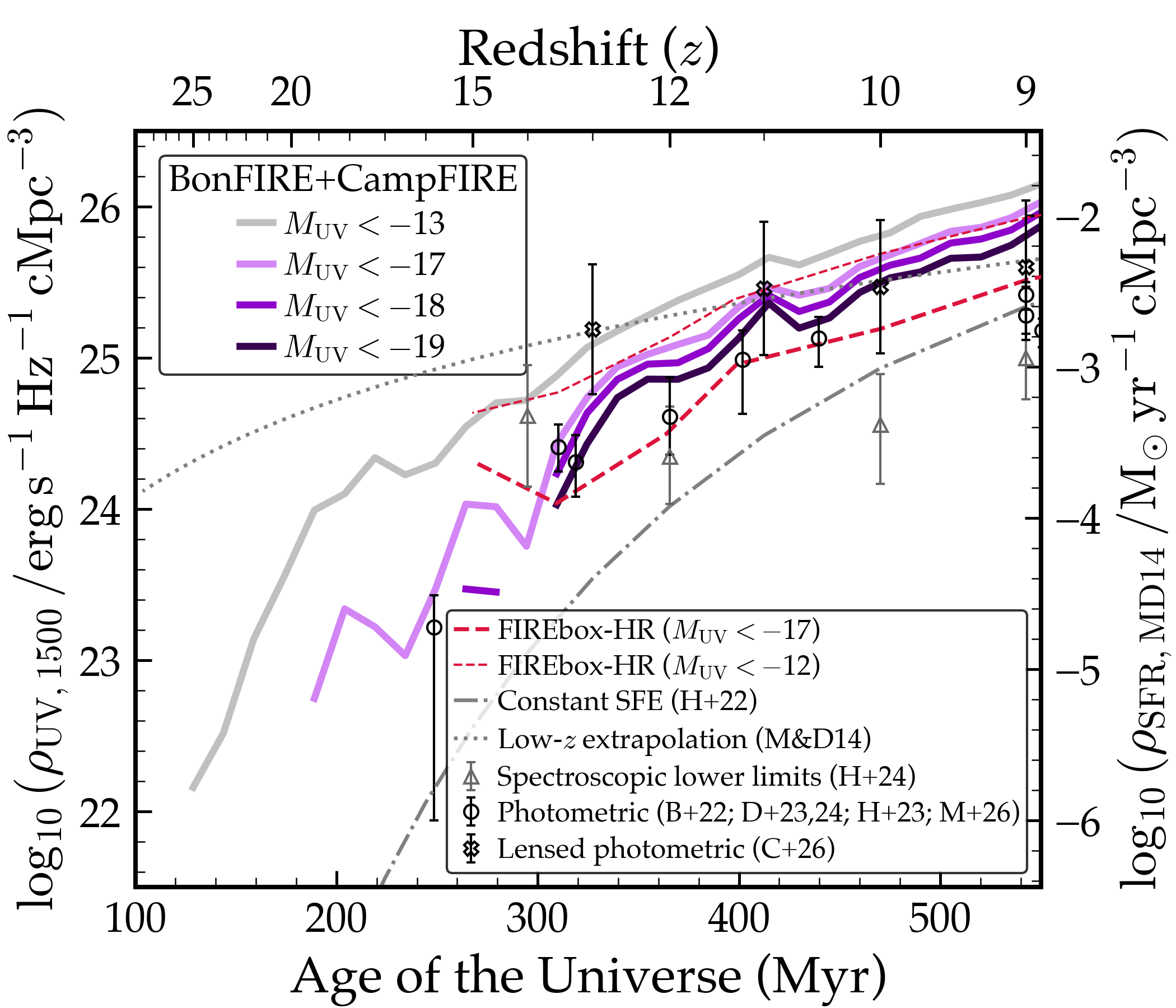}
\caption{The UV luminosity density and corresponding star formation rate density (SFRD) in BonFIRE+CampFIRE at $9\lesssim z\lesssim25$. 
We compare to theoretical results from FIREbox-HR \citep{Feldmann2025} and \citet{MadauDickinson2014}, as well as observational results from \citet{Harikane2022,Bouwens2022,Donnan2023,Donnan2024,Harikane2023,Harikane2024a,Chemerynska2026,McLeod2026}.
BonFIRE+CampFIRE is consistent with most photometric inferences, though offset to higher values relative to observations and FIREbox-HR due to the dominance of bright galaxies in our sample, most notably at late times.}
\label{fig:rhoUV}
\end{figure}

\subsubsection{UV star formation rate density}

In Figure~\ref{fig:rhoUV} we show the total UV luminosity density ($\rho_{\rm UV}$) and corresponding star formation rate density (SFRD) in BonFIRE+CampFIRE as a function of time. 
To calculate this we sum up the UV luminosity of all galaxies in BonFIRE+CampFIRE at each snapshot and divide by the BonFIRE volume. 
We convert the UV luminosity density to star formation rate density using the conversion factor $\kappa = 1.15 \times 10^{-28}~\mathrm{\Msun~yr^{-1}~(erg~s^{-1}~Hz^{-1})^{-1}}$ from \citet{MadauDickinson2014}.

We show our results for four magnitude limits, three at the bright end and one at the low end to compare to faint, lensed sources. 
The small changes to our results using the different bright limiting magnitudes indicate that the total SFR/UV luminosity density is dominated by the brightest galaxies. 
We note that our simplified dust correction (Section~\ref{sec:sims_uv}) may underestimate dust attenuation in some of the brightest galaxies, which could contribute to the overall high UV luminosity density in our simulations.

We compare to models (grey lines) from \citet{MadauDickinson2014,Harikane2022}. 
The \cite{MadauDickinson2014} UV+IR SFRD model provides a widely used empirical fit to the cosmic star formation history, characterized by a smooth rise from $z=0$ to a peak at $z\sim2$ followed by a gradual decline toward higher redshift. 
Their parameterization is anchored to pre-\JWST\ data and effectively serves as an upper envelope for the SFRD at $z\gtrsim10$ when extrapolated. 
In contrast, the \cite{Harikane2022} SFRD model is based on early \JWST\ UV luminosity functions and implies systematically lower effective star formation efficiencies at high redshift compared to \cite{MadauDickinson2014}.

These models are best compared to the $\Muv<-18$ and $\Muv<-17$ lines from BonFIRE+CampFIRE, which share similar integration limits as the other models. 
At $z\gtrsim15$, we predict a $\rho_{\rm UV}$ that lies between the models of \citet{Harikane2022} and \citet{MadauDickinson2014}, and at lower redshifts BonFIRE+CampFIRE's $\rho_{\rm UV}$ grows above the model of \citet{MadauDickinson2014}. 
Interestingly, the overall shape of BonFIRE+CampFIRE's $\rho_{\rm UV}$ versus time is similar to the model of \citet{Harikane2022}, but offset to higher values by $\sim0.5-1$ dex. 

We compare to results from FIREbox-HR in Figure~\ref{fig:rhoUV}, which we show as red dashed lines for galaxies at $\Muv<-17$ and $\Muv<-12$ \citep{Feldmann2025}.
We do not show results from FIRE-2 zoom-in simulations, but we note that they generally agree with FIREbox-HR \citep{Sun2023a}.
We compare to our lightest purple line and grey line for BonFIRE+CampFIRE, at roughly the same magnitude limits, respectively.
BonFIRE+CampFIRE overshoots FIREbox-HR by about $0.1-0.5$~dex in these regimes, which comes from the excess of bright ($\Muv\lesssim-20$) galaxies in BonFIRE+CampFIRE relative to FIREbox-HR.

We further compare to photometrically-selected high-redshift galaxy candidates \citep[black points,][]{Bouwens2022,Donnan2023,Donnan2024,Harikane2023,Chemerynska2026,McLeod2026}, and lower limits from spectroscopically-confirmed high-redshift galaxies (grey points) from \citet{Harikane2024a}.
BonFIRE+CampFIRE is consistent with all but the highest redshift lower limit.
However, BonFIRE+CampFIRE lies slightly above most photometric candidates at $z\lesssim13$, in a manner similar to our overshoot relative to FIREbox-HR, and again consistent with our bright excess in the UVLF compared to observations.
However, our predictions for $\Muv<-13$ are consistent with all but the latest redshift bin from \citet{Chemerynska2026}, which probes to much fainter luminosities using gravitational lensing.
We note that at this much fainter integration limit, FIREbox-HR is also consistent with \citet{Chemerynska2026}, due to the large population of faint galaxies present in their sample.

\begin{figure}
    \centering
    \includegraphics[width=0.47\textwidth]{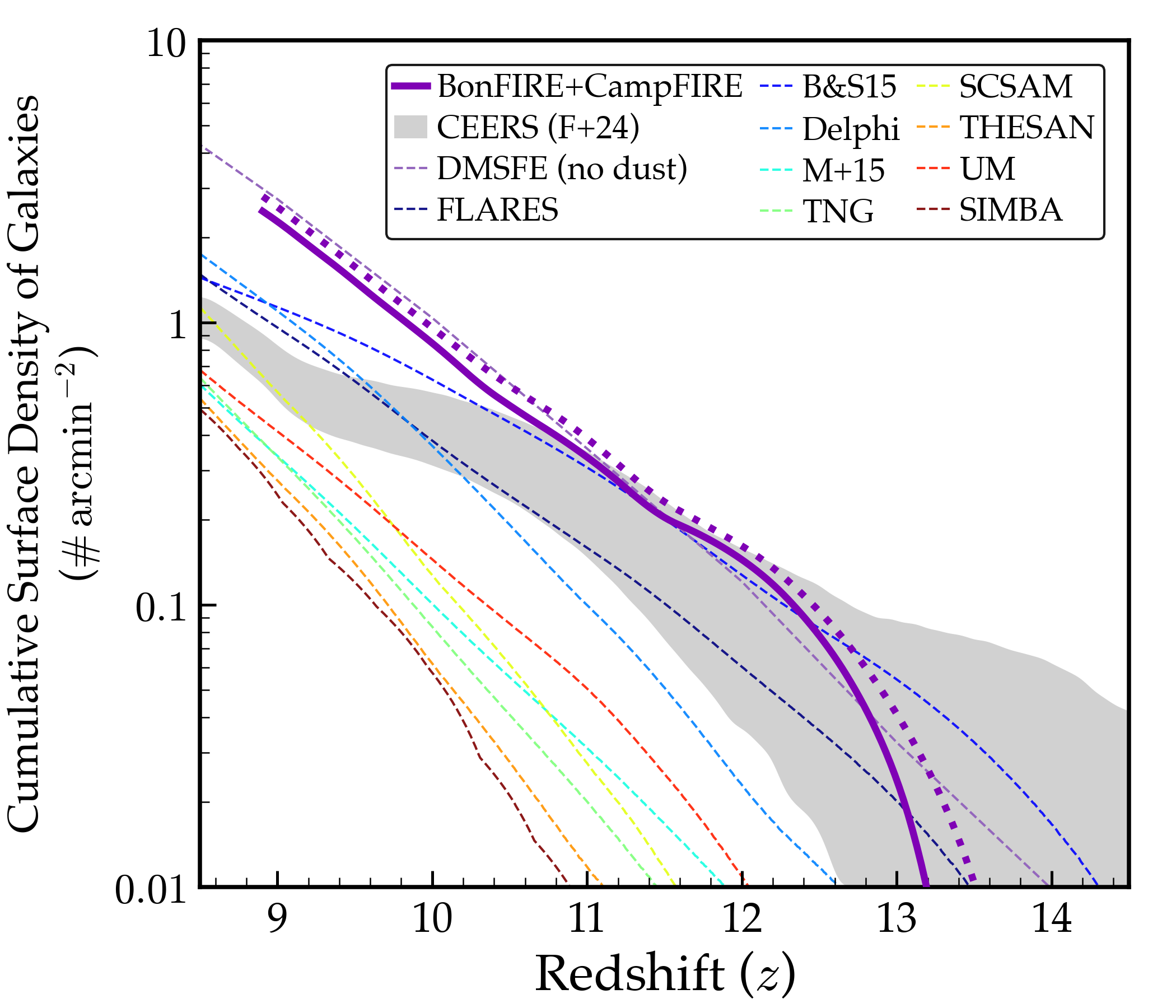}
    \caption{The cumulative surface density of galaxies at $m_{\rm F277W}<28.5$ mag. 
    We show results for BonFIRE+CampFIRE with (solid) and without (dotted) dust attenuation. 
    We compare to CEERS (shaded grey region) and several other theoretical models (see text for details). 
    At $z\gtrsim11$, BonFIRE+CampFIRE is consistent with the upper envelope of CEERS and theoretical models, but at lower redshift BonFIRE+CampFIRE predicts $\gtrsim0.5$ dex higher surface densities, similar to DMSFE \citep{Somerville2025}.}
    \label{fig:cumul_surf_density}
\end{figure}

\subsubsection{Detectability}\label{sec:results_detect}

We examine detectability of BonFIRE+CampFIRE galaxies by \JWST\ using a mock CEERS survey to measure the cumulative surface density of galaxies as a function of redshift.
We first calculate the differential (in redshift) surface density at each snapshot by counting the number of galaxies that are brighter than the limiting magnitude in CEERS ($m_{\rm F277W}=28.5$ mag) at the corresponding redshift of the snapshot, and then dividing by the projected on-sky surface area of the box and the redshift width of the box.
We then interpolate these values on a fine redshift grid ($\Delta z\approx0.03$), and integrate over redshift from $z\sim25$ to $z\sim9$.

Figure~\ref{fig:cumul_surf_density} shows our predicted cumulative surface density of galaxies as a function of redshift at $m_{\rm F277W}\leq28.5$ mag in BonFIRE+CampFIRE. 
We compare to CEERS constraints \citep{Finkelstein2024} and several theoretical models \citep{BehrooziSilk2015,Mason2015,Dayal2017,Behroozi2019,Dave2019,Yung2019,Kannan2022,Wilkins2023}. 
At $11 \lesssim z \lesssim 13$, both our dust-free and dust-attenuated predictions fall within the observational uncertainty range in CEERS. 
At $z\lesssim11$, BonFIRE+CampFIRE predicts a cumulative surface density of galaxies up to $\sim0.5$ dex higher than CEERS, consistent with our bright-end excess in the UVLF compared to observed high-redshift galaxies. 
At $z\gtrsim13$, the cumulative surface density decreases rapidly, reflecting the reduced number of luminous galaxies present at early times within the BonFIRE volume.

Relative to other models shown in Figure~\ref{fig:cumul_surf_density}, the BonFIRE+CampFIRE curves lie near the upper envelope of predicted number densities but remain within the range spanned by existing simulations and semi-analytic frameworks, closely following the DMSFE results at $z\lesssim13$ \citep{Somerville2025}. 
They are also broadly consistent with FLARES and DELPHI at intermediate redshifts, and track the higher-normalization predictions from empirical models such as \citet{BehrooziSilk2015} and \citet{Mason2015} at $z\sim9$–11. 
In contrast, models such as UniverseMachine, IllustrisTNG, and SIMBA predict systematically lower cumulative number densities across this redshift range, indicating less efficient or more delayed star formation in high-redshift halos.

\subsubsection{UV variability}\label{sec:results_uv_var}

We investigate the scatter in the UV luminosity of early galaxies in BonFIRE and CampFIRE by decomposing total scatter into two components: inter-halo scatter and intra-halo scatter. 
We first define the total variability, $\sigma_{\rm tot}$, using the instantaneous UV magnitudes of galaxies in each snapshot; we do not include a dust correction in this section to limit modeling choices. 
We divide galaxies into 0.5~dex halo mass bins and fit a linear relation between $\Muv$ and $\log_{10}(\Mhalo/\Msun)$ within each bin. 
We then compute the scatter (standard deviation) about this relation using the residuals such that our reported variability amplitudes measure scatter at fixed halo mass rather than reflecting the mean luminosity--mass trend across the bin.

To isolate longer-timescale halo-to-halo variability, we additionally compute ``de-bursted'' UV magnitudes constructed from temporally smoothed star formation histories, thereby suppressing short-timescale stochastic fluctuations associated with bursty star formation (without changing the total stellar mass). 
In our case, we reassigned the ages of star particles in a galaxy using a uniform distribution from the minimum to the 99th percentile of existing stellar ages, effectively imposing a constant SFR without choosing a specific timescale to smooth over. 
The scatter of these smoothed UV luminosities defines the inter-halo component, $\sigma_{\rm inter}$. 
We then infer the intra-halo component, $\sigma_{\rm intra}$, assuming an additive variance decomposition,
\begin{equation}
\sigma_{\rm total}^2=\sigma_{\rm inter}^2+\sigma_{\rm intra}^2.
\end{equation}
We perform this decomposition independently within each bin and for each simulation snapshot at $9<z<12$ (spanning $\sim175~\Myr$), requiring a minimum of three galaxies per halo mass bin. 
We also checked our results at higher redshift where statistics are worse, and we found little to no evolution.

\begin{figure}
    \centering
    \includegraphics[width=\columnwidth]{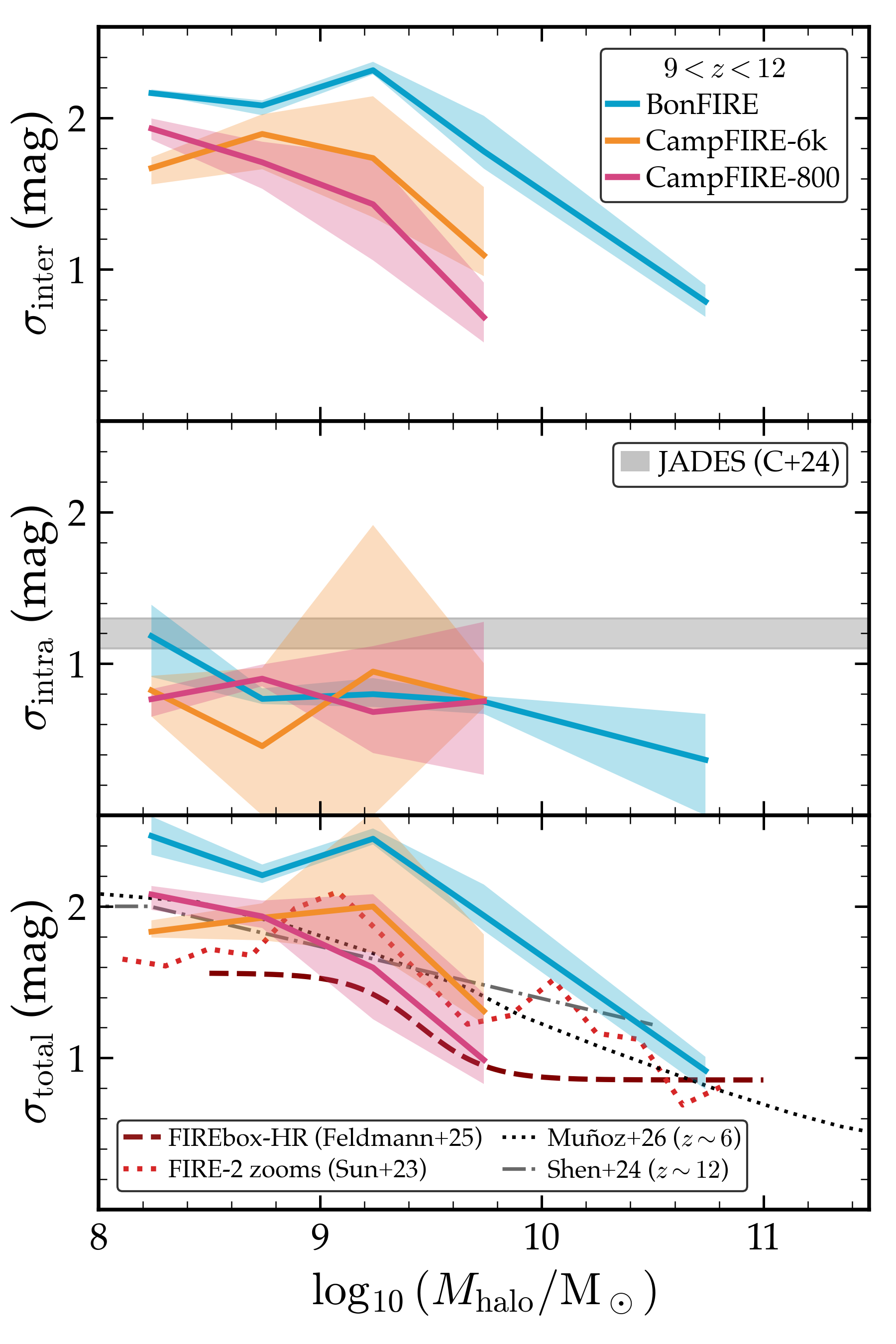}
    \caption{The UV luminosity variability of galaxies at $9<z<12$ as a function of halo mass in BonFIRE, CampFIRE-6k, and CampFIRE-800. 
    In all panels, solid colored lines show the median scatter measured in our simulations, and shaded regions indicate the 68\% scatter across snapshots. 
    \textit{Top:} The inter-halo scatter ($\sigma_{\rm inter}$, halo-to-halo variations for smooth star formation) is large in general ($\gtrsim1$~mag) and increases for lower-mass halos up to $\sim2$~mag. 
    \textit{Middle:} The intra-halo scatter ($\sigma_{\rm intra}$, inferred temporal variability within individual galaxies) is small ($\sim0.8$~mag) and relatively flat compared to $\sigma_{\rm inter}$. 
    The gray band shows an observational estimate of temporal variability ($\sim1.2$~mag) on 100~Myr timescales for JADES galaxies by \citet{Ciesla2024}.
    \textit{Bottom:} The total scatter ($\sigma_{\rm total}$, raw halo-to-halo variations including temporal variability). 
    Our results are in broad agreement with predictions from FIREbox-HR \citep{Feldmann2025}, FIRE-2 zoom-in simulations \citep{Sun2023b}, and empirical models \citep{Shen2024b,Munoz2026}, with variability decreasing toward higher halo masses.}
    \label{fig:Muv_var}
\end{figure}

Figure~\ref{fig:Muv_var} shows the median $\sigma_{\rm inter}$, $\sigma_{\rm intra}$, and $\sigma_{\rm total}$ as a function of halo mass at $9<z<12$ in each base simulation in our suite. 
The shaded regions indicate the 68\% range across snapshots within this time interval. 
We find that total UV variability decreases strongly with halo mass in all simulations, with $\sigma_{\rm total}\sim2$~mag at $\Mhalo\sim10^{8}$--$10^9~\Msun$, declining to $\sim1$~mag by $\Mhalo\sim10^{11}~\Msun$. 
This trend is driven primarily by the inter-halo component, which dominates the total variance across nearly the full mass range. 
In contrast, the inferred intra-halo variability remains comparatively modest, typically $\sigma_{\rm intra}\sim0.5$--$1$~mag, with a weaker dependence on halo mass.

In the middle panel of Figure~\ref{fig:Muv_var}, we compare our predicted intra-halo variability to results from reconstructed star formation histories for galaxies at $9\lesssim z\lesssim12$ in JADES \citep{Ciesla2024}. 
The JADES analysis finds $\sigma_{\rm UV}\sim1.2\pm0.1$~mag, consistent with the upper bounds of our intra-halo variability at the same redshifts. 
This rough agreement supports their conclusion that short-timescale variability alone cannot explain the observed abundance of bright galaxies at high redshift, and the relatively large halo-to-halo scatter in our simulations provides a reasonable mechanism for populating the bright end of the UV luminosity function. 
However, we caution the reader not to directly infer a star formation rate scatter from our results here, as they rely on significant post-processing.

In the bottom panel of Figure~\ref{fig:Muv_var}, we show that our predicted variability broadly overlaps with results from FIREbox-HR \citep{Feldmann2025}, FIRE-2 zoom-in simulations \citep[private communication,][]{Sun2023b}, and recent empirical studies \citep{Shen2024b,Munoz2026}. 
All of the FIRE simulations predict a similar trend of decreasing UV variability with increasing halo mass, with $\sigma_{\rm UV}\sim1.5$--2 mag in low-mass halos and $\sigma_{\rm UV}\sim1$ mag in more massive systems. 
BonFIRE and CampFIRE generally predict somewhat larger variability than FIREbox-HR and the FIRE-2 zooms at low masses ($\Mhalo\lesssim10^{9.5}~\Msun$), potentially reflecting differences between the FIRE-3 with simple Pop~III model and FIRE-2 physics implementations. 
Although, part of this offset may also arise from the substantially larger cosmological volume and galaxy sample in BonFIRE, which allows us to more robustly characterize the mass dependence and environmental diversity of UV variability. 
Empirical models provide an additional point of comparison: \citet{Munoz2026} and \citet{Shen2024b} similarly infer elevated total UV variability to explain the bright galaxy population at high redshift, broadly consistent with the variability predicted by our simulations. 
However, we note that the H$\alpha$-to-UV measurements used as constraints by \citet{Munoz2026} are currently limited to $4<z<6$, so their applicability at earlier times relies on assuming redshift independence of the burstiness mechanisms. 
Yet, our predicted $9<z<12$ total variability ($\sigma_{\rm total}$) and its halo-mass dependence agrees well with these empirical measurements.

Our results imply that a substantial fraction of the observed scatter in high-redshift UV luminosities may reflect genuine galaxy-to-galaxy diversity rather than purely rapid temporal variability within individual systems. 
In particular, the dominance of $\sigma_{\rm inter}$ suggests that stochastic assembly/merger histories, differing gas accretion states, and environmental variation contribute significantly to the UV luminosity distribution at fixed halo mass.

However, we note a few caveats about our analysis. 
First, the decomposition into inter- and intra-halo components assumes that the two sources of variance add independently in quadrature, which may not hold if burstiness correlates systematically with long-timescale galaxy properties. 
Second, the inferred intra-halo component depends on the adopted smoothing method used to construct de-bursted UV luminosities. 
Our fiducial smoothing method could underestimate $\sigma_{\rm intra}$ in galaxies with extremely short star formation histories, such as ultra-compact galaxies, which can dominate the galaxy population at $\Mhalo\lesssim10^{9}~\Msun$. 
However, we tested imposing a minimum width of stellar ages of 10, 50, and 100~Myr with no significant effects on our results.
Moreover, we note that a quick check of $\sigma_{\rm intra}$ estimated using explicit timescales on BonFIRE galaxies tracked with the merger tree yields compelling results; we recover $\sim0.8-1.2$~mag for 100 Myr timescales and $\sim0.5-1$~mag for 50 Myr timescales with little dependence on halo mass, comparable to our fiducial estimate of $\sim0.5-1$~mag and results from JADES.
Lastly, at fixed halo mass, higher-resolution simulations generally exhibit reduced variability amplitudes, particularly in the inter-halo component (up to $\sim0.3$--$0.4$~mag lower median $\sigma_{\rm inter}$ between the CampFIRE runs), suggesting that part of the extreme burstiness seen at lower resolution arises from unresolved star formation and feedback cycling. 
Thus, the true variability amplitudes could be lower than the values we infer here due to our finite resolution.

\subsection{Pop~III star formation}\label{sec:results_pop3}

In this section we examine the formation of low-metallicity, or Pop~III ($Z\leq10^{-5}~Z_{\odot}$), stars in the BonFIRE and CampFIRE suite.
Using a simple Pop~III model (see Section~\ref{sec:sims_pop3}), we predict the rate of Pop~III star formation across cosmic time and the UVLF for ``Pop~III galaxies".

\begin{figure}
    \centering
    \includegraphics[width=\columnwidth]{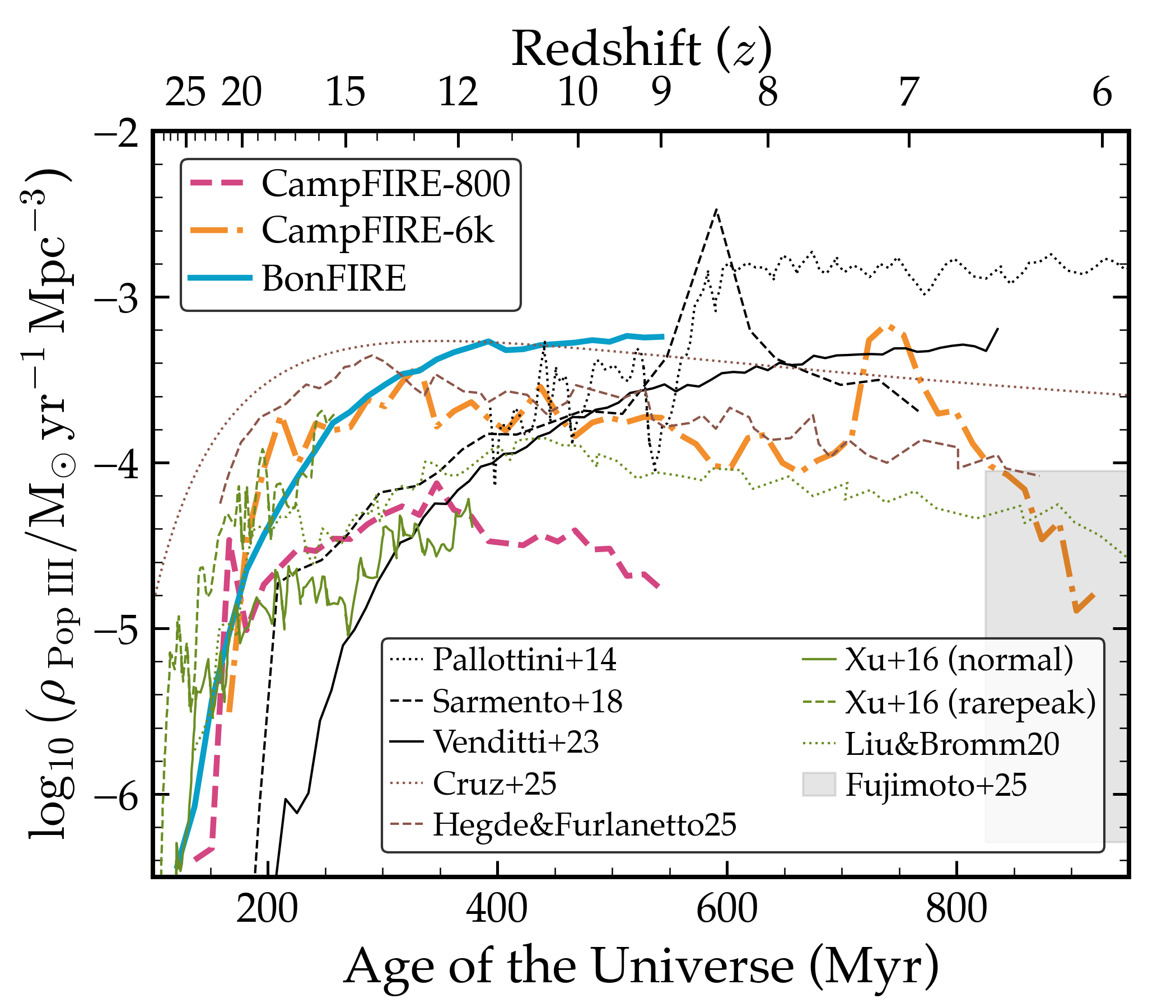}
    \caption{The Pop~III star formation rate density (SFRD) in each of the BonFIRE, CampFIRE-6k, and CampFIRE-800 simulations over time. 
    Note that the CampFIRE-6k run extends to $z=6$, allowing us to make predictions through the end of reionization. 
    We compare to various Pop~III models from the literature and to limits from recent observations of candidate Pop~III galaxies (see text for details).}
    \label{fig:pop3_frac}
\end{figure}

\subsubsection{Pop~III SFRD}\label{sec:results_pop3_sfrd}

Figure~\ref{fig:pop3_frac} shows the star formation rate density (SFRD) of Pop~III stars in BonFIRE, CampFIRE-6k, and CampFIRE-800 compared to other models. 
Overall, the BonFIRE and CampFIRE suite predicts Pop~III SFRDs that lie within the range spanned by the other models, but with important differences in both normalization and evolutionary shape.

At $z\gtrsim20$, the BonFIRE and CampFIRE suite shows a rapid rise in Pop~III SFRD, followed by a plateau and gradual decline at $z\sim10$--15. 
The onset of Pop~III star formation in our simulations is broadly consistent with other hydrodynamical models; it occurs earlier than in other large-scale cosmological simulations such as \citet{Sarmento2018} and \citet{Venditti2023}, possibly due to their limited resolution and/or volume, but somewhat later than in recent analytic/semi-analytic predictions \citep[e.g.][]{Cruz2025, Hegde2025}. 
In terms of normalization, BonFIRE and CampFIRE-6k are consistent with the upper envelope of previous simulations, including the rare-peak model of \citet{Xu2016}, while CampFIRE-800 predicts systematically lower SFRDs. 
Our results are also broadly consistent with \citet{LiuBromm2020}, which employs the same hydrodynamics solver (GIZMO), although we find somewhat higher SFRDs at $z\lesssim12$. 
The evolution of our SFRDs shows additional similarities to prior work: the peak seen in \citet{Sarmento2018} is mirrored in CampFIRE-6k, albeit at slightly later times. 
However, unlike \citet{Pallottini2014}, we do not find a sustained increase in Pop~III SFRD toward lower redshift; instead, our simulations show a flattening and gradual decline after the peak.

Compared to semi-analytic models (SAMs) and observationally motivated constraints, our Pop~III SFRDs again fall within the broad allowed range but show a more structured evolution. 
The SAM of \citet{Hegde2025} is particularly similar to CampFIRE-6k in both normalization and redshift evolution, aside from the absence of the secondary peak at $z\sim7$ seen in our simulation. 
The model of \citet{Cruz2025} predicts similarly high SFRDs but with an earlier rise and a smoother evolution in time. 
Observationally inferred constraints from \citet{Fujimoto2025b}, based on candidate Pop~III galaxies at $z\lesssim7$, span a wide range that overlaps with our predictions at similar redshifts, but do not constrain the temporal variability seen in our simulations. 
Overall, these comparisons suggest that while semi-analytic models and empirical constraints capture the typical level of Pop~III activity, our simulations exhibit additional time-dependent structure driven by inhomogeneous enrichment and the persistence of low-metallicity gas.

The normalization of our results across our three simulations suggests a strong dependence of Pop~III SFRD on resolution.
We find a difference of $\sim0.5-1~$dex between adjacent resolution levels at $z\sim9-12$, with higher resolution leading to lower Pop~III SFRD.
Interestingly, the onset of Pop~III star formation occurs at similar times across the simulations, suggesting that the primary effect of resolution is not delaying the formation of the first stars, but instead altering the subsequent star formation and enrichment history. 
At higher resolution, star formation histories and metal enrichment become more temporally resolved: gas that would form a single massive Pop~III star particle at lower resolution instead fragments into multiple star particles, likely forming over a more extended period and allowing the first stars to enrich nearby gas before subsequent stars form.  
In this picture, the reduced Pop~III SFRD at higher resolution may arise less from enhanced metal mixing and more from the ability to time-resolve local self-enrichment, causing later star formation episodes to no longer satisfy the Pop~III criterion. 
This effect may be particularly important at these redshifts, where metal transport is expected to be limited by relatively slow advection rather than fully developed turbulent mixing.

Additional factors may also contribute to the resolution dependence, including the higher density contrast of the CampFIRE zoom-in region and stochastic differences between individual simulation realizations. 
Furthermore, if lower-mass halos are more likely to host Pop~III star formation at fixed redshift, then the ability of the higher-resolution simulations to resolve such systems (as they emerge at earlier times than the resolved, higher-mass systems in lower-resolution simulations) could alter the inferred Pop~III SFRD by polluting pristine gas and thereby suppress later Pop~III formation. 
A more detailed investigation of the origin of this resolution dependence, including the relative roles of time-resolved enrichment, metal transport, and halo mass dependence, is left to future work.

\begin{figure}
\centering
\includegraphics[width=\columnwidth]{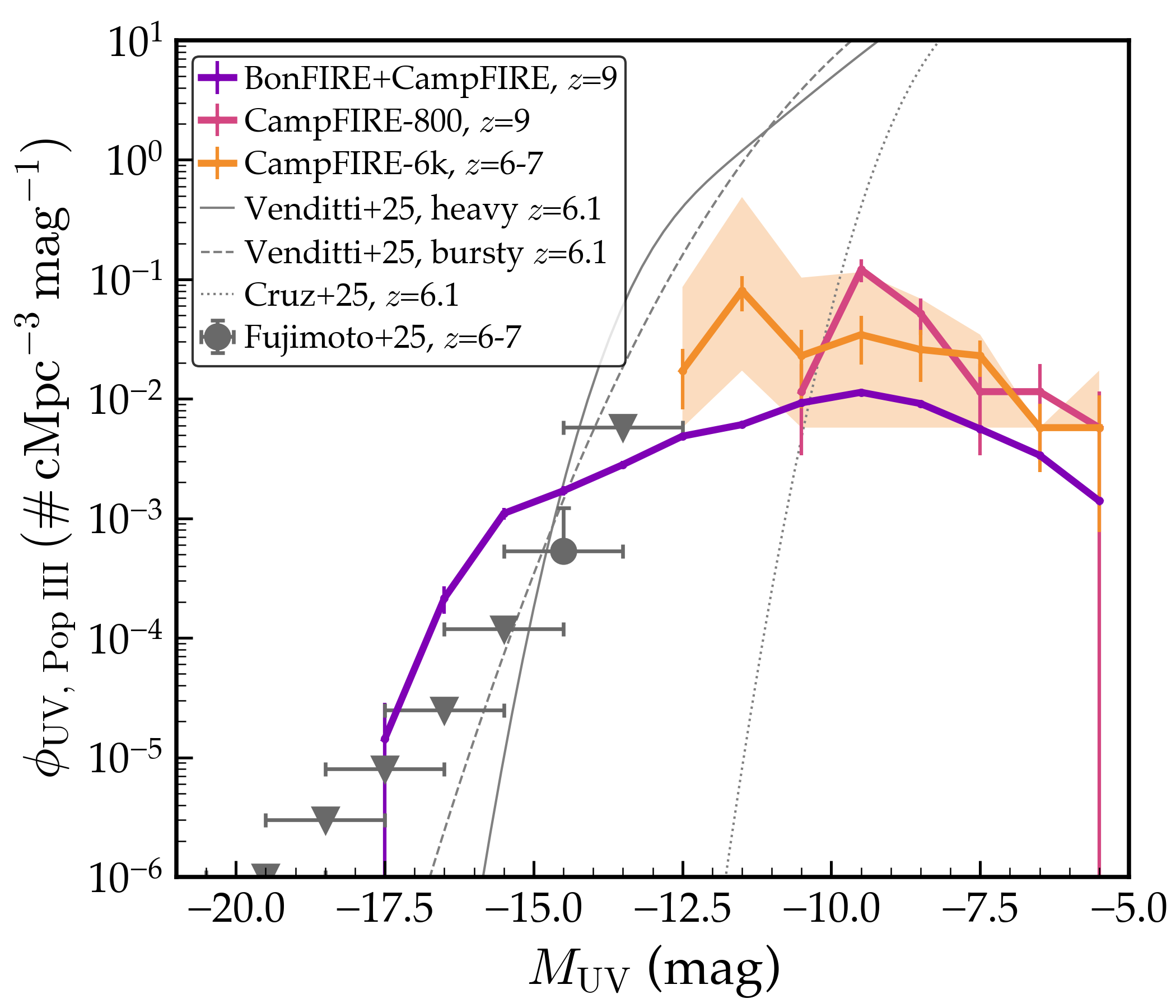}
\caption{The Pop~III galaxy UVLF for BonFIRE+CampFIRE and CampFIRE-800 at $z\sim9$, and for CampFIRE-6k at $6\leq z\leq7$.
We compare to photometric Pop~III candidates from \citet{Fujimoto2025b} at $5.6\leq z\leq6.6$, as well as the empirically calibrated models of \citet{Cruz2025} and \citet{Venditti2025}. 
The BonFIRE+CampFIRE Pop~III galaxy UVLF is similar to the observations where there is overlap between them at $\Muv\lesssim-12$, whereas, the results from both CampFIRE runs are fainter than current tentative detections.}
\label{fig:pop3_uvlf}
\end{figure}

\subsubsection{Pop~III UVLF}\label{sec:results_pop3_uv}

Recall from Section~\ref{sec:sims} that the high-resolution CampFIRE region is an overdense environment and thus does not represent an unbiased sample of the cosmic density field. 
To recover an unbiased UVLF from halos within this region, we apply a mass-dependent correction based on the large-scale halo bias, which we describe for the SMF in Sec~\ref{sec:results_mstar}. 
Note that the only direct comparison we can make with the available observations is with CampFIRE-6k, as it is the only simulation run to $z=6$, where tentative Pop III candidate galaxies have been observed. 
We still show and discuss BonFIRE and CampFIRE-800 predictions at $z\sim9$, the last available snapshot in those runs, to illustrate the Pop~III galaxy UVLF at earlier times and potential resolution effects. 
We define a Pop~III galaxy as containing at least one young ($<10~\Myr$ old) Pop~III star particle. 
This optimistic definition targets objects that might have a detectable Pop~III component, as the young, massive stars are sources of hard-spectrum tracers such as He~II lines, which fade significantly after just a few Myr \citep{BKL2001,Schaerer2002,Katz2023a}.

Figure~\ref{fig:pop3_uvlf} compares the UVLF of simulated Pop III galaxies in the BonFIRE and CampFIRE suite to observations and empirically-calibrated Pop III formation models.
We show the BonFIRE and CampFIRE-800 (bias-corrected) results at $z\sim9$.
For CampFIRE-6k, we show a shaded region highlighting the full range of the Pop~III UVLF during $z=6-7$ (12 snapshots spanning $\approx165$ Myr).
The top boundary of the shaded region roughly corresponds to the Pop~III UVLF at $z=7$ and the bottom boundary roughly corresponds to $z=6$.

We compare to recent observational searches for Pop~III galaxy candidates from \citet{Fujimoto2025b}, which relies on indirect and highly uncertain diagnostics of extremely low metallicity stellar populations. 
The inferred number densities therefore carry substantial systematic uncertainties, both in candidate identification and completeness. 
Within these uncertainties, the BonFIRE Pop~III UVLF at $z\sim9$ is broadly consistent with the observed number densities at $\Muv\lesssim-14$, despite the redshift mismatch. 
We also include the AMORE6 candidate at $z\approx5.7$ identified by \citet{Morishita2025}, which lies within the same survey footprint as \citet{Fujimoto2025b} but was not included in their original catalog, with $\Muv \approx -14.5$. 
Taken together, these comparisons suggest that our predicted abundance of relatively bright Pop~III-hosting galaxies is compatible with current observational constraints, although significantly more robust samples will be required to place meaningful limits on the Pop~III UVLF.

We further compare to theoretical models of Pop~III galaxy formation. 
While traditional models (assuming that Pop~III formation is confined within molecular-cooling minihalos, \citet[e.g.,][]{Cruz2025}) are not able to explain bright Pop~III galaxy candidates such as the AMORE6 galaxy, assuming burstier star-formation activity in Pop~III-forming galaxies than typical galaxies at this redshift (as in the ``bursty" model of \citet{Venditti2025}), and/or a residual, efficient Pop~III formation in higher-mass, atomic-cooling halos (as in their ``heavy" model) can reproduce the observed brightness. 
In contrast, the model of \citet{Cruz2025} predicts a much steeper UVLF with very few Pop III galaxies at bright magnitudes, falling well below our results at $\Muv \lesssim -11$. 
Our CampFIRE-6k predictions lie between these extremes and show a relatively flat UVLF over $-12\lesssim\Muv\lesssim-8$, indicating a broad distribution of Pop~III-hosting galaxies across faint luminosities. 
We note that the analysis of \citet{Venditti2025} uses a similar definition for Pop~III galaxies as us, whereby any amount of recent Pop~III star formation qualifies a galaxy as Pop~III.
This intermediate behavior suggests that our simulations capture both the suppression of Pop~III formation in enriched regions and its persistence in low-metallicity pockets, yielding a Pop~III galaxy population that is neither as rare nor as ubiquitous as in existing models.

\section{Discussion}\label{sec:discussion}

The BonFIRE and CampFIRE simulation suite provides several valuable insights into high-redshift galaxy formation, including the predicted existence of ultra-compact galaxies, the shape of the UVLF (including turnover and, in our case, a bright-end excess), sources of UV variability, and the persistent formation of Pop~III systems.
We discuss each of these points and caveats on our predictions in more detail below.

Ultra-compact galaxies (UCGs) are compact galaxies that form with relatively high star formation efficiencies at the centers of low-mass halos, and which are present throughout the volumes of all simulations we introduce here. 
However, their detailed properties in BonFIRE are affected by resolution limitations: once the local Jeans mass becomes unresolved, gas collapses too quickly, leading to artificially rapid and efficient star formation. 
This results in stellar systems that form too much mass in too short a time compared to the higher-resolution CampFIRE runs. 
In contrast, CampFIRE better resolves the internal structure of star-forming regions, producing more temporally extended star formation histories and thereby capturing clustered star formation and the formation of UCGs more realistically. 
These compact systems may serve as analogs to the proto–globular clusters and ultra-compact galaxies now being uncovered by \JWST\ \citep{Adamo2024,Vanzella2026}, providing a powerful theoretical framework for interpreting their formation and survival across cosmic time. 
The compact galaxies identified by \citet{Morishita2024} also exhibit elevated star formation surface densities, qualitatively consistent with the high star formation efficiencies we find for UCGs, suggesting that both populations may arise from similarly dense, single-burst-driven star formation environments.

The extreme compactness, high star formation efficiency, and single burst assembly histories of UCGs raise the possibility that these systems may also be progenitors of \textit{present-day} globular clusters and ultra-compact dwarf galaxies. 
In particular, UCGs reach stellar masses and sizes comparable to the most massive local globular clusters \citep[e.g.,][]{Baumgardt2018}. 
Although UCGs are embedded within dark matter halos in the simulations, subsequent tidal stripping or dynamical evolution could plausibly transform a subset of these objects into bound stellar systems resembling globular clusters at later times. 
Establishing such a connection will require tracking the long-term survival and dynamical evolution of UCGs to low redshift, but the results presented here suggest that the physical conditions necessary for globular cluster-like formation naturally emerge within a cosmological context.
In future work we will investigate the processes that ignite and possibly cycle or inhibit star formation in these systems following the initial burst \citep{Moreno2025}.

Turning to the excess of UV-bright galaxies in BonFIRE+CampFIRE, we find that it likely arises from a combination of substantial halo-to-halo scatter, bursty star formation, and pre-enrichment from Pop~III stars. 
As discussed in Section~\ref{sec:results_uv_var}, intra-halo variability alone is insufficient to populate the bright end of the UVLF; instead, a significant contribution comes from differences in assembly history, gas accretion, and star formation efficiency across halos of similar mass. 
Compared to FIRE-2 simulations, BonFIRE+CampFIRE generally predicts similar total UV variability, with some evidence for greater halo-to-halo diversity at low masses ($\Mhalo\lesssim10^{9.5}~\Msun$). 
Part of this difference may reflect the FIRE-3 framework with our simple Pop~III model, although the substantially larger cosmological volume of BonFIRE likely also plays an important role by sampling a broader range of environments and rare assembly histories. 
At the same time, while dust attenuation indeed moderates the bright end of our UVLFs by $\sim0.1$--0.5 dex at $\Muv\lesssim-20$, it does not fully remove the excess. 
However, a more detailed accounting of dust attenuation than our current analytical formulation could also help alleviate our bright-end excess. 
Our results therefore suggest that accurately modeling the high-redshift UVLF requires capturing not only the burstiness of star formation within galaxies, but also the diversity of galaxy growth histories across a wide range of halo masses and environments.

Moreover, the current simulations in the BonFIRE and CampFIRE suite do not model black hole sinks, accretion, or feedback from active galactic nuclei (AGN), which could significantly impact the growth and feedback processes in massive galaxies \citep{AnglesAlcazar2017b,Wellons2023,Byrne2024} and potentially modulate the galaxy UVLF at the bright end. 
Even so, we find that the properties of the most massive galaxies in BonFIRE are reasonably consistent with observations. 
For example, \citet{Tacchella2023} find that the $z=10.6$ galaxy GN-z11 has a stellar mass of $\approx 10^9~\Msun$ and star formation rate of $21^{+22}_{-10}~\Msun \mathrm{yr}^{-1}$.  
The most massive galaxy in BonFIRE at $z\sim10.7$ has a similar stellar mass of $\Mstar\approx10^{9.4}~\Msun$ and a star formation rate of $\sim 30\ \Msun \mathrm{yr}^{-1}$. 
Observations indicate that GN-z11 is very compact; \citet{Tacchella2023} measure a half-light radius of about 64~pc, with an additional extended component of $\sim200$~pc. 
By comparison, the most massive BonFIRE galaxy is more extended, with a stellar half-mass radius of $R_{\star,~1/2}\sim1~\kpc$ and only $\sim5\%$ of its stellar mass contained within the inner 60~pc. 
This structural difference may be significant, as some studies suggest that GN-z11’s high luminosity is powered, at least in part, by an AGN \citep{Maiolino2023}. 
Even though our simulations do not include AGN, we find that the UV luminosities of our most massive galaxies are comparable to those inferred for GN-z11 under purely stellar interpretations. 
For example, \citet{Tacchella2023} measure $\Muv\approx-21.6$ for GN-z11 and infer an intrinsic $\Muv\approx-21.8$, while our analogue in BonFIRE has a dust-corrected $\Muv\approx-21.1$ and an intrinsic $\Muv\approx-22.3$.

Lastly, understanding when and where Pop~III stars form is crucial for interpreting their contribution to the ionizing photon budget and early metal enrichment \citep[e.g.,][]{Bromm2013,KlessenGlover2023}. 
In particular, quantifying the fraction of stellar mass and galaxies that remain Pop~III at different epochs provides insight into the transition from primordial to metal-enriched star formation and helps identify the physical conditions under which Pop~III formation can persist to lower redshifts. 
These demographic trends also set the stage for interpreting the Pop~III contribution to the UV luminosity function, because both the abundance and timing of Pop~III star formation directly shape the observable signatures of primordial galaxies.

Our model assumes a simplified treatment of Pop~III star formation, with standard FIRE-3 supernova and stellar wind yields, which may not fully capture the complex metal enrichment processes from metal-poor stars in early galaxy evolution. 
Tracking such Pop~III-specific yields (such as from Pair-Instability Supernovae) in future simulations would allow prediction of tell-tale Pop~III enrichment patterns, for example, in deep absorption spectroscopy of the high-redshift IGM \citep{Wang_GRB2012}. 
However, we have found tentative evidence in our simulations that even our simple Pop~III model can introduce a metallicity floor in the stellar mass-metallicity relation for resolved galaxies, similar to what is observed at low-redshift (Samuel et al., in prep).
However, a detailed study of the effects of our Pop~III model and future modifications of it is needed to further assess its global impact on galaxy formation.

On a final note, the resolution of the simulations we present here is insufficient to fully resolve star formation in minihalos, which is expected to play a crucial role in the earliest phases of galaxy formation. 
We have, however, attempted to mitigate resolution effects at the faint/low-mass end as much as possible; as a result, our predictions are likely robust for galaxies with $\Mstar\gtrsim10^4$--$10^5~\Msun$, depending on the specific properties considered. 
Furthermore, although we model the effects of reionization using a combination of a uniform UV background and a simplified local radiation transport algorithm \citep[LEBRON,][]{Hopkins2018}, the absence of full radiative transfer limits the accuracy of our photoionization modeling and the resulting low-luminosity galaxy population at high redshift. 
These limitations highlight the need for future improvements in both resolution and modeling to better capture the complex processes driving early galaxy formation.

\section{Conclusions}\label{sec:conclusions}

We present first results from the BonFIRE and CampFIRE suite of simulations, three new simulations with up to $m_{\rm baryon}=800~\Msun$ mass resolution using FIRE-3 physics to model the first galaxies through the Epoch of Reionization. 
BonFIRE provides a representative view of early galaxy formation over a large $(41.2~\mathrm{cMpc})^3$ cosmological volume, while CampFIRE resolves a subregion of this volume at higher mass resolution. 
We leverage the strengths of both simulations using a novel resampling scheme that extends the resolution of CampFIRE across the full BonFIRE volume, producing a self-consistent galaxy population spanning $\Mstar\sim10^4$--$10^{10}~\Msun$. 
Our main conclusions are: 
\begin{itemize}
    \item The median $\Mstar$--$\Mhalo$ relation remains largely constant over $9\lesssim z\lesssim25$. 
    \item The halo-scale star formation efficiency ($\sfe$) increases smoothly with halo mass above the mass where it becomes resolution-limited ($\Mhalo\sim10^{8}~\Msun$ in CampFIRE and $\Mhalo\sim10^{9}~\Msun$ in BonFIRE). 
    \item Galaxy sizes ($R_{\star,~1/2}$) span $\sim5~$pc (essentially the stellar force softening) to $\sim1~$kpc, with larger galaxies also experiencing significantly more extended star formation histories. Star clusters are present in early galaxies at all masses, and ultra-compact galaxies resemble single star clusters in mass and size. 
    \item Ultra-compact galaxies have small sizes, large $\sfe$, and star formation histories dominated by a single burst relative to the general galaxy populations in all CampFIRE and BonFIRE simulations. 
    \item The median $\Muv$--$\Mhalo$ relation is stable for halos at $\Mhalo \gtrsim 10^9~\Msun$ but evolves toward fainter luminosities in lower-mass halos with increasing cosmic time, likely because of aging stellar populations. 
    \item The BonFIRE+CampFIRE UVLF broadly agrees with both observations and other theoretical models at intermediate magnitudes ($-19 \lesssim \Muv \lesssim -15$) out to $z \lesssim 17$. However, at the bright end ($\Muv \lesssim -20$), we predict an excess of $\sim0.5$~dex relative to current observational constraints. At the faint end we find evidence for a turnover at $-15 \lesssim \Muv \lesssim -13$. We provide redshift-dependent fits to the UVLF to quantify its evolving shape across cosmic time. 
    \item Total UV variability in early galaxies decreases from $\sim2$~mag in low-mass halos to $\sim1$~mag in massive systems at $9<z<12$, with the scatter dominated by halo-to-halo diversity at low masses and approximately equal contributions from halo-to-halo scatter and short-timescale burstiness at $\Mhalo\gtrsim10^{10}~\Msun$.
    \item The evolution of the total UV luminosity density follows the shape of the constant-SFE models of \citet{Harikane2024b} and \citet{Feldmann2025}, but with systematically higher normalization. 
    \item The star formation rate density of Pop~III stars (defined here as stars with $\log_{10}(Z/\Zsun)\leq-5$) is broadly consistent with previous models and tentative observed Pop~III candidate systems at $z\sim6$. 
    \item The very low metallicity or ``Pop~III galaxy" UVLF (for galaxies containing at least one young Pop~III star particle) in CampFIRE-6k predicts number densities of $\phi\sim10^{-2}$--$10^{-1}\rm{~cMpc^{-3}~mag^{-1}}$ for $-13\lesssim\Muv\lesssim-5$ at $6\leq z \leq 7$, below empirical models and fainter than current observational limits. 
\end{itemize}

Our results from the BonFIRE and CampFIRE suite demonstrate that early galaxy formation naturally produces both extended, feedback-regulated systems and compact, single-burst galaxies resembling proto–globular clusters and/or ultra-compact, low-mass galaxies. 
Our results indicate that the FIRE-3 model captures the key physical processes governing the rapid assembly of the first galaxies, while also pointing to areas for improvement in modeling and resolution. 
Altogether, our predictions provide a framework for interpreting the growing census of high-redshift galaxies from \JWST\ and for linking the earliest star formation episodes to galaxy evolution across cosmic time.

\section*{Software and third party data repository citations}\label{sec:cite}

\textit{\large Data availability}: A public version of the \textsc{Gizmo} code is available at \url{http://www.tapir.caltech.edu/~phopkins/Site/GIZMO.html}.
FIRE-2 data releases \citep{Wetzel2023,Wetzel2025} are publicly available at \url{http://flathub.flatironinstitute.org/fire}.
Data that appear in the figures are available upon reasonable request to the corresponding author.

\software{
        arXiv pre-print                       service\footnote[4]{\url{https://www.arxiv.org}},
          NASA ADS\footnote[5]{\url{https://ui.adsabs.harvard.edu}},
          astropy \citep{astropy-13, astropy-18}, 
          Gizmo Analysis\footnote[6]{\url{https://bitbucket.org/awetzel/gizmo\_analysis}} \citep{wetzel-20-gizmo},
          Halo Analysis\footnote[7]{\url{https://bitbucket.org/awetzel/halo\_analysis}} \citep{wetzel-20-halo},
          ipython \citep{ipython-07},
          matplotlib \citep{matplotlib-07}, 
          numpy \citep{numpy-20},
          scipy \citep{scipy-20},
          hmf\footnote[8]{\url{https://ascl.net/1412.006}} \citep{hmf2013}
          }

\begin{acknowledgments}

We thank L. Y. Aaron Yung and Rachel Somerville for sharing data, code, and useful insight during the preparation  of this manuscript. 

J. Samuel and MBK acknowledge support from program JWST-AR-06278 by NASA through a grant from the Space Telescope Science Institute, which is operated by the Association of Universities for Research in Astronomy, Inc., under NASA contract NAS 5-03127. 
MBK also acknowledges support from NSF grants AST-2108962 and AST-2408247; NASA grant 80NSSC22K0827; HST-GO-16686, HST-AR-17028, and JWST-GO-03788 from STScI; and the Samuel T. and Fern Yanagisawa Regents Professorship in Astronomy at UT Austin. 
PFH was supported by a Simons Investigator grant. 
GS acknowledges support from a CIERA Postdoctoral Fellowship, with additional support provided by NSF through grant AST-2307327; by NASA through grant 23-ATP23-0008; and by STScI through grant JWST-AR-03252.001-A. 
AV acknowledges funding from the Cosmic Frontier Center and the University of Texas at Austin’s College of Natural Sciences. 
XS acknowledges the support from program JWST-AR-04814 by NASA. 
AW received support via NSF CAREER award AST-2045928. 
JM is funded by a Pomona College Large Research Grant. 
JBM acknowledges support from NSF Grants AST-2307354 and AST-2408637, and NASA through grant JWST-GO-03224. 
RKC is grateful for support from the Leverhulme Trust via the Leverhulme Early Career Fellowship. 
CAFG was supported by NSF through grants AST-2108230 and AST-2307327; by NASA through grants 80NSSC22k0809, 80NSSC22K1124 and 80NSSC24K1224; by STScI through grant JWST-AR-03252.001-A; and by BSF through grant \#2024262. 
MCS acknowledges support from the NSF Graduate Research Fellowship Program under Grant No. DGE 2137420. 
J. Stern is supported by a grant from the United States-Israel Bi-national Science Foundation (BSF). 
JSB acknowledge support from NSF grant AST-2408246.
This research was also supported in part by grant NSF PHY-2309135 to the Kavli Institute for Theoretical Physics (KITP).

\section*{Large Language Model Usage Disclosure}\label{sec:llm}

We used the LLM ChatGPT to assist with analysis code drafting, debugging, and particularly optimization in dealing with the large data set presented here. 
LLM analysis suggestions were vetted and tested against slower methods and known results where possible. 
The main usage of ChatGPT appears in Appendix~\ref{app:resampling}, where we used it to develop and optimize the resampling scheme for mitigating resolution effects and to mathematically describe the algorithm. 
We also used ChatGPT to suggest formatting and improvements to the main text, which we vetted before including.

\end{acknowledgments}

%

\vspace{5mm}
\facilities{
We ran simulations using Frontera at TACC through allocations AST25003, AST21010, and AST20016 from the Advanced Cyberinfrastructure Coordination Ecosystem: Services \& Support (ACCESS) program, which is supported by U.S. National Science Foundation grants \#2138259, \#2138286, \#2138307, \#2137603, and \#2138296.
}

\appendix

\section{Halo mass functions and occupation fractions}\label{app:HOF}

\begin{figure*}
    \centering
    \begin{tabular}{cc}
    \subfigure{\includegraphics[width=0.45\textwidth]{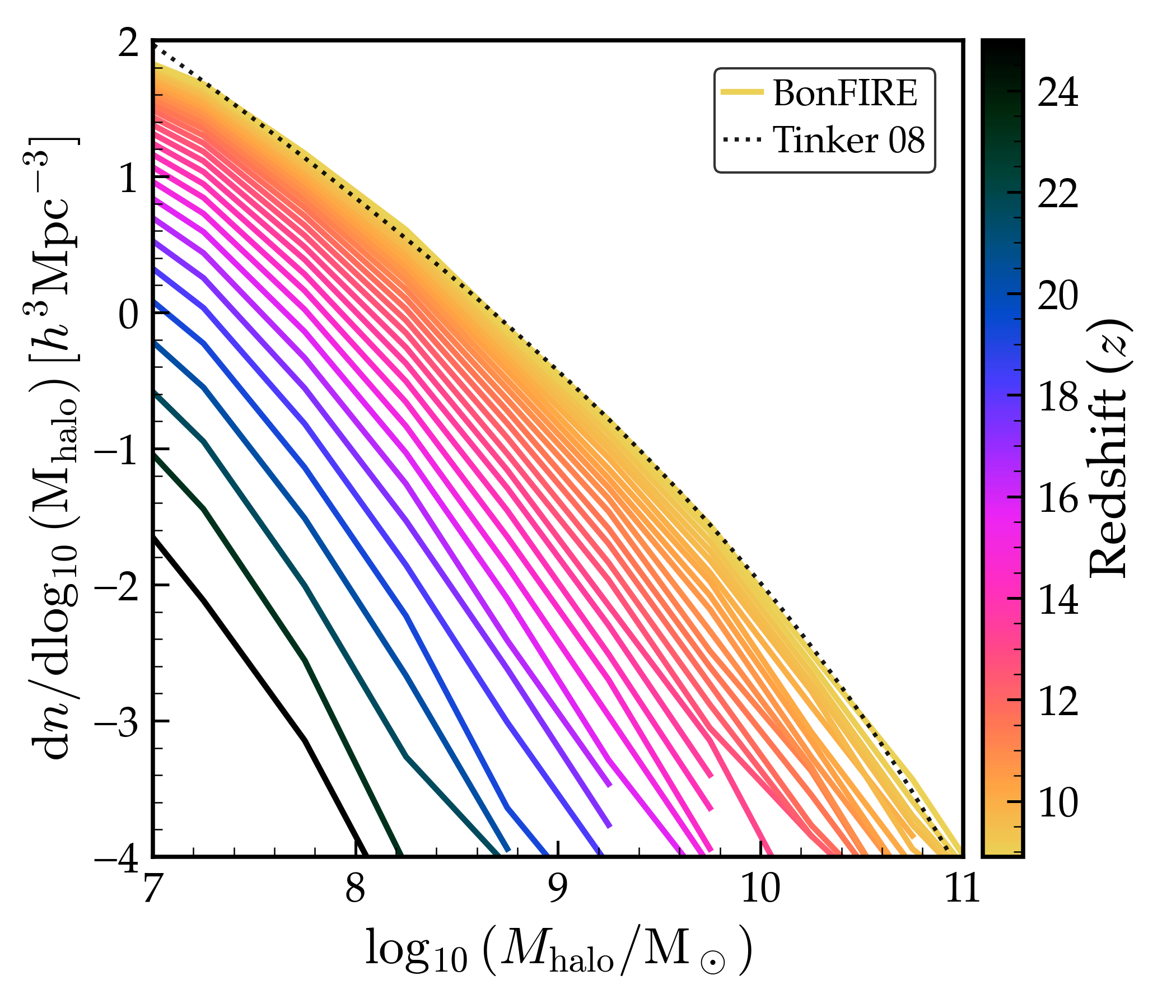}} &
    \subfigure{\includegraphics[width=0.45\textwidth]{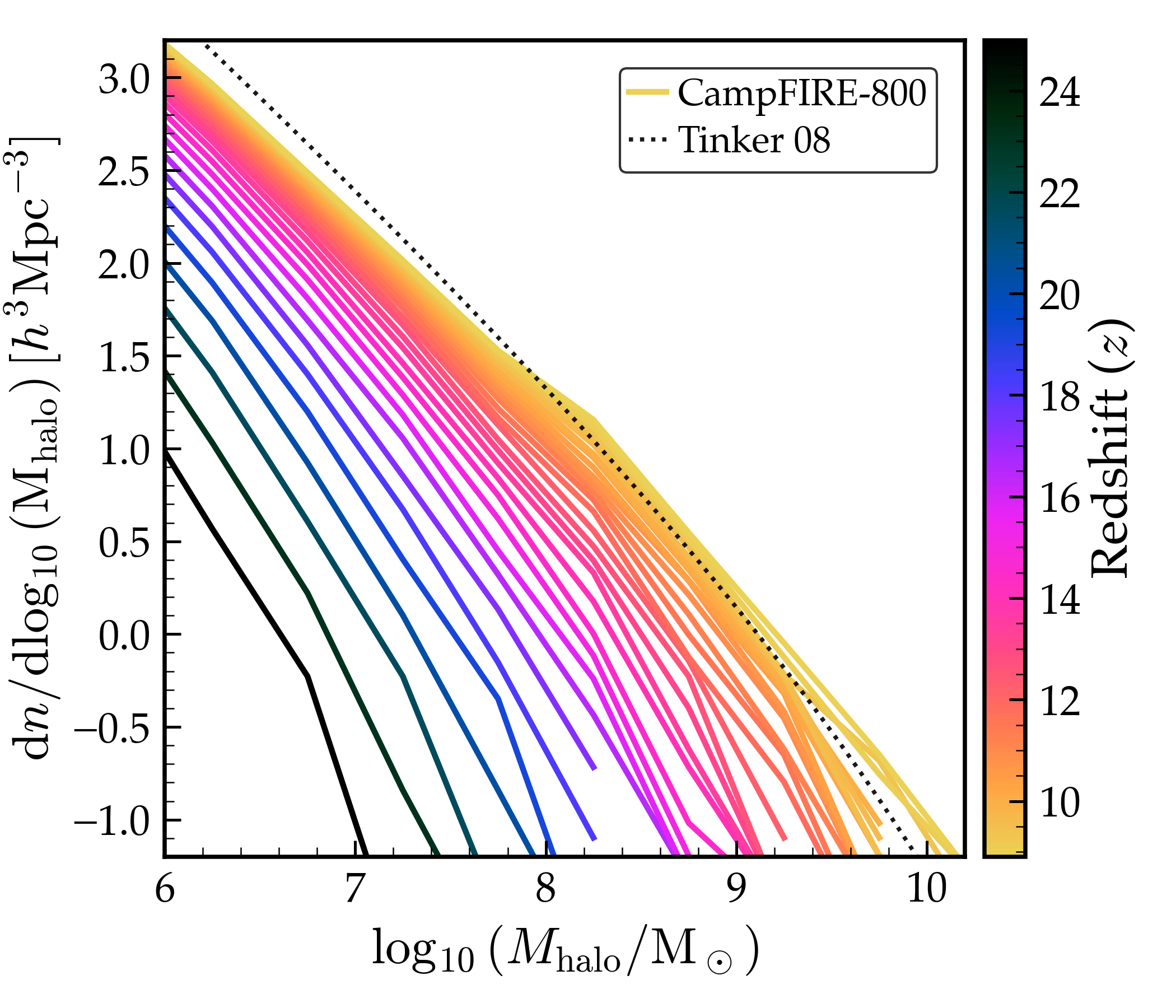}}
    \end{tabular}
    \caption{The evolution of the halo mass function in BonFIRE (left) and CampFIRE-800 (right) to $z=9$. The dotted line is a Hubble-volume averaged fit from \citep{Tinker2008} at $z=9$; in the left panel we show the Tinker fit for average matter density and in the left panel we show the fit for CampFIRE's corresponding overdensity ($\delta\approx0.4$).}
    \label{fig:HMF}
\end{figure*}

In Figure~\ref{fig:HMF} we show the evolution of the halo mass functions for BonFIRE (left) and CampFIRE-800 (right) over $z=27-9$.
The dotted line in each panel shows the analytical Tinker 2008 halo mass function at $z=9$.
Comparing to the Tinker model line, BonFIRE at $z=9$ is approximately average density or representative of the global halo mass function.
In contrast, CampFIRE is a relatively overdense volume.

\begin{figure}
\centering
\begin{tabular}{cc}
\subfigure{\includegraphics[width=0.45\textwidth]{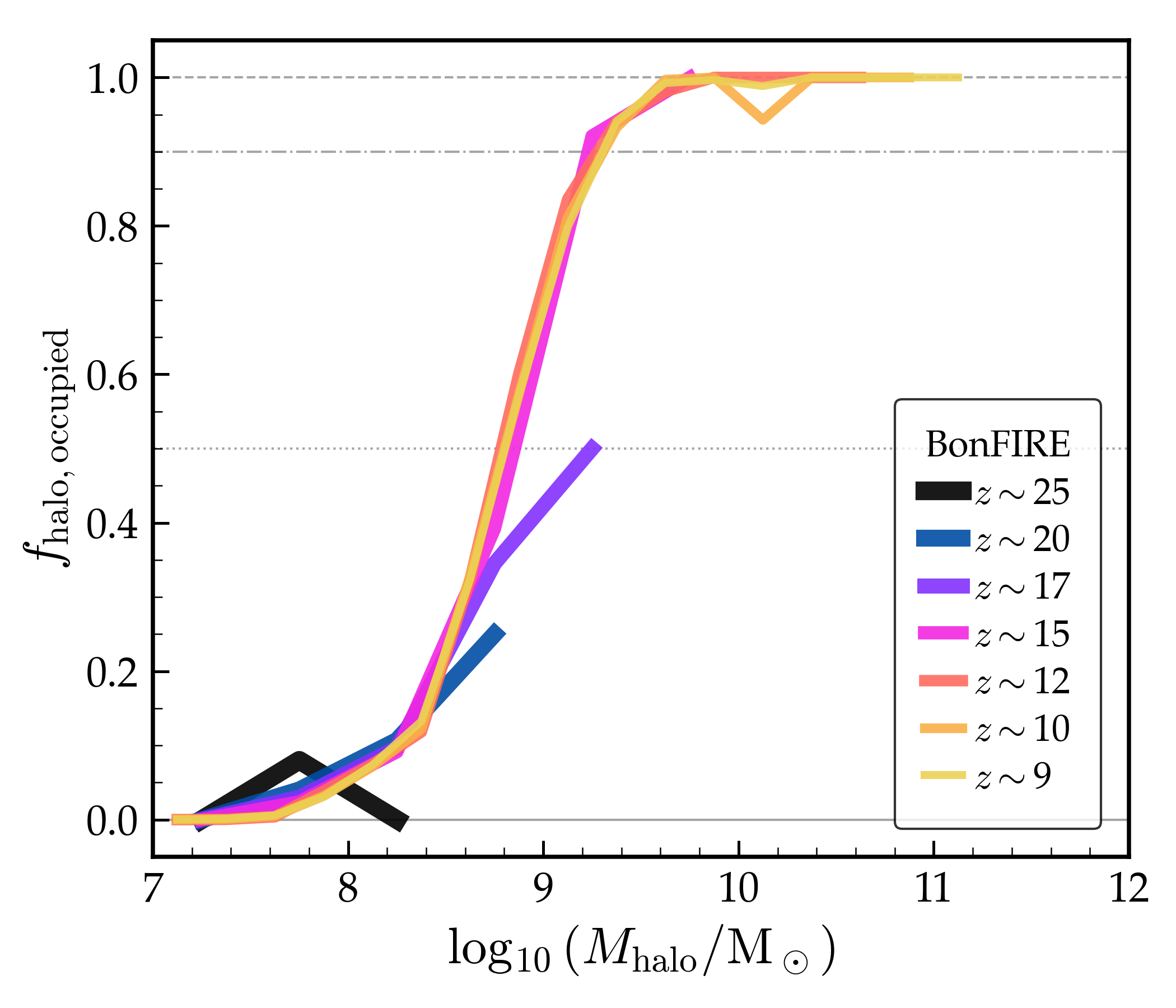}} &
\subfigure{\includegraphics[width=0.45\textwidth]{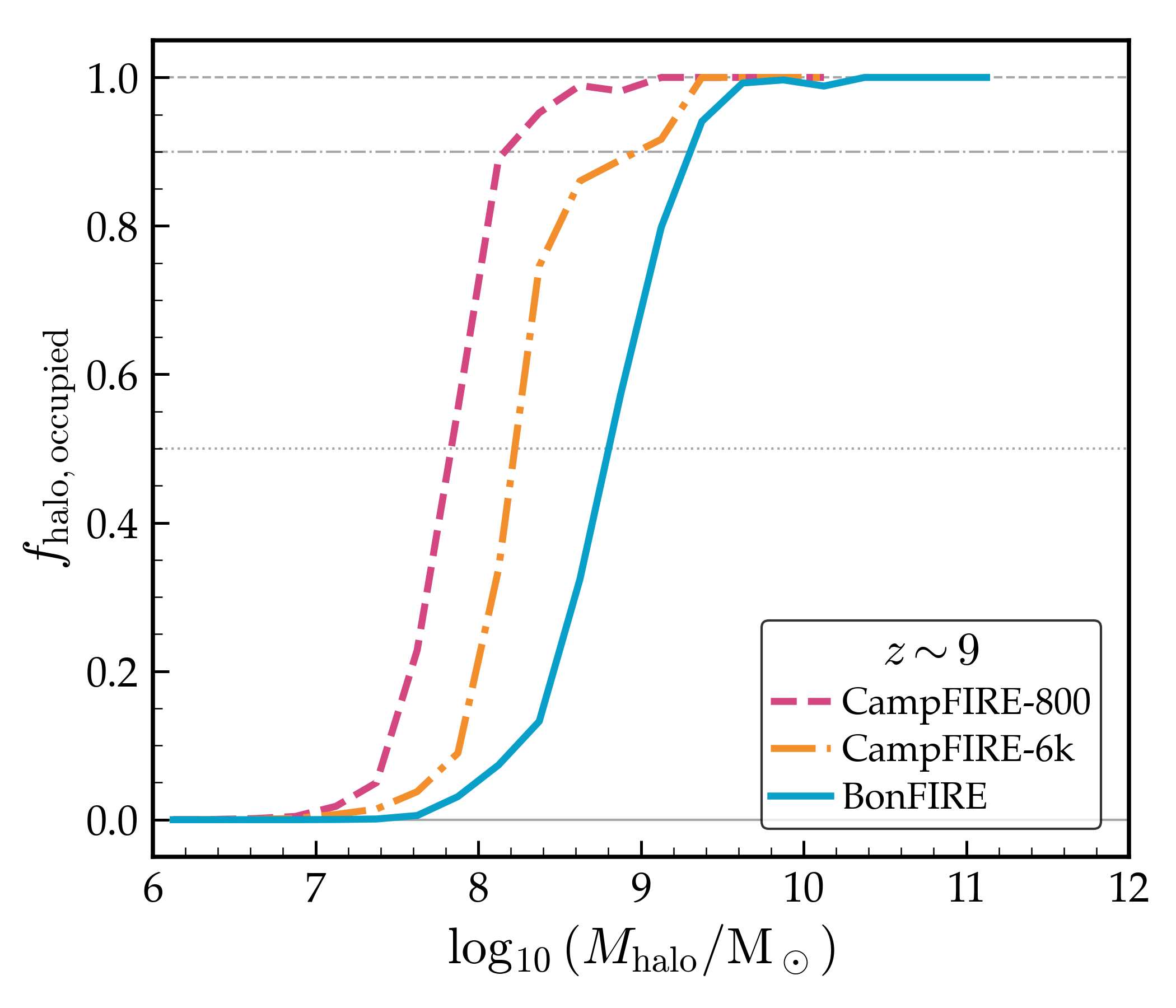}}
\end{tabular}
\caption{\textit{Left:} The fraction of halos occupied by a galaxy in BonFIRE at $z=9-25$.
The halo occupation fraction (HOF) rises steeply with halo mass, as expected, and it does not significantly evolve with redshift over $z=25$ to $z=9$. 
We note that halos at $\Mhalo<10^{7.5}~\Msun$ ($\lesssim100$ particles) are likely numerically under-resolved. 
This metric remains unchanged by our resampling procedure.
\textit{Right:} We compare the HOF in all three simulations at $z=9$.
There is a clear resolution dependence, whereby lower-mass halos are more commonly occupied at higher resolution. 
}
\label{fig:HOF}
\end{figure}

In Figure~\ref{fig:HOF} we show the evolution in the fraction of halos occupied by galaxies in BonFIRE over $z=25-9$.
We include all halos with at least one star particle in the occupied category \citep{Moreno2025}.
The halo occupation fraction combines two aspects of numerics: the resolution thresholds for galaxy formation (in this case the baryonic mass resolution $\resbf$) and dark matter halo identification.
As described in Section~\ref{sec:sims}, we identify halos with ROCKSTAR, imposing a minimum 30 dark matter particle threshold (equivalent to $\Mhalo\gtrsim10^{6.9}~\Msun$).
From the rapid decline in the occupation fraction at $\Mhalo\lesssim10^8~\Msun$, it is evident that galaxies rarely occupy halos with $\lesssim360$ dark matter particles anyway.

We find characteristic halo masses of $\log M_{50}\sim8.8$ and $\log M_{90}\sim9.3$, corresponding to 50\% and 90\% occupation fractions, respectively.
Our occupation fractions are broadly consistent with models that allow early, bursty star formation regulated by feedback such as in FIRE-2, SPHINX, THESAN, and Renaissance.
However, BonFIRE's $\log M_{50}$ and $\log M_{90}$ are slightly higher than those of the Renaissance simulations ($\log M_{50}\sim8$ and $\log M_{90}\sim8.5$), likely reflecting BonFIRE’s approximately order-of-magnitude worse resolution compared to Renaissance, as well as the more detailed treatment of reionization and Pop~III star formation in Renaissance \citep{Xu2016}.

The right panel of Figure~\ref{fig:HOF} shows the halo occupation fraction in all three simulations in the BonFIRE and CampFIRE suite at $z=9$.
Though all three simulations show similar shapes in this panel, there is an obvious resolution effect whereby the values of $\log M_{50}$ and $\log M_{90}$ decrease by about $0.4-0.8$ dex between each adjacent resolution level.
This indicates that halo occupation is sensitive to resolution, and that BonFIRE does not fully resolve the occupation of halos at $\Mhalo\lesssim10^9~\Msun$.
We correct for resolution dependence in the halo occupation fraction in Section~\ref{app:resampling}.

\section{Resolution Convergence Tests}\label{app:restests}

\begin{figure*}
    \centering
    \begin{tabular}{ccc}
    \subfigure{\includegraphics[width=0.32\textwidth]{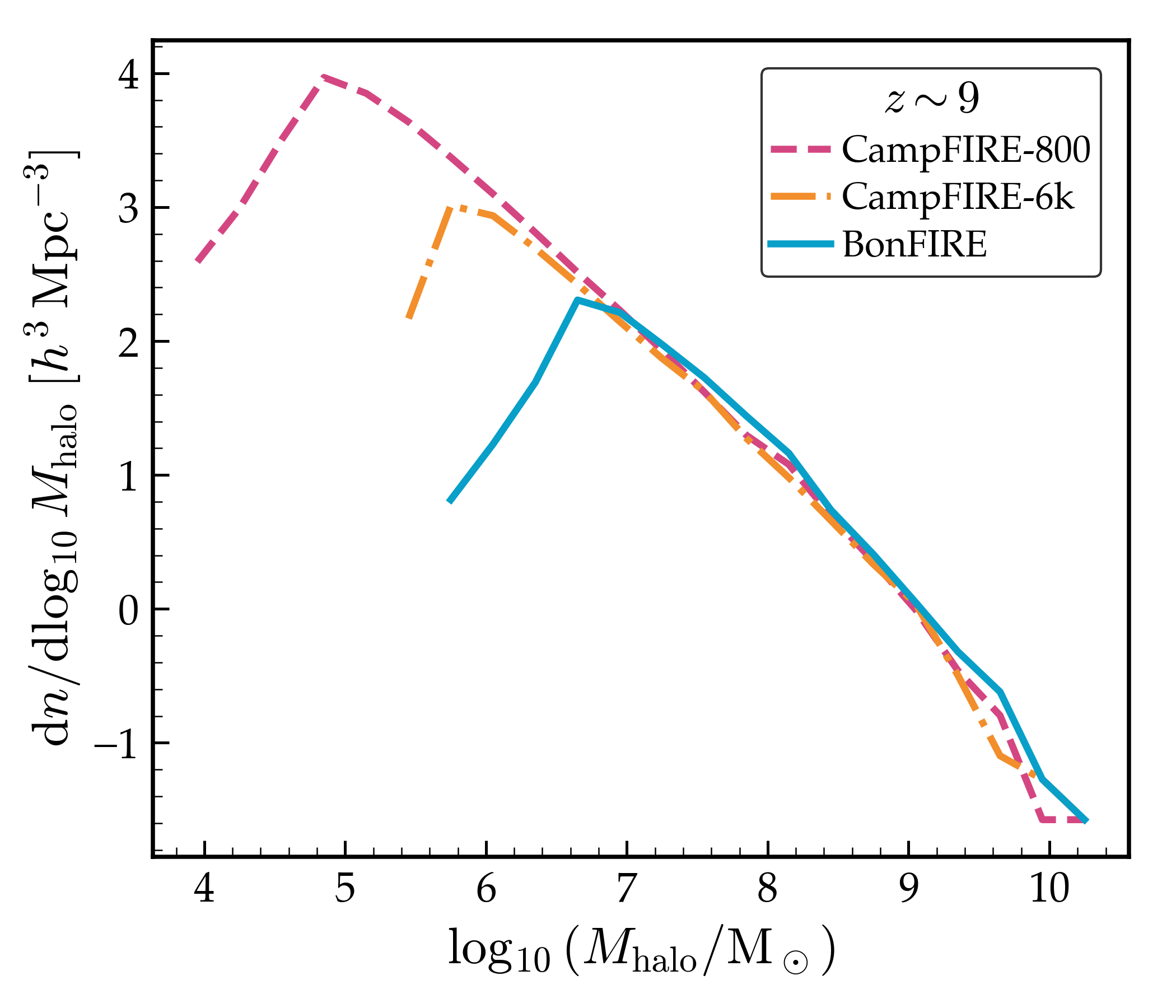}} &
    \subfigure{\includegraphics[width=0.32\textwidth]{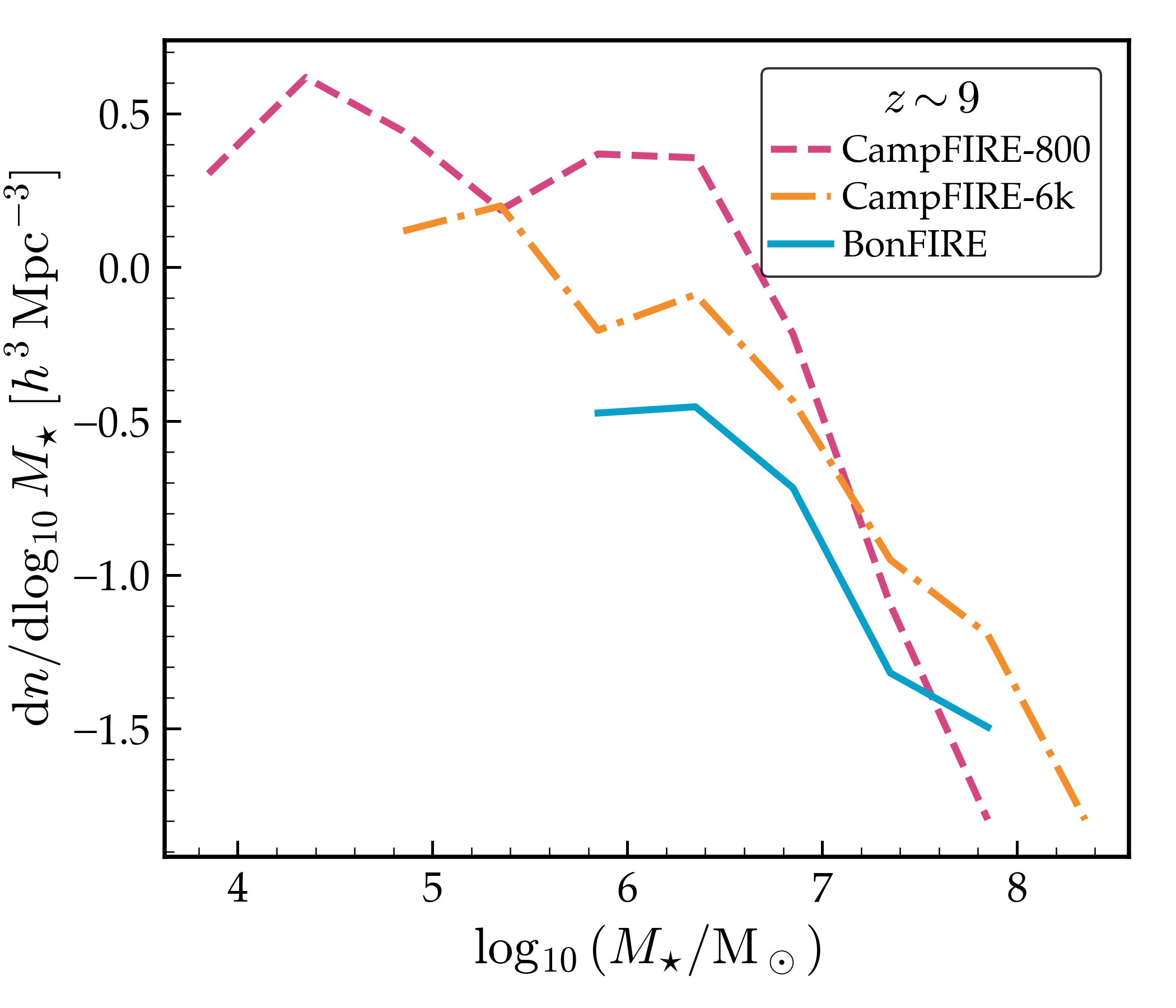}} &
    \subfigure{\includegraphics[width=0.32\textwidth]{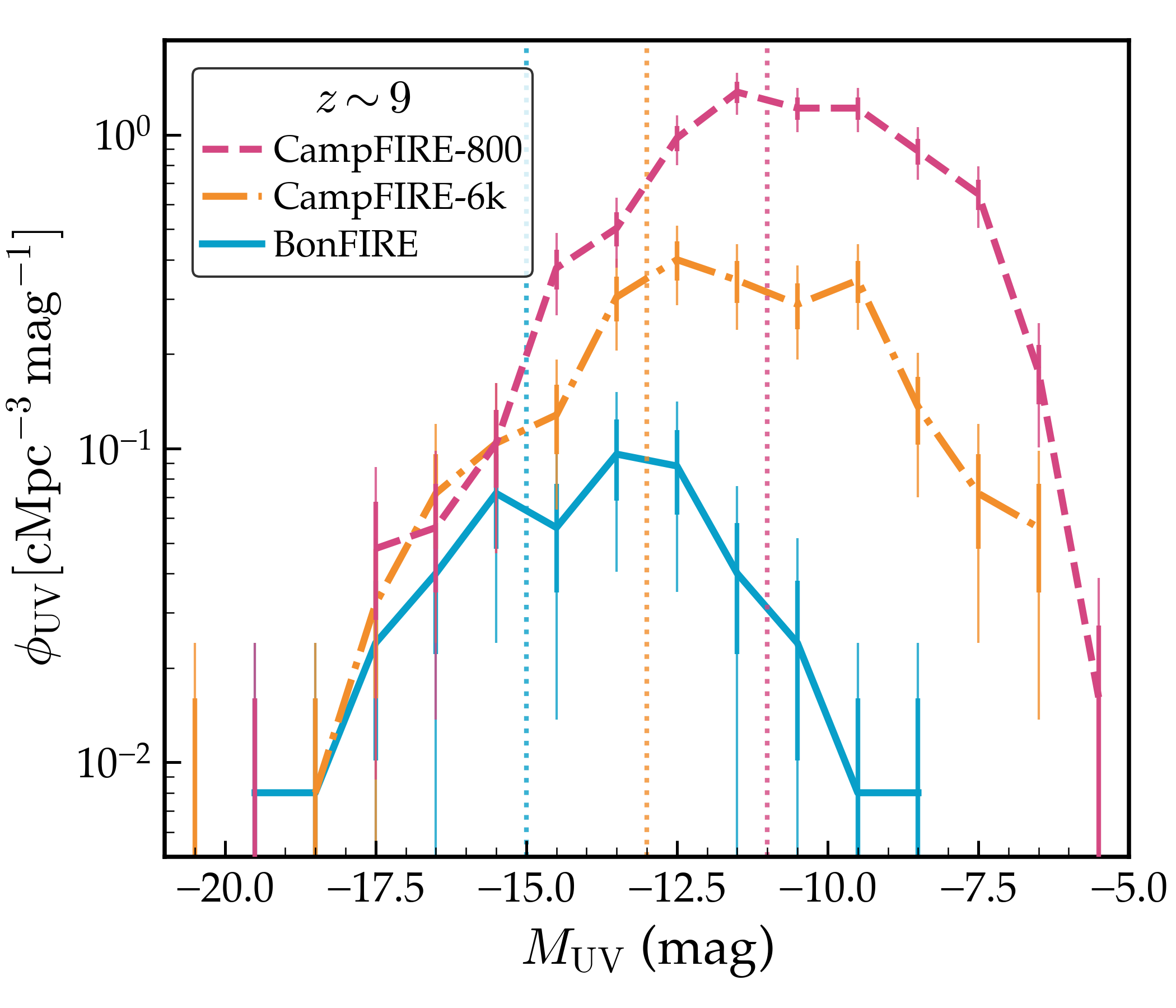}}
    \end{tabular}
    \caption{Resolution convergence tests for the halo mass function (HMF), stellar mass function (SMF), and UVLF in BonFIRE, CampFIRE-6k, and CampFIRE-800 at $z=9$. 
    We show results from only the subregion of BonFIRE corresponding to CampFIRE for a fair comparison.
    The HMF is well converged across the three simulations for $\Mhalo\gtrsim10^7~\Msun$.
    The SMFs reflect a complicated relationship between resolution and the buildup of stellar mass -- though BonFIRE and CampFIRE-6k are relatively converged at $\Mstar\gtrsim10^6~\Msun$, CampFIRE-800 has an excess of galaxies at $\Mstar\sim10^6~\Msun$ compared to them.
    The UVLFs show a clear dependence on resolution, whereby it modulates the normalization and turnover in the UVLF.
    We mark potential resolution convergence limits between the different simulations at $\Muv=-16, -13, and -10$.
    }
    \label{fig:res_test}
\end{figure*}

\begin{figure*}
    \centering
    \begin{tabular}{ccc}
    \subfigure{\includegraphics[width=0.32\textwidth]{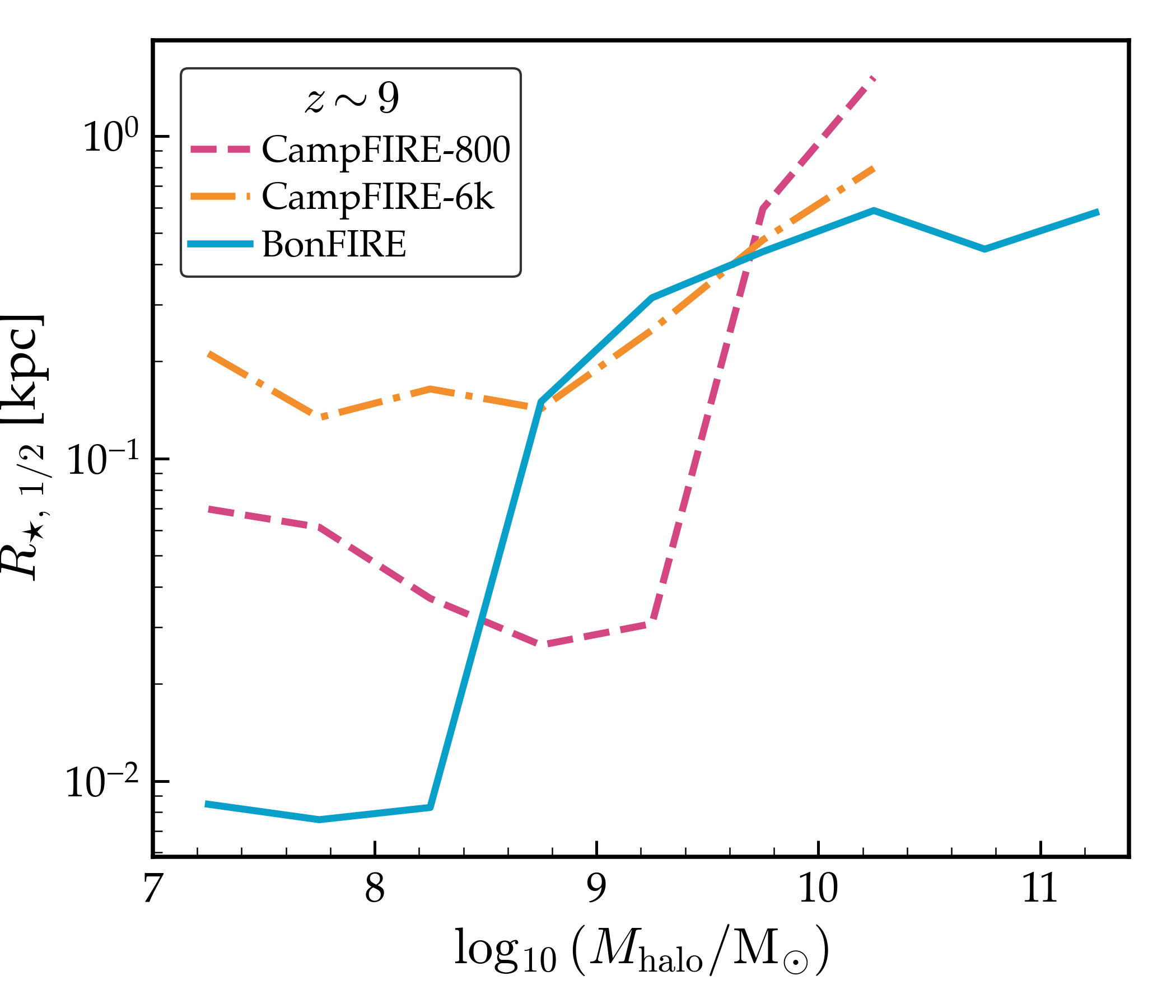}} &
    \subfigure{\includegraphics[width=0.32\textwidth]{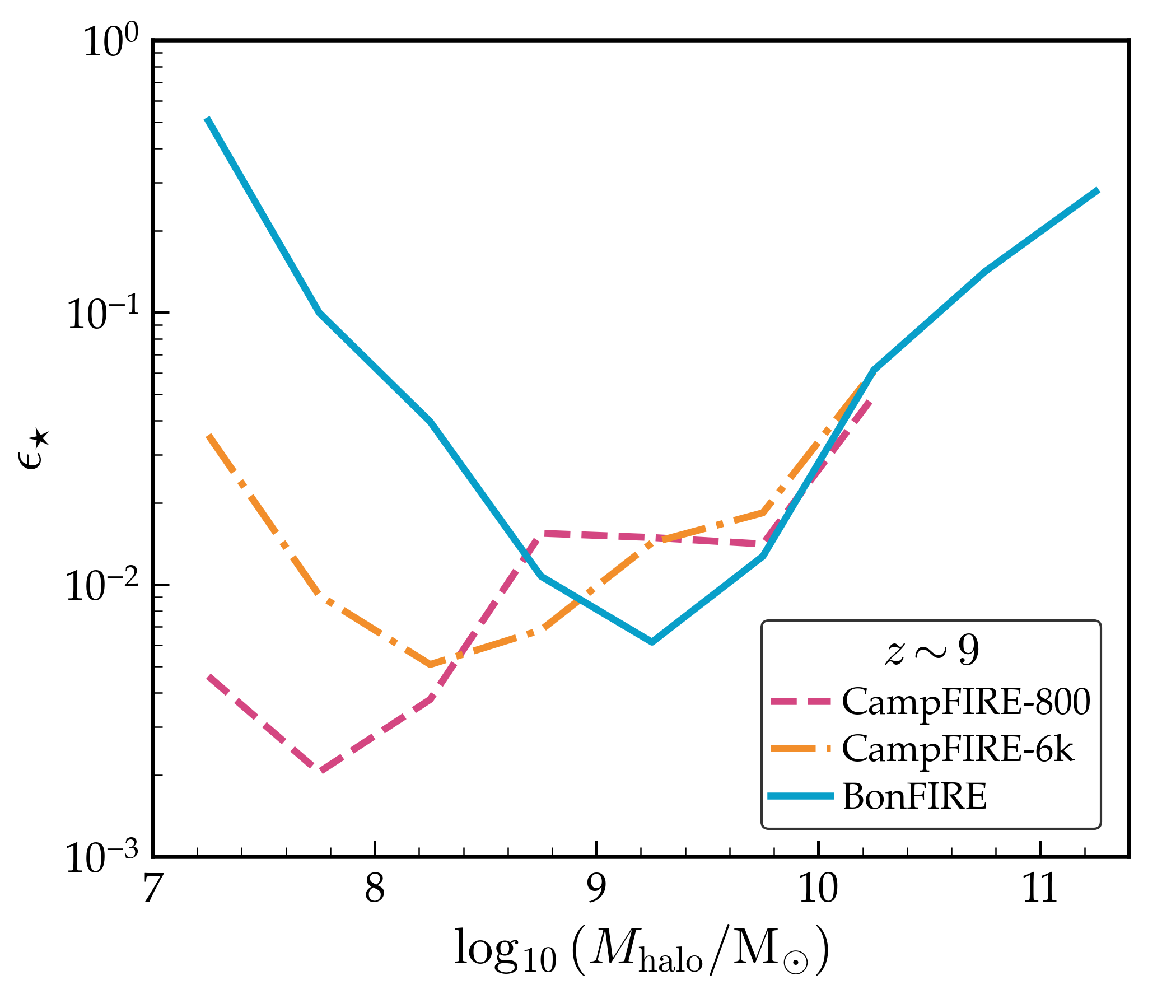}} &
    \subfigure{\includegraphics[width=0.32\textwidth]{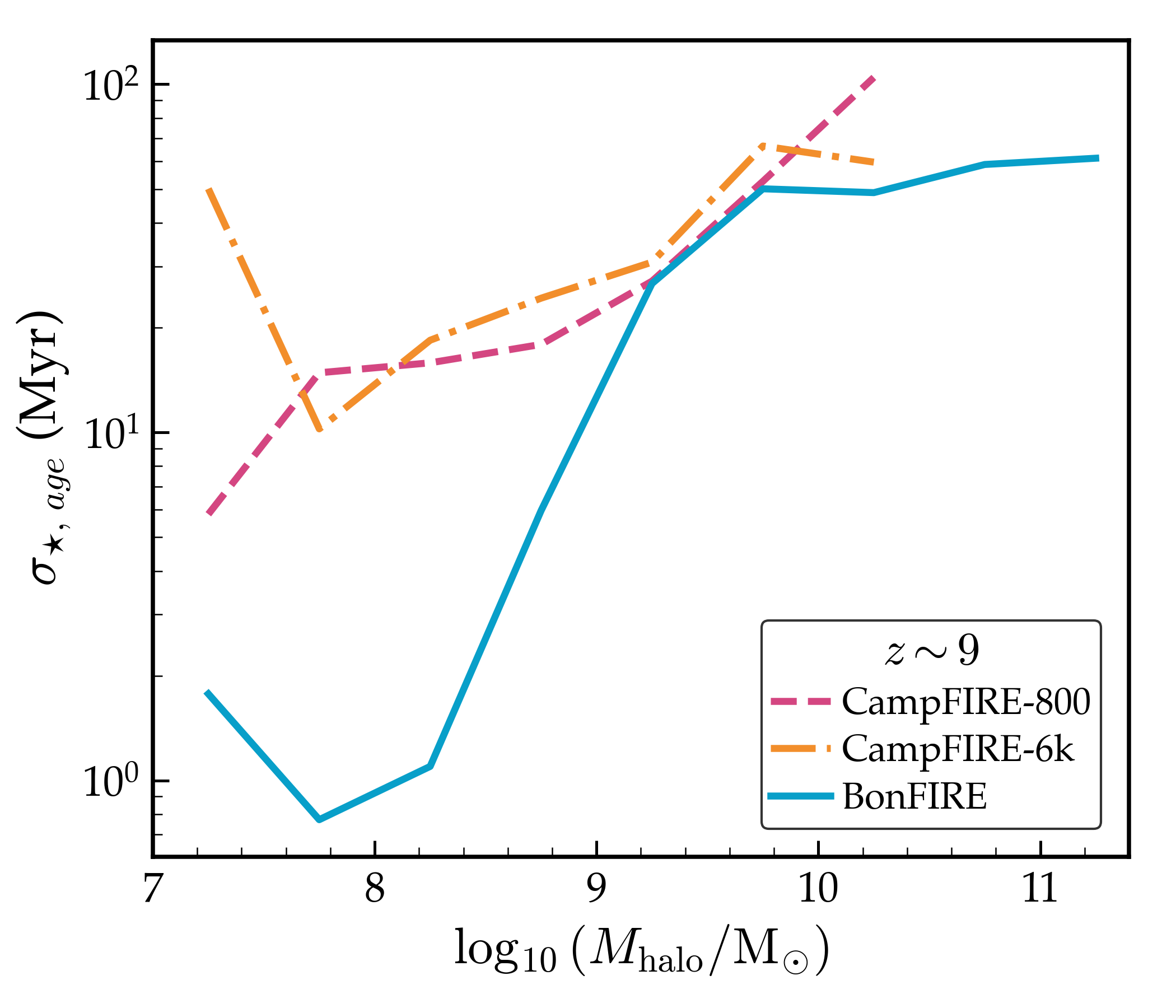}}
    \end{tabular}
    \caption{Resolution convergence tests for more nuanced galaxy properties in BonFIRE, CampFIRE-6k, and CampFIRE-800 at $z=9$.
    The size of galaxies as measured by their median 3D stellar half-mass radii ($R_{\star,~1/2}$, left) shows clear resolution dependence in BonFIRE for halos at $\Mhalo\lesssim10^{8.5}~\Msun$, such that the galaxy sizes tend towards the stellar force softening limit (8 pc).
    Whereas, galaxy sizes in the CampFIRE runs do not fall down to their respective numerical limits, and instead show smaller sizes with increasing resolution, the expected behavior.
    The median halo-scale star formation efficiency ($\sfe$, center) shows clear resolution dependence across all three simulations: there is an obvious increase in $\sfe$ at lower halo masses in each case, and the halo mass where the uptick occurs decreases with increasing resolution.
    The median age dispersion of galaxies ($\sigma_{\star,~age}$, right) shows two effects of resolution: the most prominent is the sharp decrease in $\sigma_{\star,~age}$ in BonFIRE halos at $\Mhalo\lesssim10^9~\Msun$ reflecting the formation of galaxies in low-mass halos in approximately single bursts of star formation last only $\sim1$ to a few Myr. 
    The secondary effect is an uptick in $\sigma_{\star,~age}$ in both BonFIRE and CampFIRE-6k at $\Mhalo\lesssim10^8~\Msun$, which possibly reflects staggered, unresolved star formation histories in the lowest-mass halos.
    }
    \label{fig:res_test2}
\end{figure*}

Figures~\ref{fig:res_test} and \ref{fig:res_test2} show six diagnostics for numerical convergence in galaxy formation across BonFIRE, CampFIRE-6k, and CampFIRE-800. 
In Figure~\ref{fig:res_test} we compare the halo mass functions, stellar mass functions, and UVLFs for each simulation at $z=9$, limiting the data used from BonFIRE to that which lies within the CampFIRE subregion for direct comparison.

In the left panel of Figure~\ref{fig:res_test}, we show all halos identified by Rockstar in each simulation. 
The halo mass function in BonFIRE is consistent with the other two CampFIRE runs down to $\Mhalo\approx10^7~\Msun$, indicating resolution convergence in the number of halos down to this mass scale across all three runs. 
The turnovers in the CampFIRE-6k and CampFIRE-800 HMFs at $\Mhalo\approx10^6~\Msun$ and $\Mhalo\approx10^5~\Msun$, respectively, are also consistent with our expectations given the (approximately) order of magnitude increase in particle mass resolution in each simulation. 
Combining this result with the HOFs discussed in the previous Appendix section, we conclude that the occupation of halos by galaxies down to the limit of one star particle in BonFIRE is reasonably well-resolved.

In the middle and right panels of Figure~\ref{fig:res_test}, we show galaxy stellar mass and UV magnitude in each simulation within the CampFIRE subregion. 
The middle panel shows the stellar mass functions are not as clearly converged as the HMFs, reflecting the nuanced effects of resolution in galaxy formation. 
The right panel shows some possible resolution limits identifiable in the UVLFs. 
BonFIRE converges with the two CampFIRE simulations up to $\Muv\sim-15$, and CampFIRE-6k converges with CampFIRE-800 up to $\Muv\sim-13$. 
By extrapolation, we identify $\Muv\sim-11$ as a possible resolution convergence limit in the UVLF for CampFIRE-800. 
We assume that our resampled UVLFs for BonFIRE are thus resolved up to $\Muv\sim-10$, given that they are resampled from CampFIRE-800. 
We opt to show our UVLFs in the main text up to $\Muv\sim-5$ to demonstrate the full limits of the simulations and provide a potential benchmark for future studies of resolution effects in the UVLF.

In Figure~\ref{fig:res_test2} we show the effects of resolution on three additional galaxy properties which are central to the results of this work.
We show results for all galaxies within BonFIRE rather than limiting to the CampFIRE subregion volume.
The left panel shows the median stellar half-mass radius of galaxies versus halo mass, where we find that the limited baryonic resolution of BonFIRE leads to an underestimation in galaxy sizes in halos at $\Mhalo\lesssim10^{8.5}~\Msun$.
The middle panel shows the median halo-scale integrated star formation efficiency (SFE), $\epsilon_{\star}\equiv \Mstar/(f_{\rm b}~\Mhalo)$, of galaxies versus halo mass.
Although each simulation shows a rise in SFE at low halo masses, this effect sets in at a larger halo mass in BonFIRE ($\Mhalo\lesssim10^{8.5}~\Msun$) compared to CampFIRE-6k and CampFIRE-800 ($\Mhalo\lesssim10^{8}~\Msun$).
This leads to an overestimation of SFE in low-mass BonFIRE halos of up to an order of magnitude.
Finally, the right panel shows the age dispersion of stars in galaxies as a function of halo mass.
We calculate age dispersion, $\sigma_{\star,~age}$, of a galaxy as the standard deviation of stellar ages for all stars assigned to the galaxy.
The age dispersion of BonFIRE galaxies in halos at $\Mhalo\lesssim10^{9}~\Msun$ are underestimated by up to an order of magnitude compared to their CampFIRE counterparts.
This underestimation shows that low-resolution simulations may form galaxies at low stellar masses too quickly, leading also to overestimates of the UV luminosities of these galaxies as the UV luminosity is highly sensitive to the age of the stellar population.

Overall, at $\Mhalo\lesssim10^{8.5}~\Msun$, BonFIRE galaxies exhibit systematically enhanced star-formation efficiencies, overly compact stellar sizes, and extremely narrow stellar age distributions compared to their counterparts in the higher-resolution CampFIRE-800 and CampFIRE-6k runs.
These trends highlight that BonFIRE alone cannot fully capture the multiphase ISM structure or cluster-scale fragmentation that regulate star formation in low-mass halos, leading to biased estimates of UV luminosities and galaxy morphologies in this regime.
To recover the numerically converged distribution of stellar ages, star-formation efficiencies, and galaxy sizes at the faint end, we therefore resample each low-mass BonFIRE galaxy using empirical correction functions derived from CampFIRE-800, which provides the necessary resolution to robustly model galaxy formation down to $M_{\rm halo}\sim10^{7},{\rm M_\odot}$.
This resampling procedure ensures that the resulting BonFIRE UVLF and associated population statistics reflect converged, physically meaningful predictions across the full halo mass range relevant for reionization.
We describe the resampling procedure in detail in the next section.

\section{Resampling BonFIRE Stellar Mass and UV Luminosity from CampFIRE-800}
\label{app:resampling}

We implement a non-parametric method to reconstruct the conditional distribution of galaxy properties in a low-resolution cosmological simulation using information from a higher-resolution subregion of the same volume. 
Our method combines conditional selection, bias weighting to correct for large-scale overdensity, and local density estimation. 
We first resample stellar mass conditioning on halo mass, and then resample UV magnitude conditioning on halo mass and stellar mass.

\subsection{Setup and Conditional Selection}

We let $M_{\mathrm{h}}$ denote halo mass, $\mathbf{x} = (x_1, x_2, \dots, x_K)$ a vector of conditioning variables, and $y$ the galaxy property to be resampled. We define high- and low-resolution datasets as
\begin{align}
\mathcal{D}{\mathrm{high}} &= { (M{\mathrm{h},i}^{(h)}, \mathbf{x}i^{(h)}, y_i^{(h)}) }{i=1}^{N_h}, \
\mathcal{D}{\mathrm{low}} &= { (M{\mathrm{h},j}^{(l)}, \mathbf{x}j^{(l)}, y_j^{(l)}) }{j=1}^{N_l}.
\end{align}

For each halo $j$ in the low-resolution sample, we define a tolerance-based subset of high-resolution halos,
\begin{equation}
S_j = \left\{ i \;:\; \left| x_{k,i}^{(h)} - x_{k,j}^{(l)} \right| < \epsilon_k \;\; \forall k \right\},
\end{equation}
where $\epsilon_k$ are fixed tolerances for each conditioning variable. 
In practice, we adopt $\epsilon_k = 0.25$ dex for all variables. 
If $S_j$ is empty, we relax the conditioning by dropping higher-order variables (e.g., stellar mass when resampling UV magnitudes).

\subsection{Bias Correction}

The CampFIRE subregion is overdense relative to the full BonFIRE volume, so we apply a bias correction to account for enhanced halo abundance. We assign each high-resolution halo a weight
\begin{equation}
w_i = \frac{1}{1 + b(M_{\mathrm{h},i}^{(h)}) , \delta},
\end{equation}
where $b(M_{\mathrm{h}})$ is the linear halo bias \citep{Tinker2010} and $\delta$ is the mean overdensity of the region. We normalize the weights within each subset $S_j$,
\begin{equation}
\tilde{w}i = \frac{w_i}{\sum{i \in S_j} w_i},
\end{equation}
so that $\sum_{i \in S_j} \tilde{w}_i = 1$.

\subsection{Halo Occupation Fraction Correction}

Finite resolution suppresses the fraction of halos hosting resolved galaxies in BonFIRE relative to CampFIRE-800. We model the halo occupation fraction as
\begin{equation}
f_{\rm occ}(M_{\rm h}) = \frac{N_{\rm gal}(M_{\rm h})}{N_{\rm halo}(M_{\rm h})},
\end{equation}
which we measure in consistent halo mass bins in both simulations. We then define the correction factor as
\begin{equation}
R_{\rm occ}(M_{\rm h}) = \frac{f_{\rm occ}^{(h)}(M_{\rm h})}{f_{\rm occ}^{(l)}(M_{\rm h})}.
\end{equation}

For each BonFIRE halo that does not host a resolved galaxy, we perform a stochastic activation step. We promote an unoccupied halo of mass $M_{\rm h}$ to host a galaxy with probability
\begin{equation}
p_{\rm act}(M_{\rm h}) =
\frac{f_{\rm occ}^{(h)}(M_{\rm h}) - f_{\rm occ}^{(l)}(M_{\rm h})}
{1 - f_{\rm occ}^{(l)}(M_{\rm h})},
\end{equation}
so that the corrected occupation fraction matches that of the high-resolution sample. We then assign galaxy properties to activated halos using the resampling procedure described below.

\subsection{Conditional Density Estimation and Sampling}

For each subset $S_j$, we estimate the conditional probability density function of $y$ using a weighted Gaussian kernel density estimator,
\begin{equation}
\hat{P}(y ,|, \mathbf{x}j^{(l)}) =
\frac{1}{Z_j} \sum{i \in S_j} \tilde{w}_i ,
K_h\left(y - y_i^{(h)}\right),
\end{equation}
where $K_h$ is a Gaussian kernel with bandwidth $h = \alpha \sigma_y$, with $\alpha = 0.2$ and $\sigma_y$ the standard deviation of $y$ within $S_j$. We perform the KDE in logarithmic space for stellar mass and in flux space for UV luminosity.

We obtain the resampled property by drawing
\begin{equation}
y_j^{(l,\mathrm{resampled})} \sim \hat{P}(y ,|, \mathbf{x}_j^{(l)}).
\end{equation}

To improve computational efficiency, we discretize the conditioning variables onto a coarse grid and reuse KDEs for halos within the same grid cell.

\subsection{Sequential Resampling}

We apply the resampling procedure sequentially. First, we resample stellar masses conditioning only on halo mass. We then resample UV magnitudes conditioning on both halo mass and the resampled stellar mass. This yields
\begin{equation}
P(y_1, y_2 ,|, M_{\rm h}, \mathbf{x})
= P\left(y_1 ,|, M_{\rm h}, \mathbf{x}\right),
P\left(y_2 ,|, M_{\rm h}, \mathbf{x}, y_1\right),
\end{equation}
and preserves correlations between galaxy properties while incorporating the bias correction through the weights $w = [1 + b(M_{\rm h})\delta]^{-1}$. If the second-stage conditioning fails, we fall back to $P(y_2|M_{\rm h},\mathbf{x})$.

The resulting corrected low-resolution catalog,
\begin{equation}
\mathcal{D}_{\mathrm{low}}^{\mathrm{corr}} =
\left\{ (M_{\mathrm{h},j}^{(l)}, \mathbf{x}_j^{(l)}, y_j^{(l,\mathrm{resampled})}) \right\},
\end{equation}
reproduces the conditional distributions of galaxy properties in the high-resolution sample while correcting for both large-scale bias and resolution-dependent selection effects.

\section{Dust correction}\label{app:dust}

\begin{figure}
\centering
\includegraphics[width=0.45\textwidth]{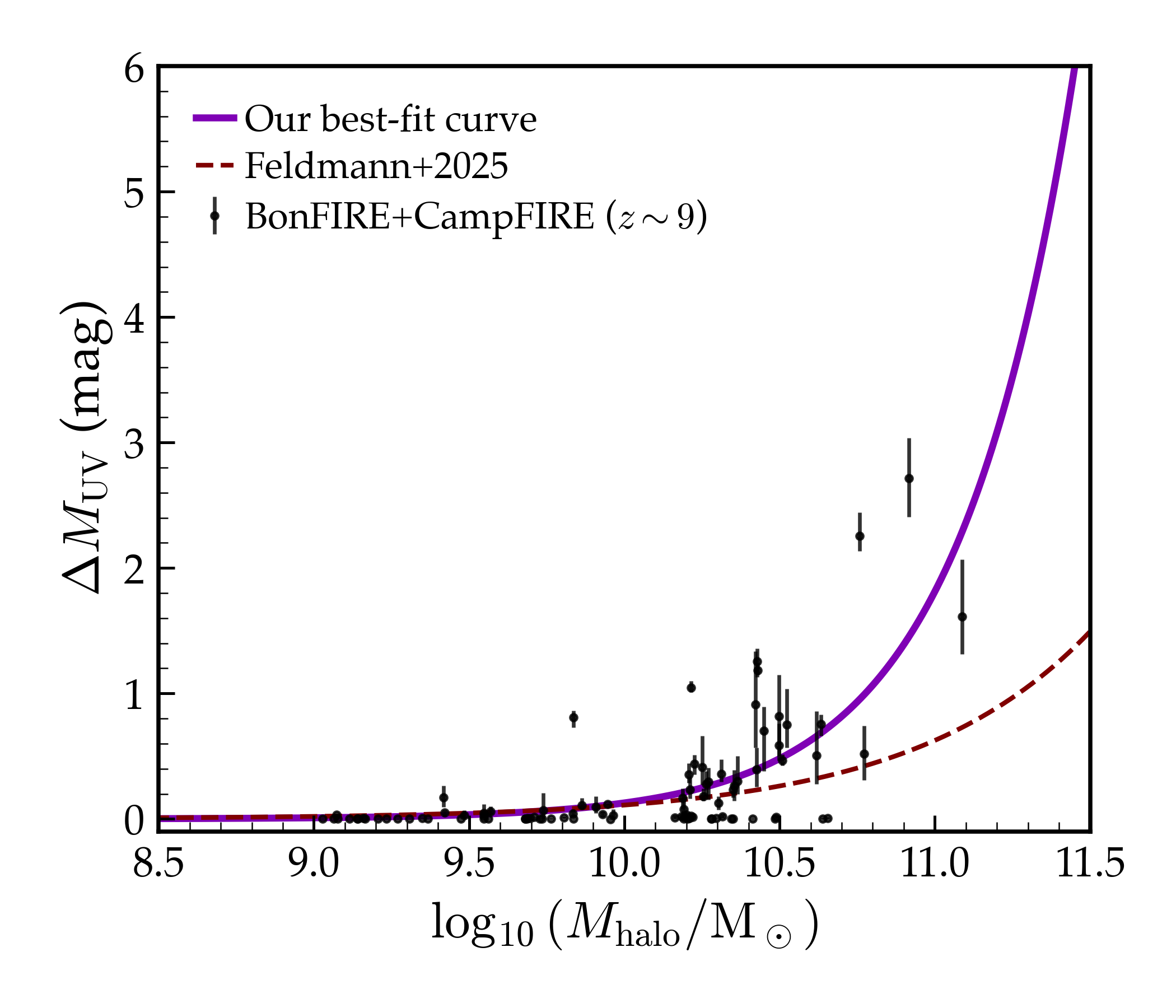}
\caption{Dust attenuation in the UV, $\Delta \Muv$, as a function of halo mass at $z\sim9$ from \textsc{SKIRT} radiative transfer calculations of BonFIRE galaxies (points). 
The solid curve shows our best-fit exponential model, while the dashed curve shows the fit from \citet{Feldmann2025}. 
Our model predicts a steeper rise in attenuation with halo mass and a lower characteristic mass scale for the onset of significant dust extinction.}
\label{fig:dust}
\end{figure}

We estimate the effect of dust attenuation on the UV luminosities of BonFIRE galaxies using radiative transfer calculations with \textsc{SKIRT}. 
We post-process a sample of $\sim100$ galaxies at $z=9$ spanning $\Mhalo\sim10^{9}$–$10^{10},\Msun$, including an additional set of halos selected to more uniformly sample this mass range. 
In these calculations, we assume a constant dust-to-metal mass ratio of 0.2 and adopt the \citet{Weingartner2001} dust model with a Milky Way grain size distribution. 
Figure~\ref{fig:dust} shows the resulting attenuation, expressed as $\Delta \Muv$, as a function of halo mass. 
The attenuation increases steeply with halo mass, reaching $\Delta \Muv\gtrsim2$ mag for $\Mhalo\gtrsim10^{11},\Msun$, while remaining negligible below $\Mhalo\lesssim10^{9.5},\Msun$. 
We fit these results with an exponential function of the form
\begin{equation}
\Delta M_{\rm UV} = 1.03~ \mathrm{exp}\left[\frac{\log \Mhalo - 10.79}{0.38}\right],
\end{equation}
which provides a good description of the median trend and is shown as the solid curve in Figure~\ref{fig:dust}. 
Compared to the model of \citet{Feldmann2025} (dashed curve), our fit exhibits a steeper rise and a lower characteristic mass scale for the onset of significant attenuation. 
We emphasize that this correction depends only on halo mass and neglects variations in dust geometry and galaxy properties; it should therefore be regarded as a simplified, empirical prescription.

\section{Alternative UVLF fits}\label{app:fits}

We also explored a generalized turnover model in which the suppression factor is parameterized as 
$[1+10^{0.4\beta(\Muv-M_{\rm turn})}]^{-1}$, allowing the sharpness of the faint-end turnover to vary. 
The addition of the shape parameter $\beta$ leads to fits that turn over too sharply compared to our binned data, and we further assess its effects on our fits using standard information criteria. 
We find that the data do not generally favor the more flexible model: at most redshifts, the improvement in goodness-of-fit is insufficient to offset the penalty for the additional parameter, with $\Delta\mathrm{AIC}<0$ and $\Delta\mathrm{BIC}<0$ indicating a preference for the simpler $\beta=1$ model. 
However, at $z\sim14$, the free-$\beta$ model is strongly preferred, with $\Delta\mathrm{AIC}\approx12$ and $\Delta\mathrm{BIC}\approx11$, suggesting a statistically significant deviation from our fiducial turnover shape of $\beta=1$. 
At $z\sim12$, the evidence for a better fit with the free-$\beta$ model is more marginal, with moderate improvements in both AIC and BIC. 
These results indicate that while a fixed-turnover shape provides an adequate description of the UV luminosity function across most redshifts probed in this work.

Briefly, we explored fitting other functional forms to our data, namely a double power law and also a modified Schechter function from \citealt{Bouwens2017,Feldmann2025}. 
As the double power law cannot capture the faint-end turn over in our UVLFs, it produced poor fits in general unless we disregarded a large fraction of our galaxy sample. 
However, the modified Schechter function includes a multiplicative term to model the faint-end turnover of the form $10^{-0.4\delta(\Muv+16)^2}$. 
We ran our same fitting pipeline with the modified Schechter function and found reasonable fits to our data in all but the highest redshift stack, but when comparing fits, our Schechter function with turnover (Equation~\ref{eq:schechter}) returned significantly lower AIC in all redshift stacks, and thus provides a better fit to our data in general.

\bibliography{references}{}
\bibliographystyle{aasjournalv7}


\end{document}